\documentclass[aps,prd,twocolumn,amsfonts,amssymb,amsmath,showpacs,letterpaper,superscriptaddress,nofootinbib,altaffilletter,fontenc]{revtex4-2}
\usepackage[utf8]{inputenc}
\usepackage[T1]{fontenc}

\usepackage{array}
\usepackage{graphicx}	% Including figure files
\usepackage{caption}
\usepackage{subcaption}
\usepackage{amsmath}	% Advanced maths commands
\usepackage{amssymb}	% Extra maths symbols
\usepackage{tabularx}
\usepackage{enumerate}
\usepackage{enumitem}
\usepackage{comment}
\usepackage{multirow}
\usepackage{float}
\usepackage[citecolor=blue,linkcolor=blue,colorlinks=true,urlcolor=blue]{hyperref}
\usepackage[misc]{ifsym}
\usepackage{orcidlink}

\usepackage{acronym}
\usepackage[normalem]{ulem}
\newacro{adm}[ADM]{Arnowitt-Deser-Misner}
\newacro{bbh}[BBH]{binary black hole}
\newacro{bh}[BH]{black hole}
\newacroplural{bh}[BHs]{black holes}
\newacro{bhns}[BHNS]{black hole-neutron star}
\newacro{bhpt}[BHPT]{black hole perturbation theory}
\newacro{bns}[BNS]{binary neutron star}
\newacro{bf}[BF]{Bayes' factor}
\newacro{cbc}[CBC]{compact binary coalescence}
\newacro{ce}[CE]{Cosmic Explorer}
\newacro{da}[DA]{data analysis}
\newacro{et}[ET]{Einstein Telescope}
\newacro{eob}[EOB]{Effective-One-Body}
\newacro{eom}[EOM]{equations of motion}
\newacro{fd}[FD]{frequency domain}
\newacro{fft}[FFT]{Fast Fourier transform}
\newacro{gw}[GW]{gravitational-wave}
\newacroplural{gw}[GWs]{Gravitational-waves}
\newacro{gr}[GR]{general relativity}
\newacro{grb}[GRB]{gamma-ray burst}
\newacro{grhd}[GRHD]{general-relativistic hydrodynamics}
\newacro{gwosc}[GWOSC]{Gravitational Wave Open Science Center}
\newacro{gwtc1}[GWTC-1]{the first gravitational-wave transients catalog}
\newacro{gsf}[GSF]{Gravitational Self Force}
\newacro{hm}[HM]{Higher mode}
\newacroplural{hm}[HMs]{Higher modes}
\newacro{ifo}[IFO]{interferometer}
\newacro{imr}[IMR]{inspiral-merger-ringdown}
\newacro{im}[IMR]{inspiral-to-merger}
\newacro{kagra}[KAGRA]{Kamioka Gravitational Wave Detector}
\newacro{ligo}[LIGO]{Laser Interferometer Gravitational-Wave Observatory}
\newacro{lisa}[LISA]{Laser Interferometer Space Antenna}
\newacro{lr}[LR]{Light Ring}
\newacro{lso}[LSO]{Last Stable Orbit}
\newacro{lvc}[LVC]{LIGO-Virgo Collaboration}
\newacro{lvk}[LVK]{LIGO-Virgo-Kagra Collaboration}
\newacro{lo}[LO]{leading order}
\newacro{ns}[NS]{neutron star}
\newacroplural{ns}[NSs]{neutron stars}
\newacro{nr}[NR]{numerical relativity}
\newacro{nqc}[NQCs]{Next-to-quasicircular corrections}
\newacro{nlo}[NLO]{next-to-leading order}
\newacro{nnlo}[NNLO]{next-to-next-to-leading order}
\newacro{n3lo}[N3LO]{next-to-next-to-next-to-leading order}
\newacro{n4lo}[N3LO]{next-to-next-to-next-to-next-to-leading order}
\newacro{ode}[ODE]{Ordinary Differential Equation}
\newacroplural{ode}[ODEs]{Ordinary Differential Equations}
\newacro{pn}[PN]{post-Newtonian}
\newacro{pm}[PM]{post-Minkowskian}
\newacro{pe}[PE]{parameter estimation}
\newacro{psd}[PSD]{power spectral density}
\newacro{pa}[PA]{post-adiabatic}
\newacro{qnm}[QNM]{quasi-normal mode}
\newacro{qc}[QC]{quasi-circular}
\newacro{qk}[QK]{quasi-Keplerian}
\newacro{snr}[SNR]{signal-to-noise ratio}
\newacro{spa}[SPA]{stationary-phase approximation}
\newacro{sxs}[SXS]{Simulating eXtreme Spacetimes}
\newacro{td}[TD]{time domain}
\newacro{ng}[NG]{Nect Generation}
\newacro{rift}[\texttt{RIFT}]{Rapid Iterative FiTting}

\definecolor{dodgerblue}{rgb}{0.12, 0.56, 1.0}

\definecolor{grassgreen}{RGB}{63, 155, 11}

\definecolor{tangerine}{RGB}{255, 148, 8}

\usepackage{ragged2e}
\begin{document}
%\linenumbers

\title{The Cost of Circularity: Quantifying Eccentricity-Induced Biases in Binary Black Hole Inference}

\author{Tamal RoyChowdhury }
\affiliation{Leonard E. Parker Center for Gravitation, Cosmology, and Astrophysics, University of Wisconsin–Milwaukee, Milwaukee, WI 53201, USA}

\author{V. Gayathri}
\affiliation{Leonard E. Parker Center for Gravitation, Cosmology, and Astrophysics, University of Wisconsin–Milwaukee, Milwaukee, WI 53201, USA}
\thanks{E-mail: gayathribuvana1991@gmail.com}
\author{Rossella Gamba}
\affiliation{Institute for Gravitation \& the Cosmos, Department of Physics \& Department of Astronomy and Astrophysics,
The Pennsylvania State University, University Park, Pennsylvania 16802, USA}
\affiliation{Department of Physics, University of California, Berkeley, CA 94720, USA}
\author{Shubhagata Bhaumik}
\affiliation{Department of Physics, University of Florida, PO Box 118440, Gainesville, FL 32611-8440, USA}
\author{Imre Bartos}
\affiliation{Department of Physics, University of Florida, PO Box 118440, Gainesville, FL 32611-8440, USA}
\author{Jolien Creighton}
\affiliation{Leonard E. Parker Center for Gravitation, Cosmology, and Astrophysics, University of Wisconsin–Milwaukee, Milwaukee, WI 53201, USA}
%\author{ABC}
%\affiliation{Leonard E. Parker Center for Gravitation, Cosmology, and Astrophysics, University of Wisconsin–Milwaukee, Milwaukee, WI 53201, USA}
 
%\date{}
%\pubyear{\the\year{}}

%\label{firstpage}

\begin{abstract}
Dynamically assembled binary black holes are expected to retain measurable orbital eccentricity in the LIGO-Virgo-KAGRA band, but most parameter estimation analyses still assume quasi-circular inspirals. This raises a critical question: how strongly does unmodeled eccentricity bias the inferred properties of BBH mergers? We address this by injecting eccentric signals generated with \textsc{TEOBResumS--Dal\'{\i}} and recovering them using the circular, precessing \textsc{IMRPhenomXPHM} waveform model. Across $20$-$80 \, M_\odot$ and eccentricities up to $e=0.5$, we find that circular waveform models remain reliable only for very small eccentricities. Above $e\sim0.2$ at 10 Hz, recovered masses, spins, inclination, and distances begin to show significant systematic offsets. Circular precessing templates mimic eccentric amplitude and phase modulations by introducing artificial precession, highlighting a major degeneracy between these effects. For high-mass, moderately eccentric mergers, circular models misestimate parameters at a level that would bias astrophysical interpretation and population studies. Our results establish the parameter-space boundaries where eccentric waveform models become essential for accurate inference in current and next-generation detectors.

%We investigate systematic effects in parameter estimation for eccentric binary black hole (BBH) systems using simulated gravitational-wave signals. The study focuses on sources with total masses in the range $20$--$80\,M_\odot$ and orbital eccentricities spanning $e=0$--$0.5$. Waveforms are generated with the \textsc{TEOBResumS-Dal\'i} model and subsequently analyzed using the \textsc{IMRPhenomXPHM} waveform approximant for recovery, both with and without spin-precession effects. These analyses are carried out using Bayesian inference, and we quantify biases in key astrophysical parameters, such as masses, spins, and distances, arising from the non-inclusion of the eccentric model. %\RG{The first part of this sentence reads a bit weird?}. 
%The results highlight the extent to which eccentricity and precession interplay impact the fidelity of parameter recovery, and provide insight into the robustness of current waveform models for upcoming gravitational-wave observations. Our findings underline the importance of improving waveform systematics to enable accurate characterization of eccentric BBH mergers in the era of third-generation detectors.% Tamal \RG{... of improving waveforms to avoid systematics and enable...}
\end{abstract}

\date[\relax]{Dated: \today }

\maketitle

\begin{comment}
    \begin{keywords}
gravitational waves -- eccentric binary black holes -- dynamical formation
\end{keywords}
\end{comment}

\section{Introduction}
\label{sec:intro}

The fourth observing run (O4) of Advanced LIGO \citep{TheLIGOScientific:2014jea}, Advanced Virgo \citep{TheVirgo:2014hva}, and KAGRA \citep{KAGRA:2018plz} recently concluded, with analysis {of the collected data ongoing.} Prior to this, three observing runs were successfully completed, resulting in $91$ confident detections of compact binary coalescences (CBCs) \citep{2019PhRvX...9c1040A,2021PhRvX..11b1053A,GWTC2.1,LIGOScientific:2025rsn,2023PhRvX..13d1039A}. From the first part of the {fourth observing run, named O4a,} $128$ new candidate events have been released, encompassing binaries with a wide range of masses and mass ratios, including previously unobserved types of binary systems \citep{GWTC-4.0}. Among these confirmed detections, we have detected two intermediate mass black hole binary systems, namely GW190521 \citep{PhysRevLett.125.101102} and GW231123 \citep{LIGOScientific:2025rsn}.  The event GW231123 was detected in the first half of the fourth observation run and has a total mass of about $190-265\, M_{\odot}$, making it the most massive binary black hole observed through GWs to date. We expect to see more of these types of extraordinary binaries in the current and future observing runs \citep{KAGRA:2013rdx,Kiendrebeogo:2023hzf}.

{The current parameter estimation (PE) analyses in the LIGO-Virgo-KAGRA (LVK) collaboration have significantly improved our understanding of the astrophysical properties of CBCs. The fourth observing run has yielded more precise measurements of component masses, spins, luminosity distances, and orbital orientations by incorporating advanced waveform models and Bayesian inference techniques \cite{2023PhRvX..13d1039A, GWTC2.1,GWTC-4.0}. They now allow exploration beyond the quasi-circular approximation. The improvements in waveforms have enabled the study of various subdominant effects such as orbital eccentricity, spin precession, and deviations from general relativity.  Such extensions are crucial for avoiding biases in parameter recovery. They also help to better constrain the astrophysical formation scenarios that are expected to produce compact binaries.}

%These improvements are very important for parameter estimation algorithms now to explore beyond the quasi-circular approximation, allowing for the inclusion of subdominant effects such as orbital eccentricity and spin precession. Such extensions are essential to avoid biases in parameter recovery and to better constrain the astrophysical formation scenarios of compact binaries.}

{Dedicated eccentric binary black hole (eBBH) searches have been performed in the first, second, and third observing runs \citep{Abbott_2019_ebbh, O3_eBBH_Collab}.  These studies did not report any confident detections of binaries with non-negligible orbital eccentricity. This work assumes that all binaries have circular orbits and estimates their parameters accordingly \citep{2023PhRvX..13d1039A}. Several studies have been conducted to estimate the orbital eccentricity of detected Binary Black Hole (BBH) systems \citep{Iglesias:2022xfc,Gupte:2024jfe}. Notably, GW190521 has been identified as an eccentric binary based on results from analyses using different parameter estimation methods and waveform models \citep{GayathrieBBH,2020ApJ...903L...5R,2021arXiv210605575G}. Nonetheless, certain events show mild deviations from circularity, motivating the development and application of eccentric waveform families such as \textsc{SEOBNRv5EHM}~\cite{Gamboa:2024imd, Gamboa:2024hli} and \textsc{TEOBResumS-Dal\'i} \citep{Nagar2021_TEOBResumS,Nagar2024_TEOBResumS_Dali, albanesi2025effectiveonebodymodelinggenericcompact} for inferring properties of such events. These waveform models facilitate a joint inference of masses, spins, and eccentricities. This allows us to distinguish between dynamical formation channels and isolated binary evolution, as a non-zero eccentricity is a direct confirmation that the event originated via the dynamical formation channel \citep{Romero-Shaw:2019itr,PhysRevD.103.023015,Zevin:2021rtf}. Accurate eccentricity estimation remains challenging due to waveform degeneracies and the need for high signal-to-noise ratios.}

%Eccentricity offers a key probe of astrophysical origin: dynamical formation channels in globular clusters, galactic nuclei, Active galactic nuclei and hierarchical triple systems can preserve measurable eccentricity into the LIGO band, whereas isolated binary evolution predicts efficient circularisation long before merger \citep{Peters:1963ux,Antonini:2016gqe,Samsing:2017xmd}. Accurate eccentricity measurement therefore directly informs the relative contributions of isolated versus dynamical formation.

Eccentricity measurements have strong astrophysical significance because eccentric orbits are indicative of dynamical formation scenarios. Binaries formed through dynamical interactions in dense stellar environments, such as globular clusters, galactic nuclei, active galactic nuclei(AGN), or hierarchical triples undergoing Kozai-Lidov oscillations, can retain measurable eccentricities in the LIGO band \citep{Antonini:2016gqe, Samsing:2017xmd}. The AGN formation channel, characterized by dense environments, can produce eccentric binaries through binary-single interactions \cite{ Samsing:2020tda}. In the case of binaries formed through isolated stellar evolution, they are expected to circularize due to gravitational-wave emission long before merger \citep{Peters:1963ux}. Thus, a reliable detection of eccentricity can serve as a crucial probe of the relative contributions of various formation channels.

%{Both orbital eccentricity and spin-induced precession introduce complex amplitude and phase modulations in the gravitational-wave signal \cite{Apostolatos:1994mx, Smith:2012du}. Eccentricity leads to repeated bursts of radiation near periapsis and modifies the inspiral timescale, while precession arises from misaligned component spins, causing the orbital plane to wobble. Neglecting these effects in the parameter estimation of detected BBHs can lead to biased recovery of intrinsic parameters such as masses and spins.}
Both eccentricity and spin precession introduce complex amplitude and phase modulations in the waveform \cite{Apostolatos:1994mx, Smith:2012du}. %The resulting degeneracy between the two effects complicates interpretation, particularly for high-mass or moderate SNR events, where precession signatures can mimic features induced by mild eccentricity \cite{Huerta:2014eca}. This underscores the need for waveform models that incorporate both effects self-consistently.
{A notable challenge in such analyses is the degeneracy between eccentricity and precession. Certain precession-induced modulations in the waveform can mimic the features produced by moderate eccentricity, complicating their disentanglement, particularly for low-SNR events \cite{Huerta:2014eca} and high mass ones \cite{Romero_Shaw_2023}. Addressing this degeneracy requires the development of accurate waveform models that jointly incorporate both effects, as well as advanced statistical inference methods. A better understanding of these phenomena is key to improving the accuracy of parameter estimation and constraining the astrophysical origins of gravitational-wave sources.}

%However, waveform models employed in parameter estimation are predominantly calibrated to quasi-circular binaries, and deviations from these assumptions can lead to systematic biases in the inferred masses, spins, and luminosity distances~\cite{Divyajyoti:2023rht, Huez:2025npe, 2025arXiv251004332D}. The simultaneous presence of orbital eccentricity and spin precession further complicates the recovery of source parameters, potentially impacting astrophysical interpretations and population inferences. These considerations underscore the necessity of a comprehensive assessment of systematic effects in parameter estimation for eccentric binary black holes. Such an analysis enables a rigorous evaluation of the fidelity of current waveform models and informs the development of improved modeling techniques, which are essential for the accurate characterisation of eccentric BBH mergers in the era of advanced and third-generation gravitational-wave observatories.
%\RG{I think there needs to be also a paragraph dedicated to comparisons to previous (existing) work, saying what the novelty here is (more systematic study, \dots) \addnew{add new paragraph about this}}

Several studies have investigated how parameter-estimation pipelines behave when analyzing eccentric gravitational-wave signals using waveform models that assume quasi-circular orbits. Because circular templates cannot reproduce the characteristic modulation, higher harmonics, and phase evolution of eccentric binaries, {it leads to systematic biases in the recovery of parameters.} Simulation studies injecting eccentric signals and recovering them with circular waveform models (\cite{Huerta2017, Lower2018, RamosBuades2020, OSheaKumar2021,2025arXiv251004332D}) show that the inferred masses are typically biased toward lower mass ratios, spins are often misestimated -- sometimes interpreted as artificial anti-alignment, and the luminosity distance is overestimated due to a loss in matched-filter efficiency. These analyses consistently find that even mild eccentricity ($e \geq 0.1$ at $10$ Hz) produces noticeable {deviations} in the posteriors, while moderate eccentricity can cause the sampler to converge on incorrect modes altogether. Such studies highlight that quasi-circular models remain reasonably robust only for very low residual eccentricity, and they emphasize the need for eccentric waveform models in PE pipelines to avoid systematic errors in source characterization.

{In this study, we investigate the systematic effects of orbital eccentricity on parameter estimation in the eBBH parameter space. We perform a series of equal mass non-spinning eccentric BBH injections across a range of masses and eccentricities, while keeping other parameters fixed, by using an eccentric waveform model. We use a quasi-circular waveform with spins for parameter recovery and do a qualitative study of the biases observed in the recovered parameters. We aim to find the parameter space in which the inclusion of eccentricity for parameter estimation is important and cannot be ignored.}

The paper is organized as follows. In Sec.~\ref{sec:detection_rate_formation} we discuss the detection prospects, astrophysical merger rates, and formation channels of eBBH systems.  In Sec.~\ref{sec:waveforms}, we describe the waveform models used for eBBH systems, including precessing, aligned-spin and no-spin configurations. Section~\ref{sec:bayesian_pe} provides an overview of the Bayesian parameter estimation methodology used in our analysis. Section~\ref{sec:results} presents our results, including the impact of eccentricity and spin on parameter recovery. Finally, Sec.~\ref{sec:conclusion} summarizes our conclusions and highlights prospects for future studies.

%\paragraph{similar work in other groups}
%\paragraph{motivation}

%\paragraph{summary about this work}

%\begin{itemize}
%    \item Introduce the latest parameter estimation results from the LIGO data release.
%    \item Discuss the PE results from eccentricity searches.
%    \item Motivation to understand the presence of eccentricity in GW signals (dynamical formation channels, globular clusters, triple systems, etc.)
%    \item Effects of eccentricity and precession on GW signals.
%    \item Degeneracy between eccentricity and precession and motivation to understand it.
    
%\end{itemize}

\section{Ebbh: detection, astrophysical rate and formation channels} \label{sec:detection_rate_formation}

% \begin{itemize}
%     \item formation channels - \textit{Notes}: dynamical vs isolated smoking gun. Discuss how other source parameters can be used but overlap. Different dynamical channels cite eccentric distribution papers. Rates for formation channels
%     \item Briefly discuss the objective detection - \textit{Notes}: Briefly discuss detection strategies so far (collab paper, pycbc paper), results on sensitivity vs eccentricity. \ToDo{Add discussion about GW190521. Precessing vs eccentric.}
%     \item astrophysical rate
% \end{itemize}

Detecting gravitational waves from eBBHs presents several unique challenges. The waveform for an eBBH merger can significantly deviate from the quasi-circular inspiral--merger--ringdown (IMR) models that are currently used in traditional matched-filter searches, particularly for lower masses and moderate to high eccentricities. Matched-filter pipelines, such as PyCBC \cite{Allen:2005fk, Usman:2015kfa, Nitz:2017svb} and GstLAL \cite{PhysRevD.95.042001, Sachdev:2019vvd, Joshi:2025nty}, that use aligned-spin, quasi-circular templates are therefore expected to have reduced sensitivity to eccentric mergers \citep{2010PhRvD..81b4007B}. There have been recent efforts to incorporate eccentricity into template bank searches for binary neutron star, neutron star-black hole mergers \citep{Dhurkunde:2023qoe, Phukon:2024amh}, and more recently, BBH systems \citep{Wang:2025yac}. Additionally, a recent study performed a targeted template based eccentric search for the neutron star-black hole binary GW200105 \citep{Phukon:2025cky}, a binary system for which several works have suggested the presence of significant residual eccentricity \citep{Morras:2025xfu, Tiwari:2025fua, Jan:2025fps}. However, these searches remain computationally expensive and are typically limited to moderate eccentricities.

As a result, eBBH detection efforts have increasingly relied on model-agnostic search algorithms that are less sensitive to waveform mismatches and can hence capture a broader range of signal morphologies. In particular, the coherent WaveBurst (cWB) algorithm \citep{Klimenko:2005xv, cWB_WDM, cWB_MaxL}, which is designed to identify coherent excess power across multiple detectors without strict reliance on waveform templates, is demonstrated to have comparable sensitivity in recovering signals from eBBH systems \citep{Abbott_2019_ebbh, O3_eBBH_Collab}. While cWB does not perform parameter estimation in the same way as Bayesian inference pipelines, it plays an important role in identifying signatures of eccentricity \citep{Das:2024zib} as well as constraining the merger rates \citep{Abbott_2019_ebbh, O3_eBBH_Collab}. Furthermore, injections of numerical relativity and semi-analytic eccentric waveforms into detector noise have demonstrated that such unmodeled searches can recover eBBH signals even when eccentricity is moderate to high \citep{Bhaumik:2024cec}.

Estimating the astrophysical merger rate of eccentric binary black holes remains a challenge, due to the lack of confirmed detections and the complexity of modeling such systems. Current estimates of the eBBH merger rate \citep{O3_eBBH_Collab} have been derived from the sensitivity of unmodeled search pipelines \citep{Biswas:2007ni} using data from the first three observing runs of LVK, with upper limits remaining several orders of magnitude higher than the rate inferred for circular binaries under the assumption that no eccentric binaries were detected \citep{Abbott_2019_ebbh}. These constraints place bounds on the population of highly eccentric systems and can be used to validate the predictions of various dynamical formation models by ruling out astrophysical models that predict mergers rates that are greater than the upper limits derived from the search sensitivities \citep{Abbott_2019_ebbh}. 

Several recent works have also focused on the detectability of eBBH mergers with current gravitational-wave detectors. Ref.~\citep{Singh:2025ojp} investigates the observability of eBBH mergers with current detector sensitivity in a population of simulated BBH mergers from globular cluster. Refs.~\citep{Zeeshan:2024ovp,Zeeshan:2026pga} show the effect of not including eccentricity in population inference and show no significant evidence of eccentricity in population inference using real events. %\TRC{Do we need to cite some paper here to address the comment?}

%\RG{Probably we should also cite ~\cite{Dhurkunde:2023qoe} somewhere right?}

\section{Waveforms for eBBH system} 
\label{sec:waveforms}

While quasi-circular binary systems have been extensively studied in both analytical and numerical frameworks thoughtout the years, 
their eccentric counterparts have received comparatively limited attention.
This disparity is largely due to the astrophysical expectation that eccentricity
is efficiently radiated away during the inspiral of comparable-mass
compact binaries~\cite{Peters:1963ux}.

% NR
%
From the \ac{nr} perspective, the volume of publicly available simulations of
eccentric binaries is modest. 
The \ac{sxs} collaboration has just recently released 
{more than $200$} new simulations of \acp{bbh} with initial eccentricities up to $\sim 0.7$~\cite{Ramos-Buades:2022lgf,SXS_Collab_2019, Scheel:2025jct}; 
the Rochester Institute of Technology(RIT) group has contributed
roughly eight hundred short-duration \ac{bbh} simulations~\cite{Healy:2022wdn}, which, while limited in length, 
are still informative for modeling the merger regime~\cite{Carullo:2023kvj, Carullo:2024smg}. 
Additional simulations include $\sim100$ cases from the \texttt{Maya} catalog~\cite{Ferguson:2023vta} and a
small number of eccentric \acp{bns} evolutions available through the CoRe database~\cite{Gonzalez:2022mgo}.
These figures contrast starkly with the thousands of quasi-circular (or quasi-spherical) \ac{bbh} simulations, including those with spin precession, highlighting the relatively under-explored nature of bound, non-circular configurations in \ac{nr}.

% PN
%
On the analytical side,early developments in \ac{pn} calculations date back to \cite{1976ApJ...210..764W}. The \ac{qk} parameterization formulated by Damour and Deruelle in the 1980s~\cite{1985AIHPA..43..107D} highlighted
foundational issues such as the non-uniqueness and gauge dependence of eccentricity in general
relativity. The \ac{qk} formalism has since been extended up to 4\ac{pn} order for
conservative dynamics~\cite{Damour:1988mr, 1993PhLA..174..196S, Memmesheimer:2004cv, Cho:2021oai}, and to 3\ac{pn} for radiation-reaction contributions~\cite{Placidi:2021rkh, Placidi:2023ofj, Gamboa:2024imd}.
However, fully analytical waveform models remain limited to the inspiral phase~\cite{Klein:2018ybm, Klein:2021jtd, Arredondo:2024nsl, Morras:2025nlp, Morras:2025nbp}, though merger-ringdown completions can be attached~\cite{Hinder:2017sxy, Cho:2021oai, Paul:2024ujx, Manna:2024ycx, Chattaraj:2022tay}.

% EOB
%
In the context of the \ac{eob} formalism, since
the effective Hamiltonian is based on the re-mapping of full \ac{pn}-expanded ADM results, non-circular contributions to the conservative dynamics have always been accounted for, starting from the original work by Buonanno and Damour~\cite{Buonanno:1998gg}.
Extending models to eccentric binaries, then, primarily involves the generalization of radiation-reaction forces and the related waveform multipoles. 
Initial work in this direction includes the studies by Bini and Damour~\cite{Bini:2012ji} and Hinderer and Babak~\cite{Hinderer:2017jcs}, the latter also proposing a reparameterization of the \ac{eob} dynamics in terms of orbital elements. More sophisticated models have since emerged, such as the SEOBNRE family~\cite{Cao:2017ndf, Liu:2021pkr, Liu:2023dgl}, \textsc{TEOBResumS--Dal\'{\i}}~\cite{Chiaramello:2020ehz, Nagar:2021gss, Nagar:2021xnh, Nagar:2023zxh}, and the SEOBNRv4EHM and SEOBNRv5EHM models~\cite{Ramos-Buades:2021adz, Gamboa:2024imd, Gamboa:2024hli}.
%
% Phenom
%
Phenomenological models have also begun development, providing a less accurate description of the waveform (with respect to \ac{eob}-based waveforms) in exchange for increased speed in waveform generation~\cite{Setyawati:2021gom, Planas:2025feq}.
Finally, \ac{nr} surrogates have been built from both \ac{eob}
waveforms as well as simulations, though for the moment mainly limited to non-spinning binary systems~\cite{Islam:2024tcs, Islam:2021mha, Islam:2025bhf, Nee:2025nmh}.

In this work, we employ \textsc{TEOBResumS--Dal\'{\i}}, the latest extension of the \textsc{TEOBResumS} family tailored for eccentric inspirals. In this model, eccentricity is explicitly introduced only through the construction of initial conditions, after which the system evolves using the canonical \ac{eob} variables. Non-circular effects in both radial and azimuthal radiation-reaction components are implemented following the approach described in Ref.~\cite{Chiaramello:2020ehz, Nagar:2024oyk}. 
The model was validated against a large and diverse set of \ac{nr} simulations,
including quasi-circular, eccentric, and precessing \acp{bbh} and \ac{bns}~\cite{albanesi2025effectiveonebodymodelinggenericcompact},
finding it more than $99\%$ accurate over a considerable portion of the parameter space.
Additional studies have further investigated its accuracy, limiting its applicability to systems with
initial eccentricity $\sim 0.6-0.7$~\cite{Gamboa:2024hli, Bhaumik:2024cec}.

\section{Overview of Bayesian parameter estimation method}
\label{sec:bayesian_pe}
%\begin{itemize}
%    \item LIKELIHOODS, PRIORS, AND POSTERIORS
%    \item MODELS, EVIDENCE AND ODDS
%    \item SAMPLERS
%    
%\end{itemize}

The physical parameters are estimated for the detected CBC signal, including the masses, spins, luminosity distance, sky position, and other parameters. This estimation is commonly carried out using Bayesian stochastic sampling methods \cite{Thrane_2019}. The primary aim is to construct a posterior distribution $p(\boldsymbol{\theta}|\mathbf{d})$, where $\boldsymbol{\theta}=\{\theta_1,\theta_2,...,\theta_p\}$ is the set of model parameters $m=15$ ( for a CBC system) and $\mathbf{d}=\{d_1,d_2,...,d_n\}$ is the strain data from a $n$ network of gravitational-wave detectors. Based on Bayes' theorem, the posterior distribution is given by
\begin{equation}\label{eq:1}
    p(\boldsymbol{\theta}|\mathbf{d}) = \frac{{\cal{L}}(\mathbf{d}|\boldsymbol{\theta})\pi(\boldsymbol{\theta})}{p(d|{\cal H})} = \frac{{\cal{L}}(\mathbf{d}|\boldsymbol{\theta})\pi(\boldsymbol{\theta})}{\int {\cal{L}}(\mathbf{d}|\boldsymbol{\theta})\pi(\boldsymbol{\theta}) d\boldsymbol{\theta}} %\propto {\cal{L}}(\mathbf{d}|\boldsymbol{\theta})\pi(\boldsymbol{\theta})
\end{equation} 
where ${\cal{L}}(\mathbf{d}|\boldsymbol{\theta})$ is the likelihood function, the conditional probability density function of the data given the parameters $\boldsymbol{\theta}$. The likelihood function assuming Gaussian noise is taken to be

\begin{equation}
{\cal{L}}(\mathbf{d}|\boldsymbol{\theta}) \propto \exp \left( -\frac{1}{2} \left< \mathbf{d}-h(\boldsymbol{\theta})|\mathbf{d}-h(\boldsymbol{\theta}) \right>\right)
\end{equation}
where $h(\boldsymbol{\theta})$ is the model at given $\boldsymbol{\theta}$, $\left<\cdot | \cdot\right>$ is a scalar product defined as $\left<a|b\right> = 4 \Re  \int_0^{\infty} {\frac{a^*(f) b(f)}{S_n(f)}df}$ with $S_n$ being the noise power spectral density of the given GW detector. $a^*$ is the complex conjugate of $a$, and $\pi(\boldsymbol{\theta})$ is the prior probability density function. {Therefore, this is simply a normalization constant with respect to the posterior distribution of $\boldsymbol{\theta}$}.
%where $h(\boldsymbol{\theta})$ is model at given $\boldsymbol{\theta}$, $\left<\cdot | \cdot\right>$ is a scalar product \footnote{The scalar product $\left<a|b\right> =4 \int_0^{\infty} {\frac{a^*(f) b(f)}{S_n(f)}df}$, where $S_n$ is the noise power spectral density of given GW detector, $a^*$ the conjugate of $a$.}and $\pi(\boldsymbol{\theta})$ is the prior probability density function and is therefore simply a normalization constant with respect to the posterior distribution of $\boldsymbol{\theta}$. 

{Within the Bayesian framework, all inference is performed conditional on a specific model hypothesis $\mathcal{H}$, which defines the underlying assumptions about the signal morphology and parameter space ($m$). %The posterior probability density function for the parameters $\boldsymbol{\theta}$ given the data $d$ and model $\mathcal{H}$ is expressed through Bayes' theorem as $p(\boldsymbol{\theta}|d,\mathcal{H}) \propto p(d|\boldsymbol{\theta},\mathcal{H}) \pi(\boldsymbol{\theta}|\mathcal{H})$, where $p(d|\boldsymbol{\theta},\mathcal{H})$ is the likelihood of observing the data given parameters $\boldsymbol{\theta}$ under model $\mathcal{H}$, and 
Where $\pi(\boldsymbol{\theta}|\mathcal{H})$ is the prior probability density function that encodes our prior knowledge about the parameters within the context of model $\mathcal{H}$. The model waveform is denoted as $h(\boldsymbol{\theta}|\mathcal{H})$, representing the predicted gravitational wave signal for given parameters $\boldsymbol{\theta}$ under the assumptions of $\mathcal{H}$. This conditional formulation naturally extends to model comparison, where we can evaluate the relative support for different hypotheses by computing Bayes factors between competing models.}
 
 Here, $m$ represents the dimension of the parameter space. As $m$ increases, computing marginal distributions of the Bayesian posterior becomes increasingly difficult because it involves solving high-dimensional integration problems. In order to solve this problem, we can use the simulation-based computational techniques such as Markov Chain Monte Carlo or nested sampling. To get posteriors for free parameters $\boldsymbol{\theta}_f$, we need to marginalize over other parameters $\boldsymbol{\theta}_o$ as $p(\boldsymbol{\theta}_f|\mathbf{d}) = \int p(\boldsymbol{\theta}|\mathbf{d}) d \boldsymbol{\theta}_o $.  
In this analysis, we use the Bilby parameter estimation pipeline ~\cite{Ashton:2021anp}. This pipeline is widely used in the GW community due to its flexibility and extensible framework for Bayesian parameter estimation. Built upon modular components that interface with waveform models from LALSuite ~\cite{PhysRevD.91.042003} and other gravitational wave software libraries, Bilby allows users to define likelihood functions and priors while accessing a variety of nested sampling and MCMC techniques through backends such as Dynesty~\cite{dynesty}, CPNest~\cite{cpnest}, \texttt{emcee}~\cite{emcee}, and PyMultiNest~\cite{pymultinest}. This allowed us to utilise the eBBH waveforms, which are accessible outside the LALSuite framework. Bilby's strength lies in its ability to marginalise over calibration errors, include arbitrary priors, and efficiently parallelise large sampling jobs across HPC environments.
%Definition of likelihood function assuming Gaussian noise,
The adoption of Bilby in recent observation runs (e.g., GWTC-3)~\cite{2021arXiv211103606T} underscores its reliability and interpretability, especially when used in conjunction with stochastic samplers that map the high-dimensional likelihood surfaces characteristic of binary black hole or neutron star mergers.
\begin{table}[h]
\begin{tabular}{cl}
  \hline
  \hline
  $\rm{Parameter}$ & $\rm{Value}$   \\
\hline
$q$                & $1$                            \\
$D_\mathrm{L}$         & $410 \mathrm{Mpc}$              \\
$\theta_\mathrm{jn}$      & $0.4 \;\rm{rad}$               \\
$\psi$             & $2.659 \;\rm{rad}$             \\
$\phi$             & $1.3 \;\rm{rad}$               \\
$\rm{RA}$          & $1.375 \;\rm{rad}$             \\
$\rm{DEC}$         & $-1.2108 \;\rm{rad}$           \\
$t_\mathrm{geocent}$   & $1126259642.413 \;\mathrm{s}$      \\
\hline

\end{tabular}\vspace{-0.6cm}
\caption{
List of parameters that are fixed for all the injected parameters. 
}
\label{tab:paramlist}
\end{table}
PE analyses that employ eccentric waveform models typically introduce two additional parameters compared to non-eccentric models, such as eccentricity and mean anomaly. %\RG{May want to expand on the parameters and define them?} 
Here we use nested sampling~\cite{Speagle_2020}, which is implemented in the Bilby sampler. 
%We can also calculate the Bayes factor, which can be calculated between recoveries with eccentric and quasi-circular models
We can also calculate the Bayes factor, which is defined as the ratio of the Bayesian evidence for two competing hypotheses $\mathcal{H}_1$ and $\mathcal{H}_2$, and it is given by $\mathcal{B}_{12}
= \frac{p(d \mid \mathcal{H}_1)}{p(d \mid \mathcal{H}_2)}$, where $p(d \mid \mathcal{H})$ is the Bayesian evidence, obtained by marginalising the likelihood over the model parameters, see Eq. \eqref{eq:1}. A Bayes factor greater than unity indicates preference for $\mathcal{H}_1$, whereas a value less than unity favours $\mathcal{H}_2$. In the case of systematic studies, we compute the Bayes factor between different configuration setups to understand which configuration has strong evidence for the injected eccentric system. %recoveries with quasi-circular precessing versus quasi-circular non-spining  models. 

\begin{figure*}[htbp!]
\centering
    \includegraphics[width=0.32\textwidth]{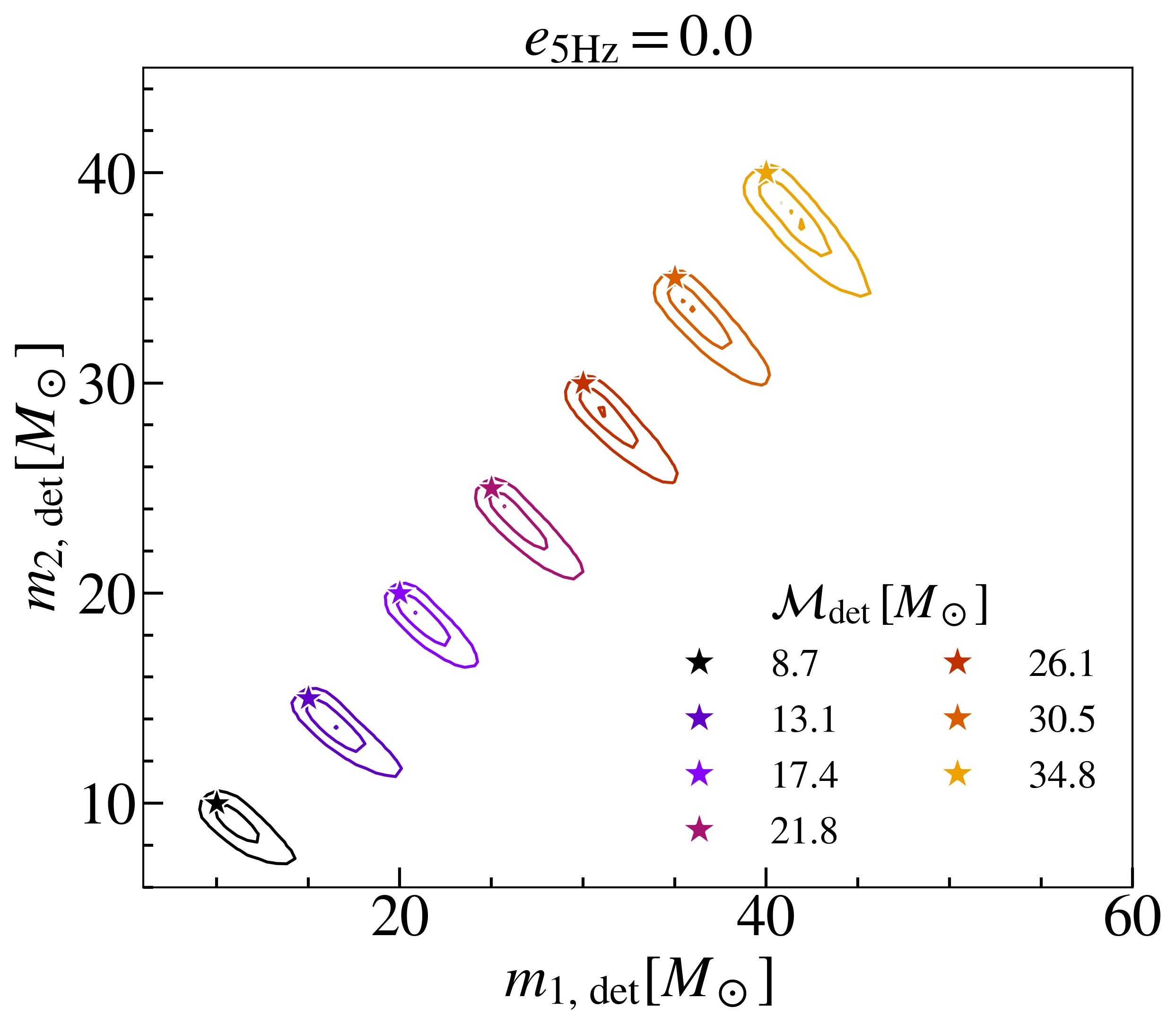}
    \includegraphics[width=0.32\textwidth]{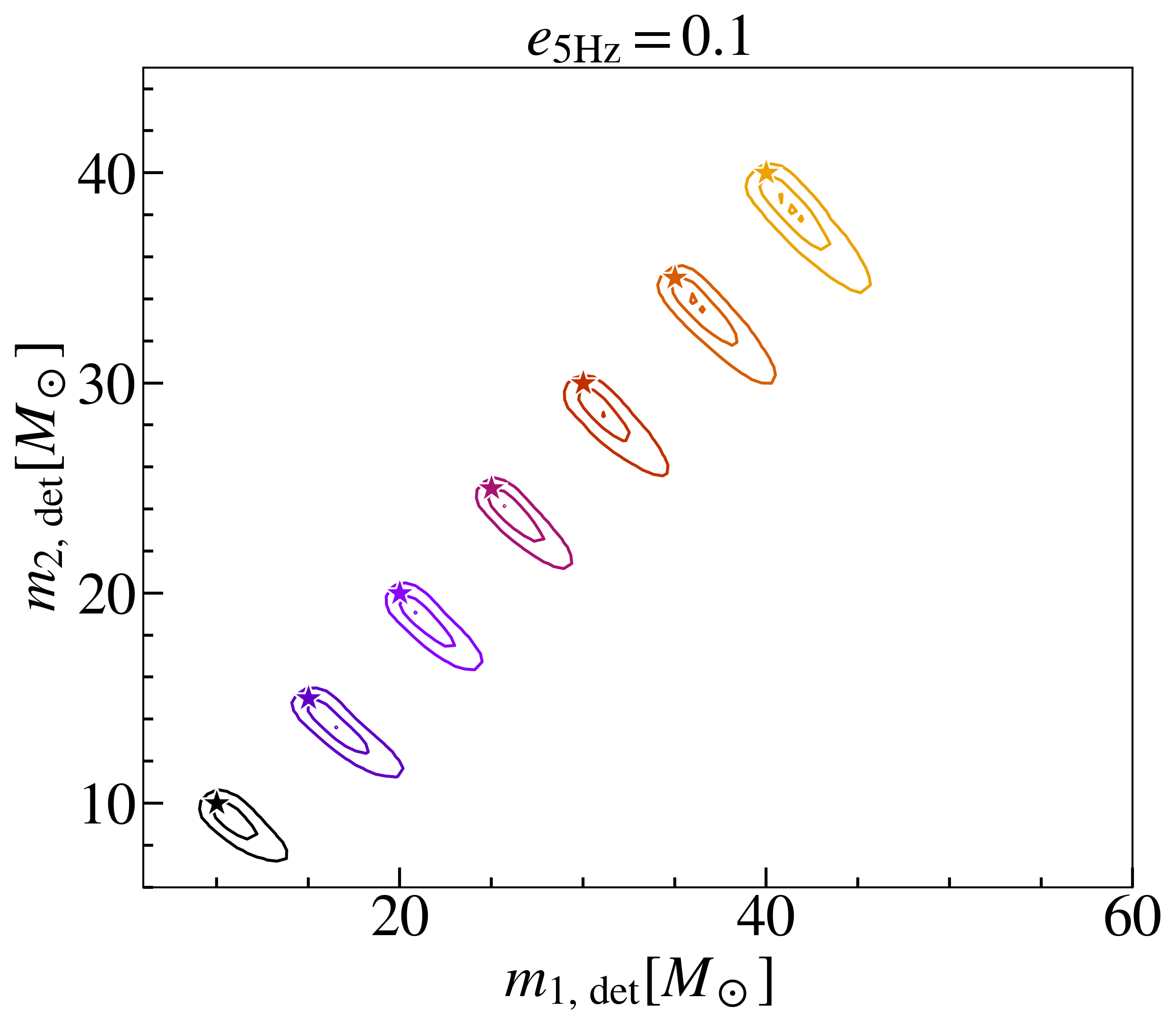}
    \includegraphics[width=0.32\textwidth]{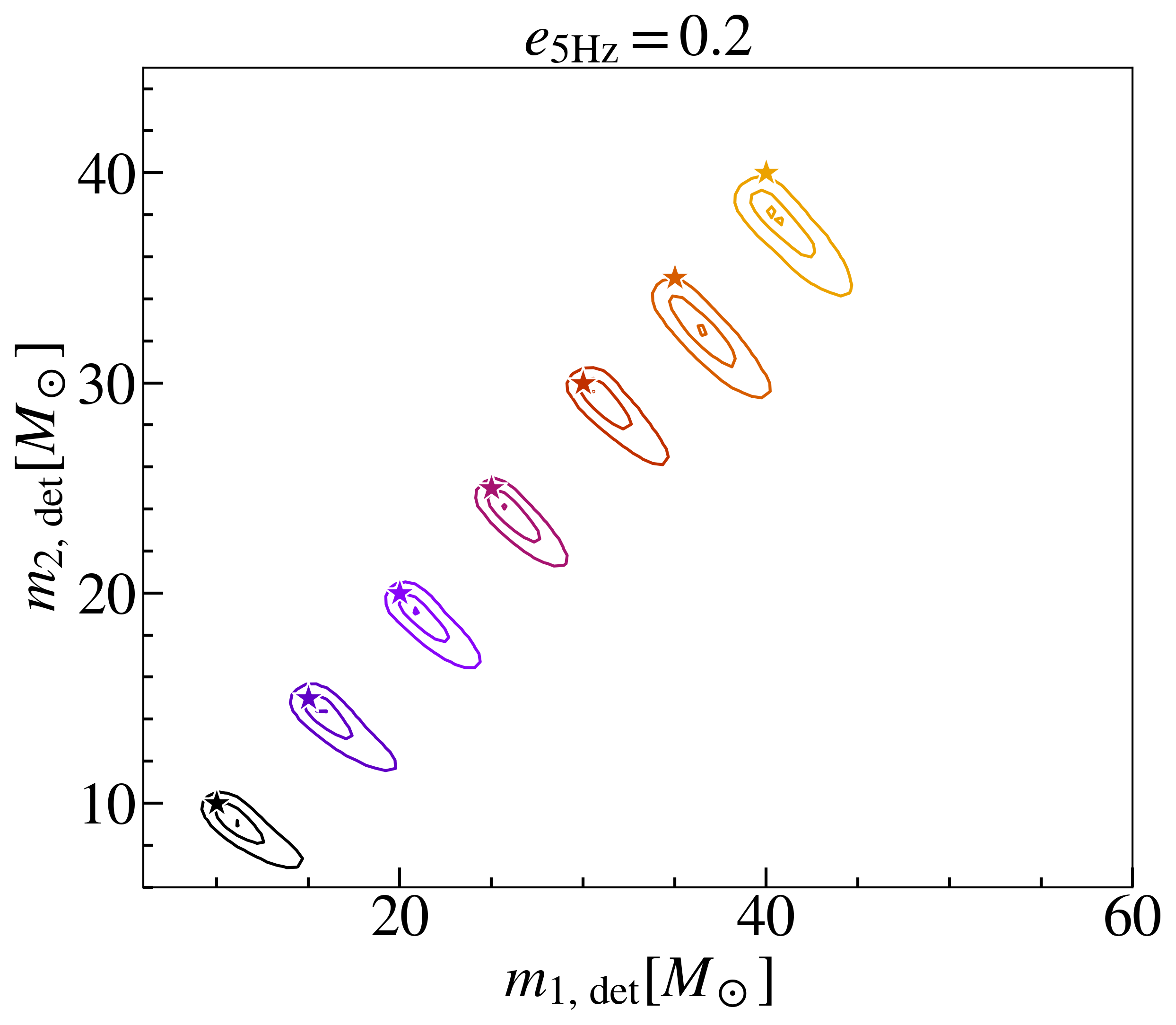}
    \includegraphics[width=0.32\textwidth]{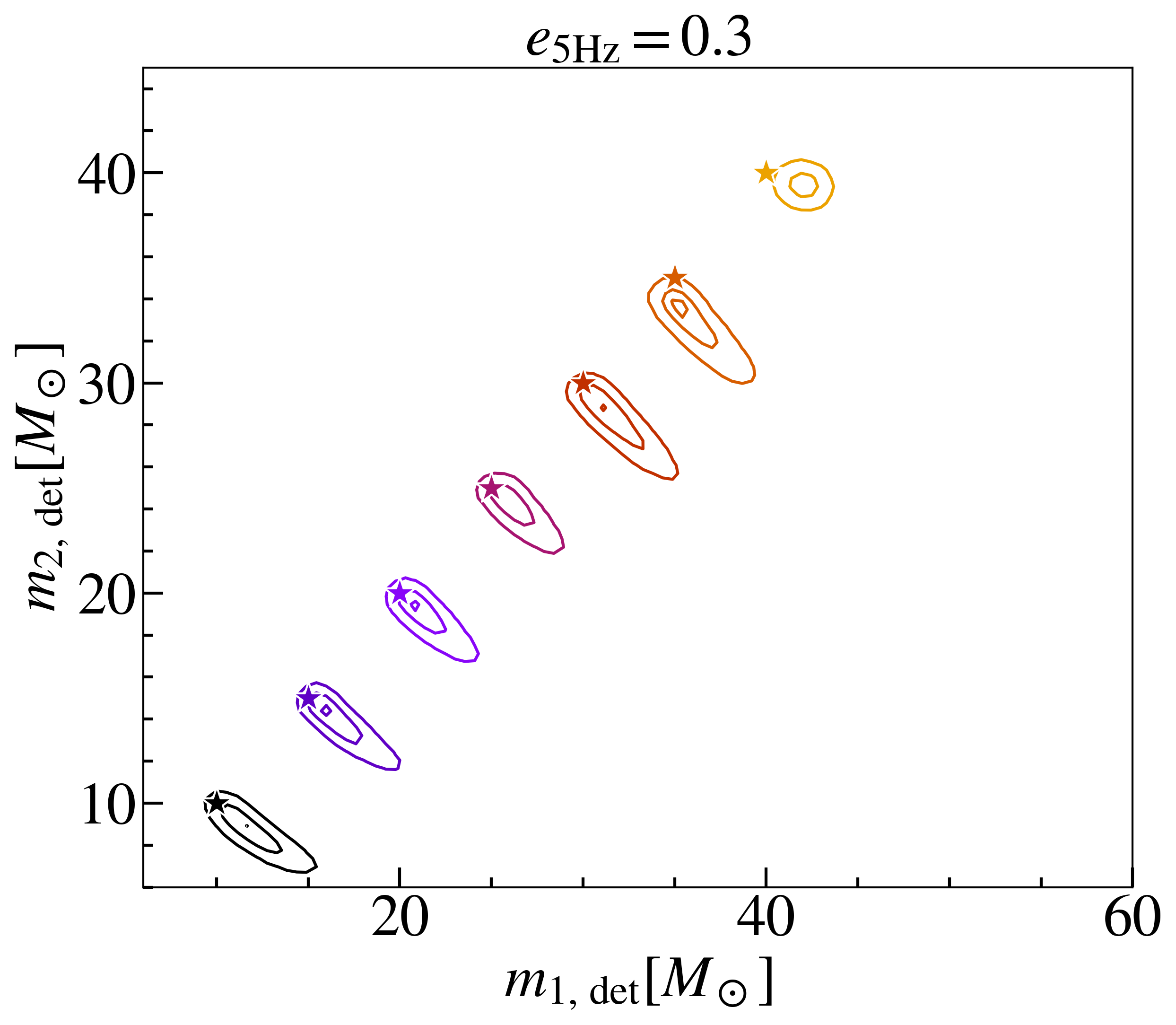}
    \includegraphics[width=0.32\textwidth]{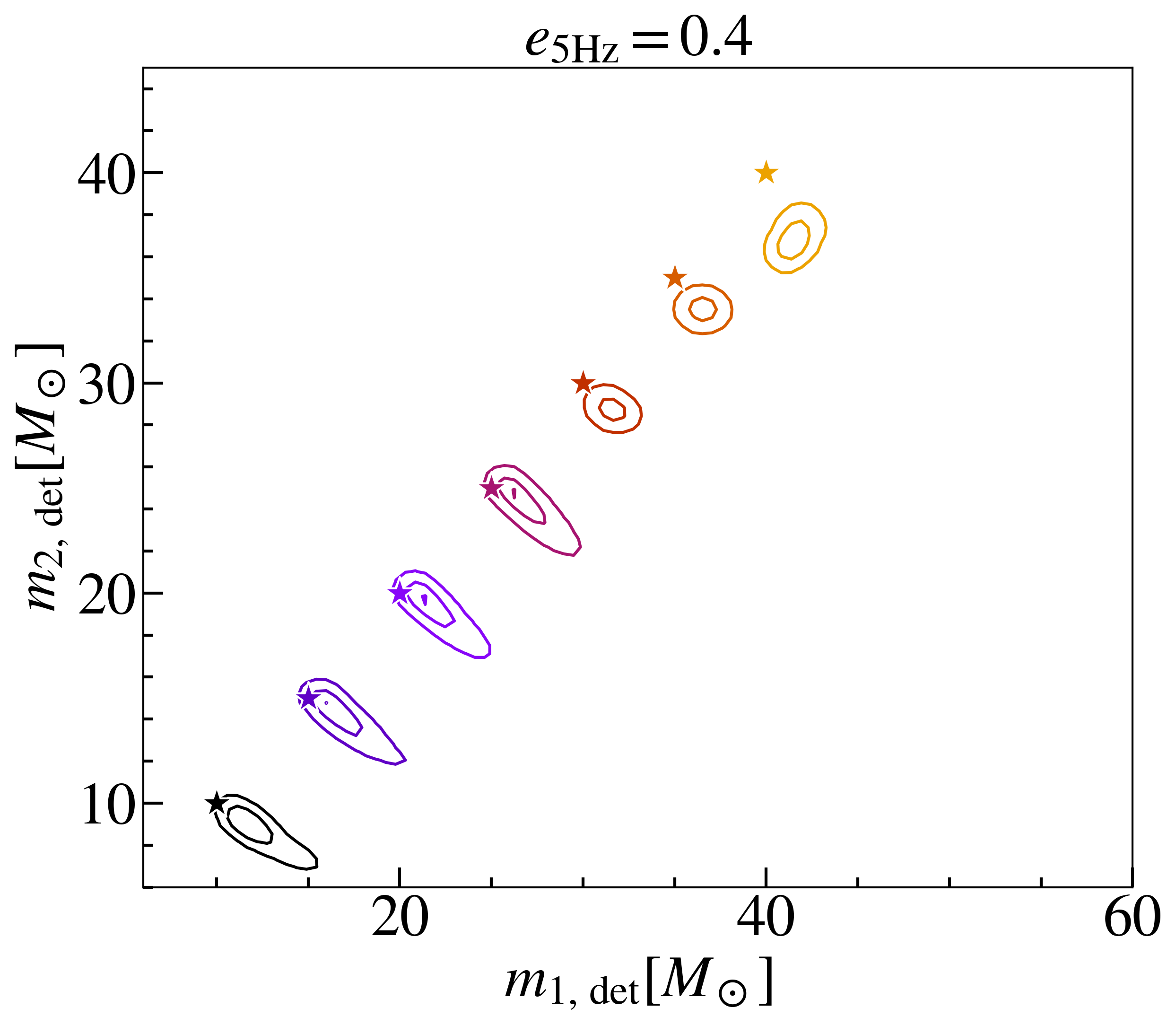}
    \includegraphics[width=0.32\textwidth]{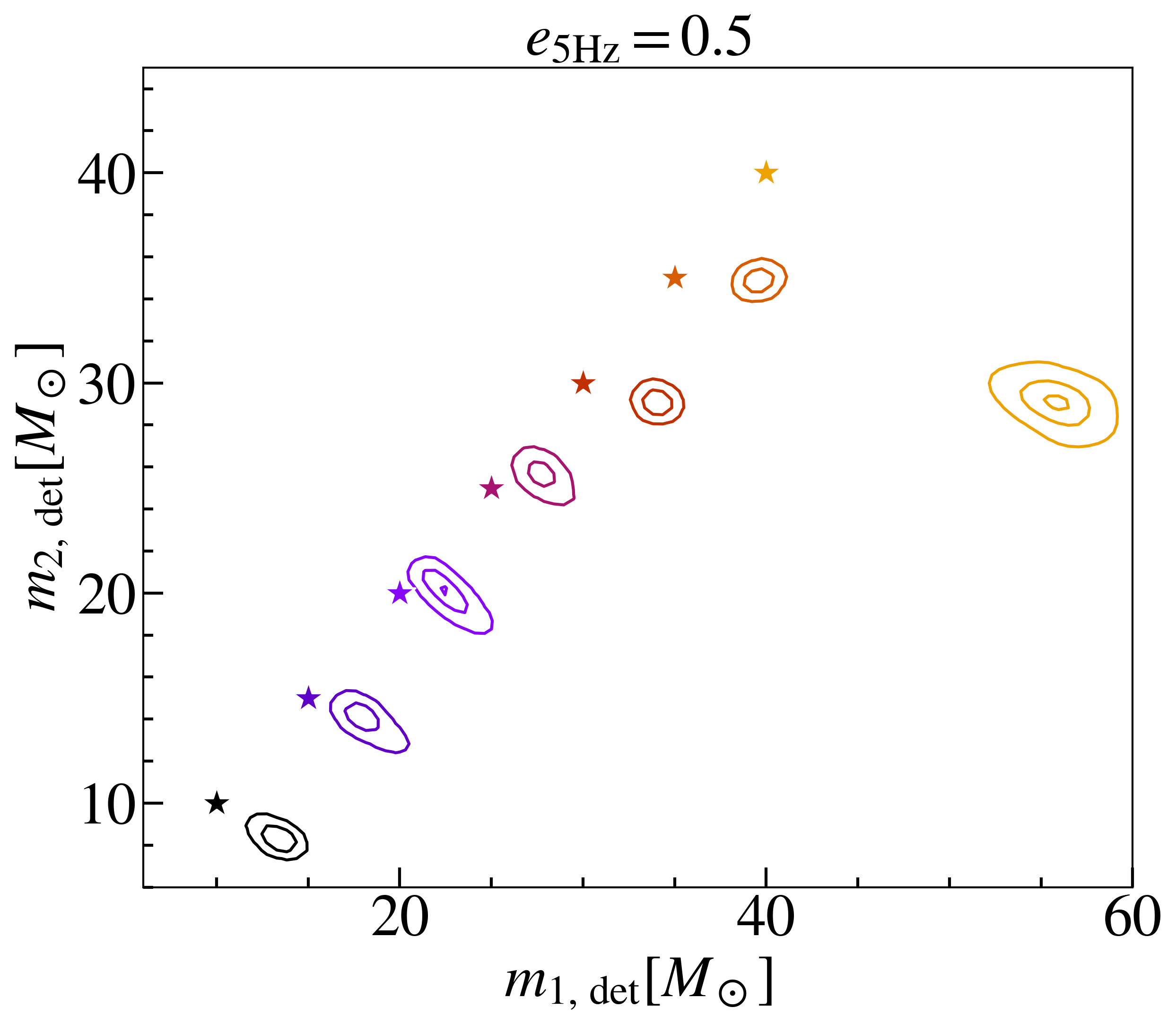}
    
    \caption{
    Posterior distributions of detector-frame component masses $m_1$ versus $m_2$ from parameter estimation using a precessing-spin waveform model, for injections with varying initial eccentricities. The injected signals are for eccentric, non-spinning systems, while the recovery uses non-eccentric waveforms with non-aligned component spins, that lead to spin-precession. Each subplot corresponds to an injected eccentricity, ordered increasingly. Star markers indicate the injected $m_1$ and $m_2$ values. The 90\%, 50\%, and 10\% credible regions are shown in the plots, and the injected detector-frame chirp mass $\mathcal{M}$ is listed in the legend.}
    %Posterior plots for detector frame masses $m_1$ versus $m_2$ of precessing spin systems for varying initial eccentricities of the injected signal. Each subplot corresponds to one injected eccentricity, which is arranged in increasing order. Markers star denote the injected mass $m_1$ and $m_2$ values.  We plot the 90\%, 50\%, and 10\% credible intervals obtained from parameter estimation. We have noted that the injected value of chirp mass $\cal{M}$ in the detector frame is in the legend.}
    \label{fig:comp_mass_precess}
\end{figure*}

%\begin{figure*}[htbp!]
%\centering
%    \includegraphics[width=0.32\textwidth]{plots/comp_mass_aligned_set_0.png}
%    \includegraphics[width=0.32\textwidth]{plots/comp_mass_aligned_set_1.png}
%    \includegraphics[width=0.32\textwidth]{plots/comp_mass_aligned_set_2.png}
%    \includegraphics[width=0.32\textwidth]{plots/comp_mass_aligned_set_3.png}
%    \includegraphics[width=0.32\textwidth]{plots/comp_mass_aligned_set_4.png}
%    \includegraphics[width=0.32\textwidth]{plots/comp_mass_aligned_set_5.png}
    
%    \caption{ Posterior plots for detector frame masses $m_1$ versus $m_2$ of aligned spin systems for varying initial eccentricities of the injected signal. Each subplot corresponds to one injected eccentricity, which is arranged in increasing order.  Markers star denote the injected mass $m_1$ and $m_2$ values.  We plot the 90\%, 50\%, and 10\% credible intervals obtained from parameter estimation. We have noted that the injected value of chirp mass $\cal{M}$ in the detector frame is in the legend.  }
%    \label{fig:comp_mass_align}
%\end{figure*}

\begin{figure*}[htbp!]
\centering
    \includegraphics[width=0.32\textwidth]{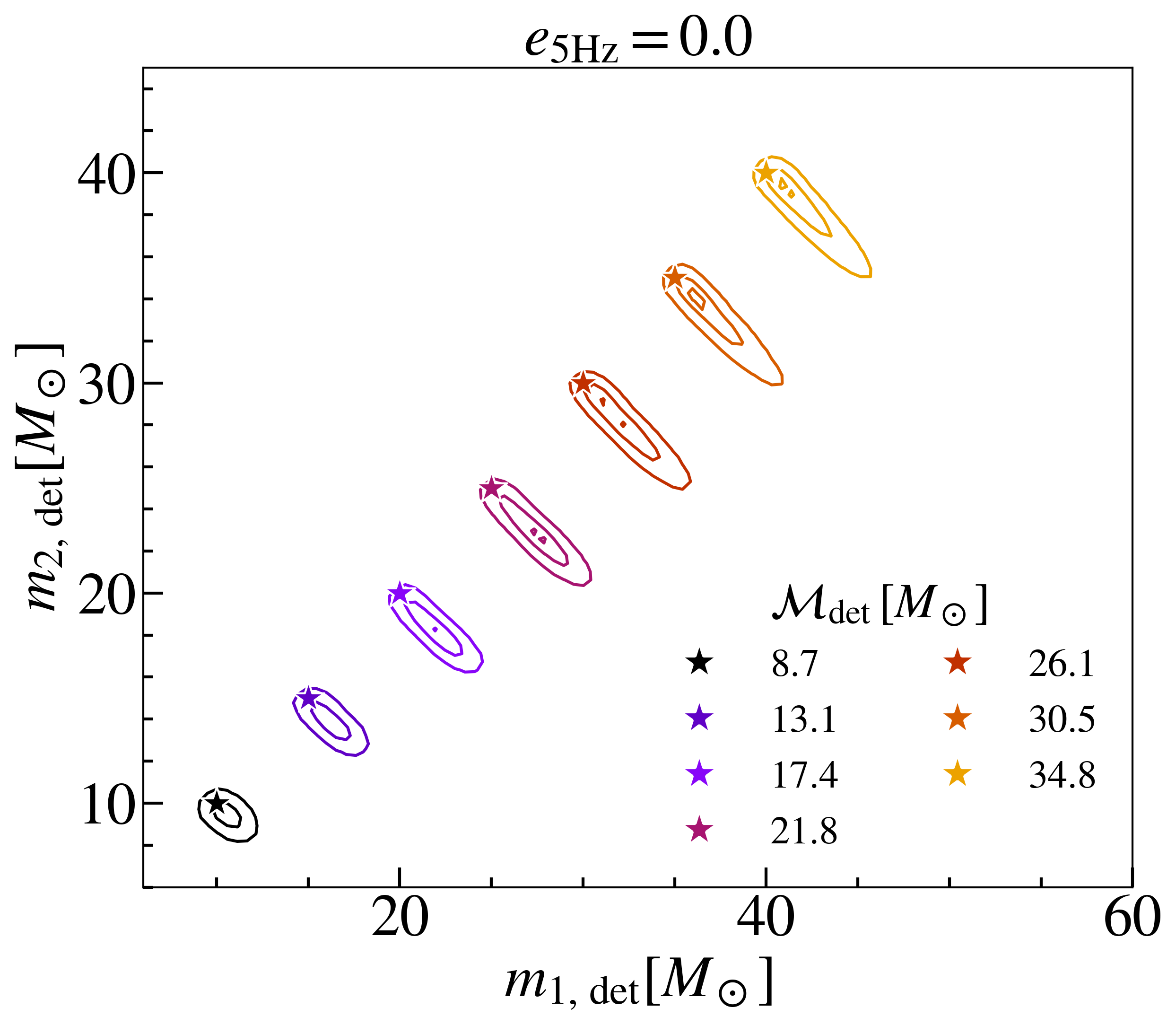}
    \includegraphics[width=0.32\textwidth]{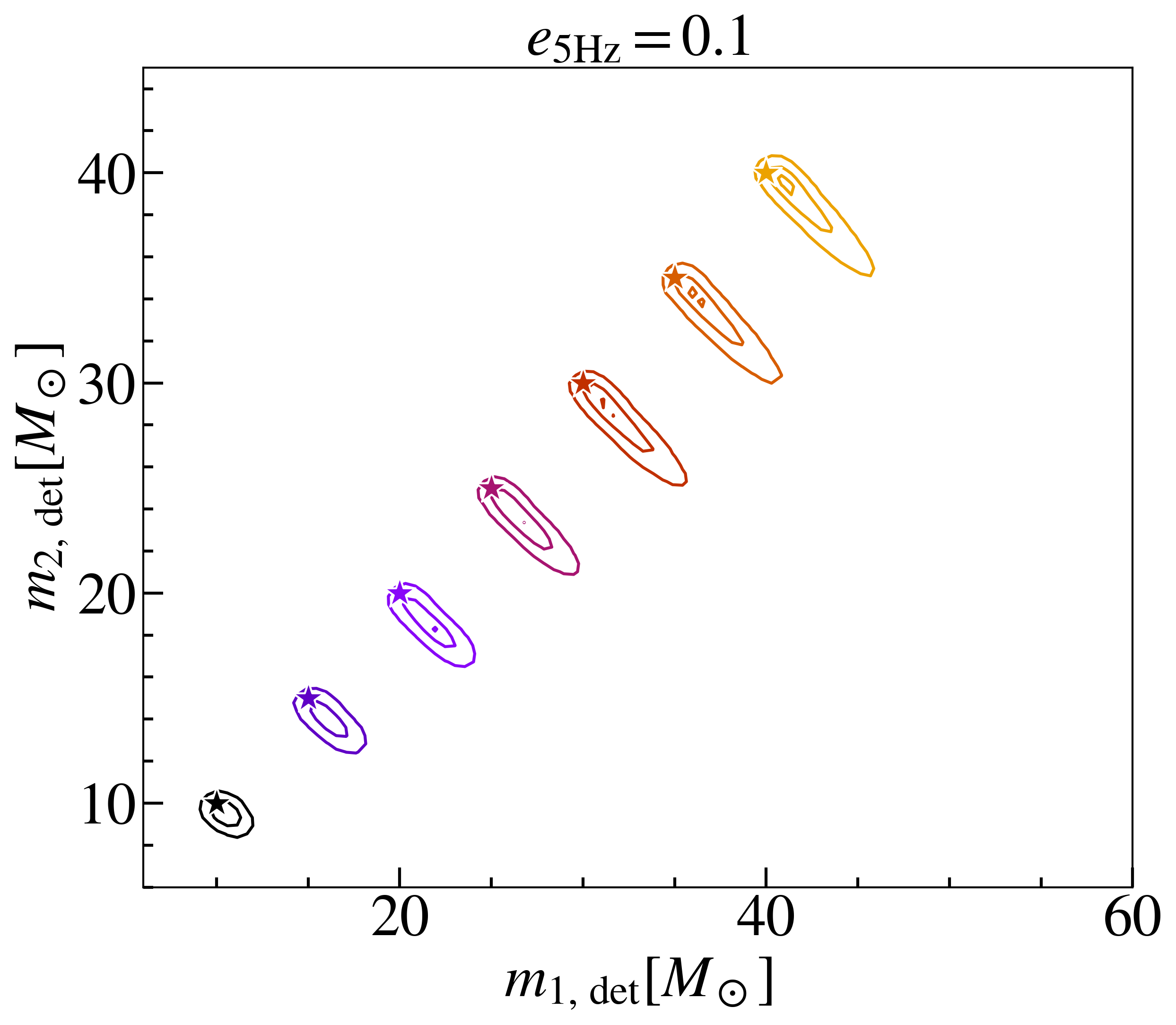}
    \includegraphics[width=0.32\textwidth]{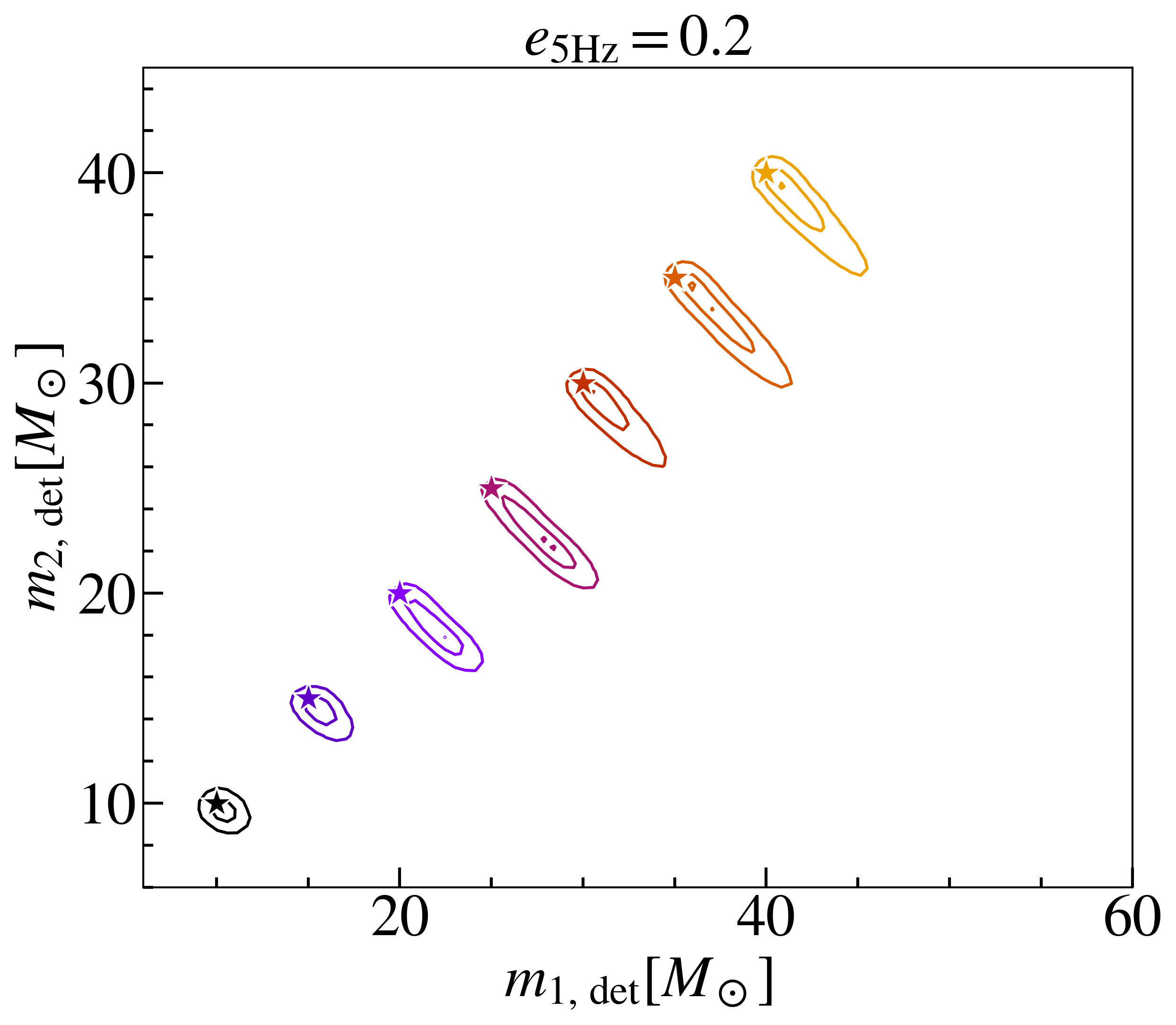}
    \includegraphics[width=0.32\textwidth]{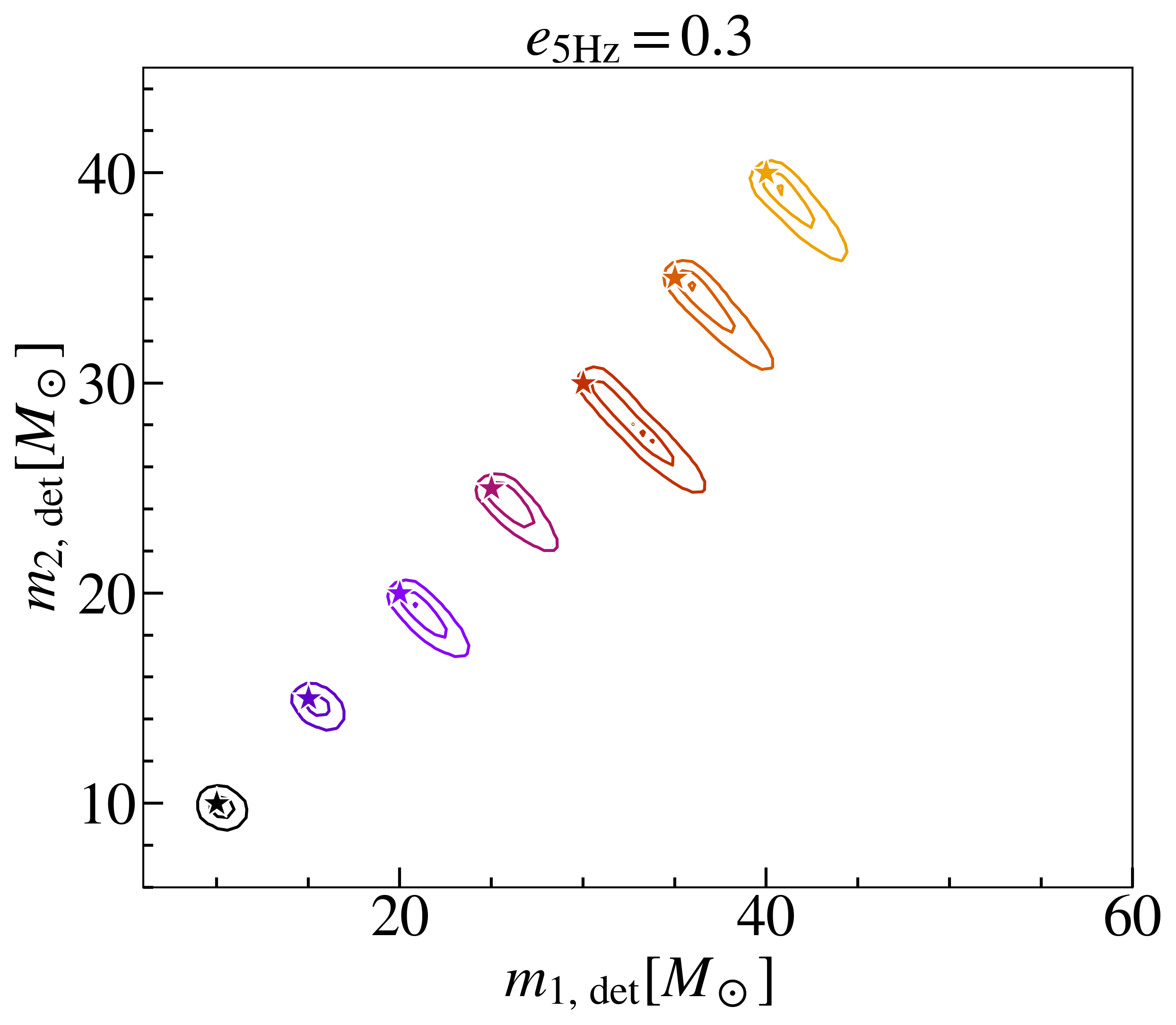}
    \includegraphics[width=0.32\textwidth]{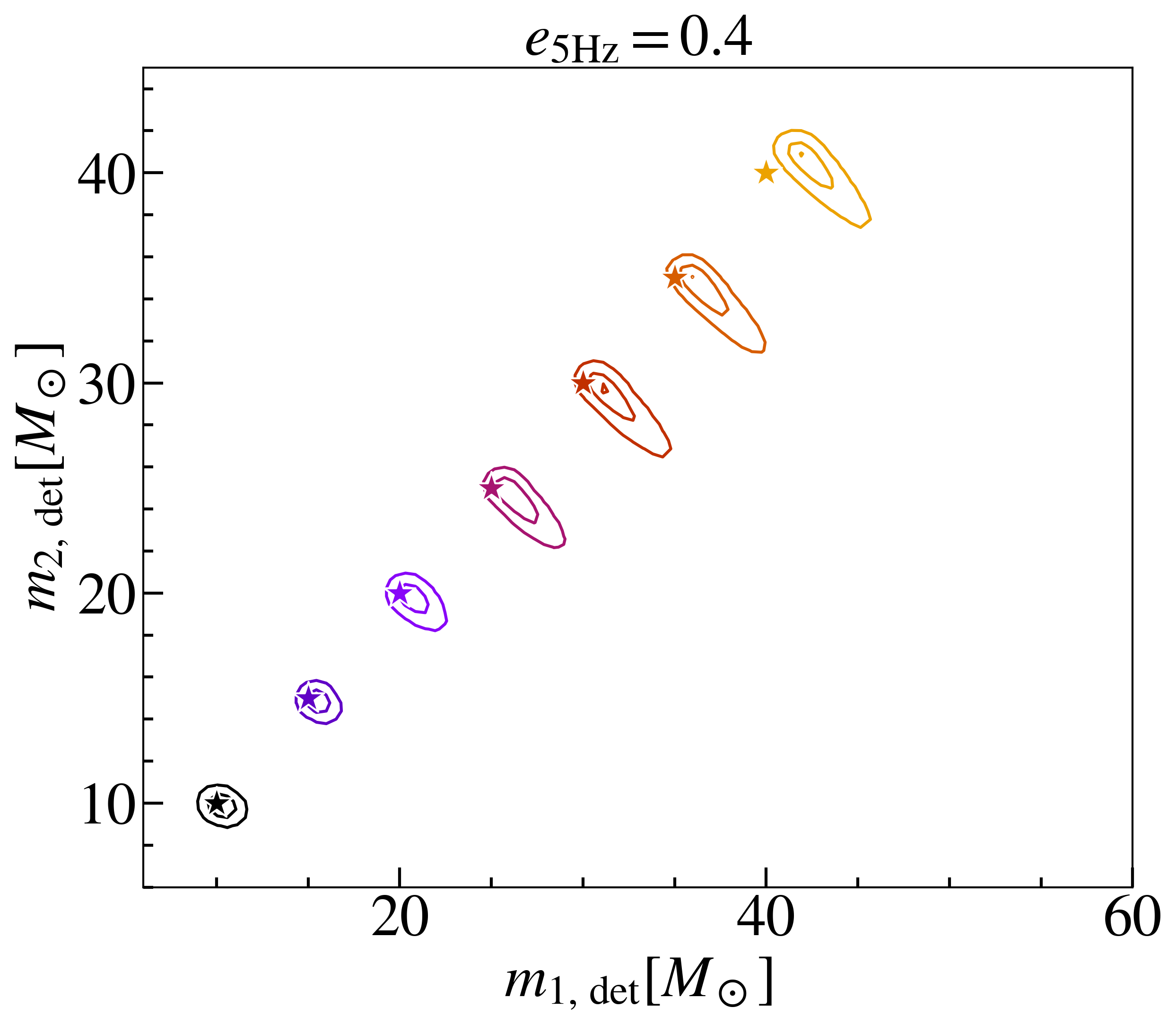}
    \includegraphics[width=0.32\textwidth]{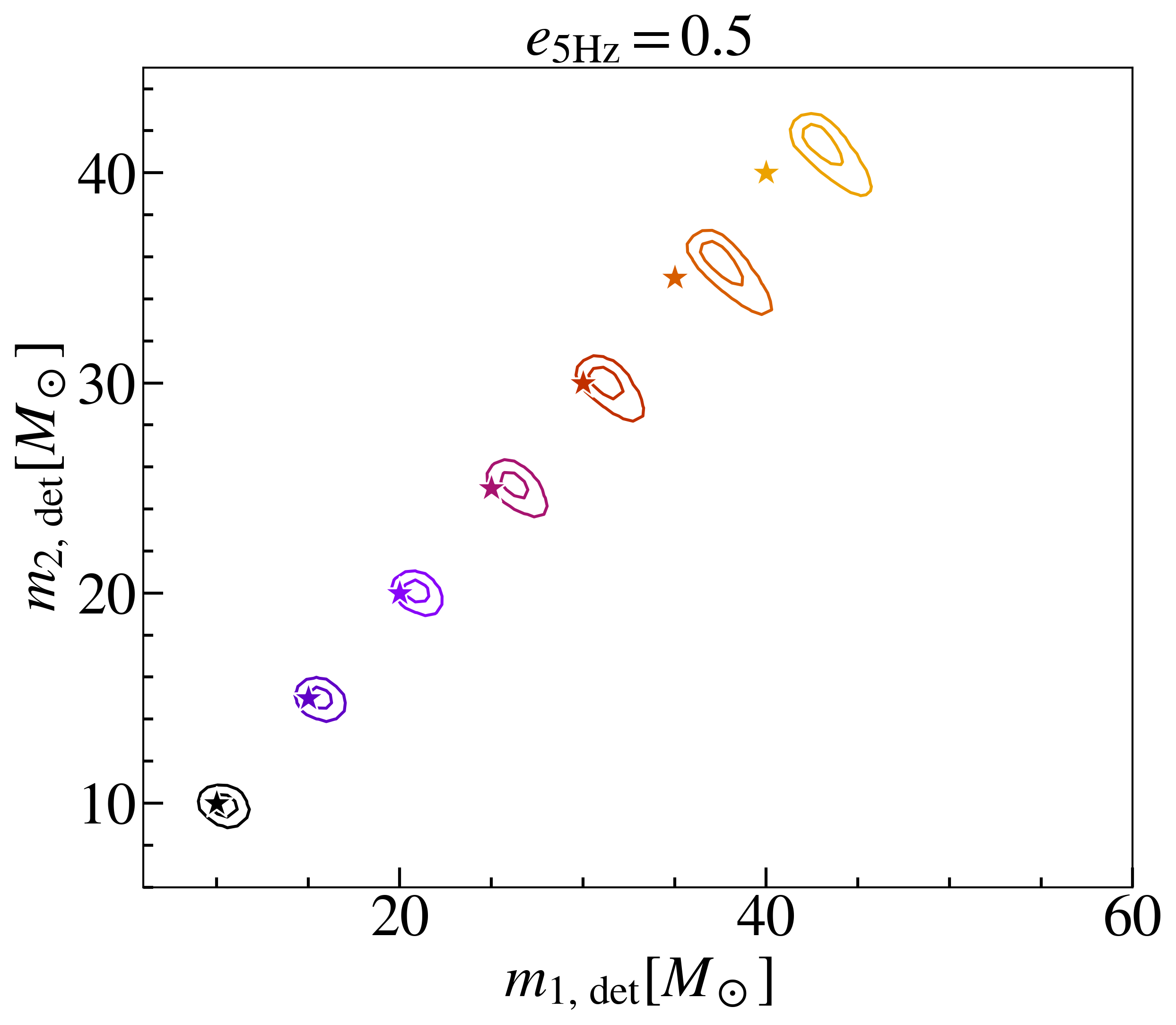}
    
    \caption{
    Posterior distributions of detector-frame component masses $m_1$ versus $m_2$ from parameter estimation using a non-spinning waveform model, for injections with varying initial eccentricities. The injected signals are eccentric and non-spinning, while the recovery uses non-eccentric waveforms with no component spins. Each subplot corresponds to an injected eccentricity, ordered increasingly. Star markers indicate the injected $m_1$ and $m_2$ values. The 90\%, 50\%, and 10\% credible regions are shown, and the injected detector-frame chirp mass $\mathcal{M}$ is listed in the legend.}

    %Posterior plots for detector frame masses $m_1$ versus $m_2$ of non-spinning systems for varying initial eccentricities of the injected signal. Each subplot corresponds to one injected eccentricity, which is arranged in increasing order.  Markers star denote the injected mass $m_1$ and $m_2$ values.  We plot the 90\%, 50\%, and 10\% credible intervals obtained from parameter estimation. We have noted that the injected value of chirp mass $\cal{M}$ in the detector frame is in the legend.  }
    \label{fig:comp_mass_nospin}
\end{figure*}

\begin{figure*}[htbp!]
\centering
    \includegraphics[width=0.32\textwidth]{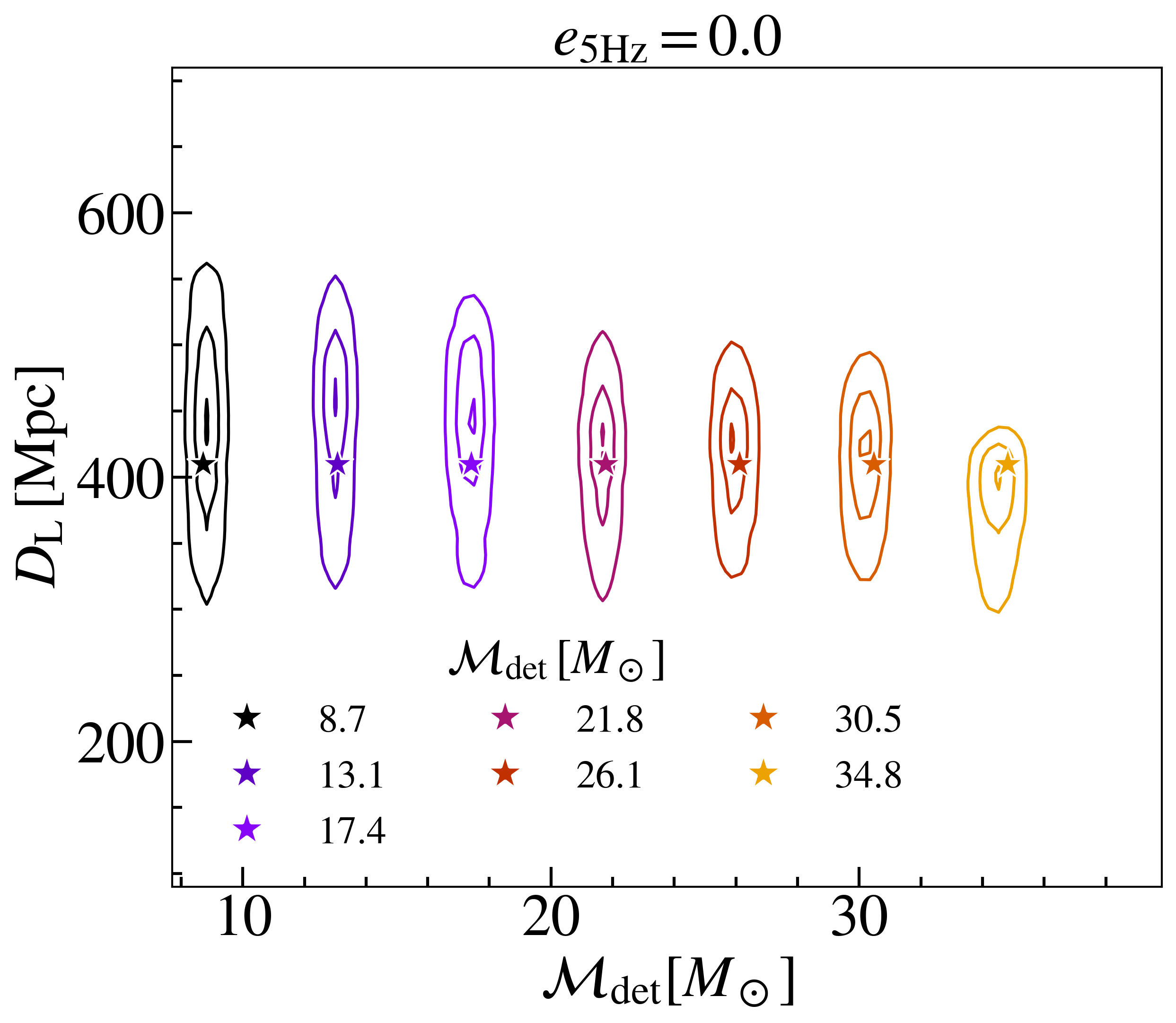}
    \includegraphics[width=0.32\textwidth]{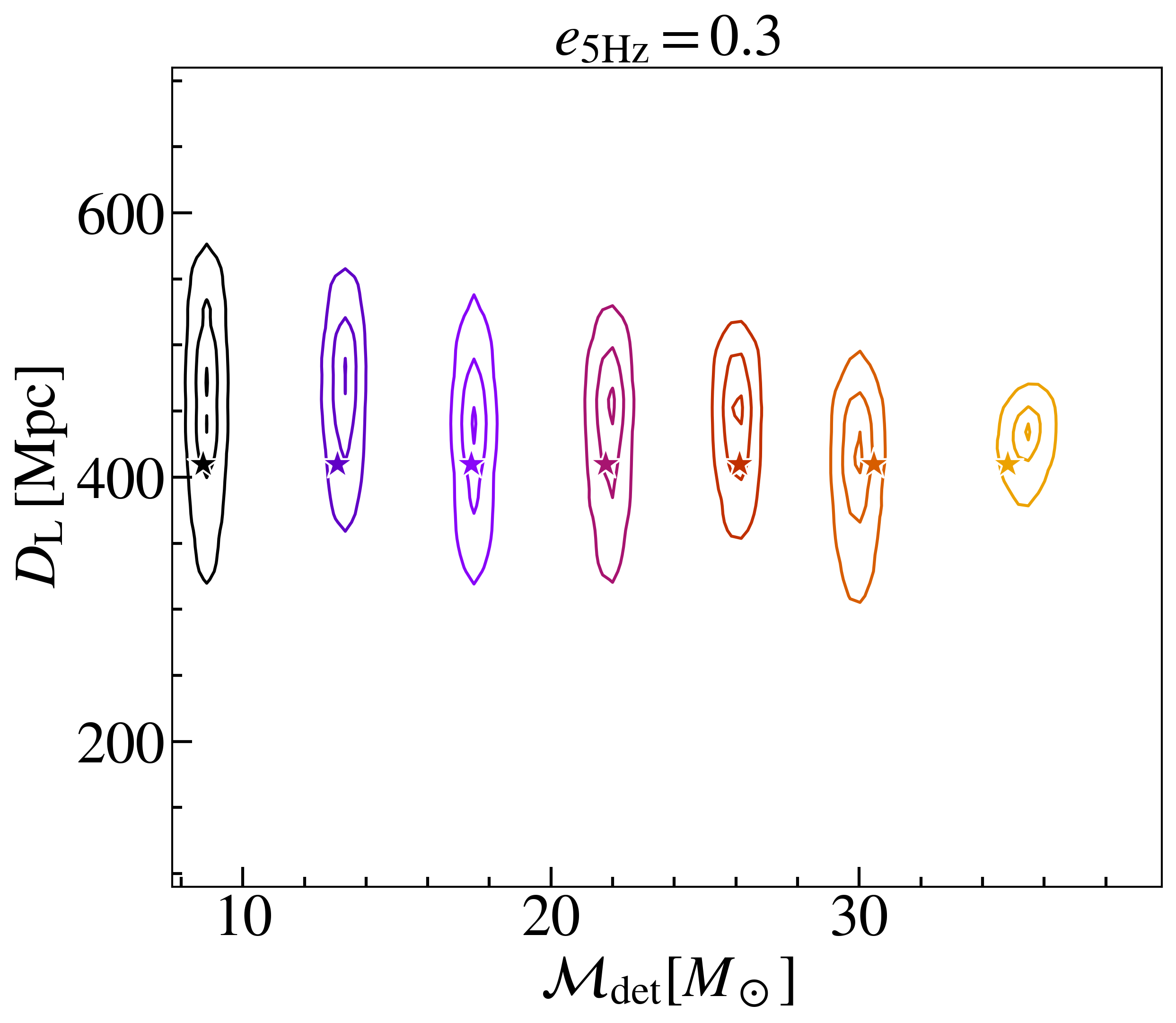}
    \includegraphics[width=0.32\textwidth]{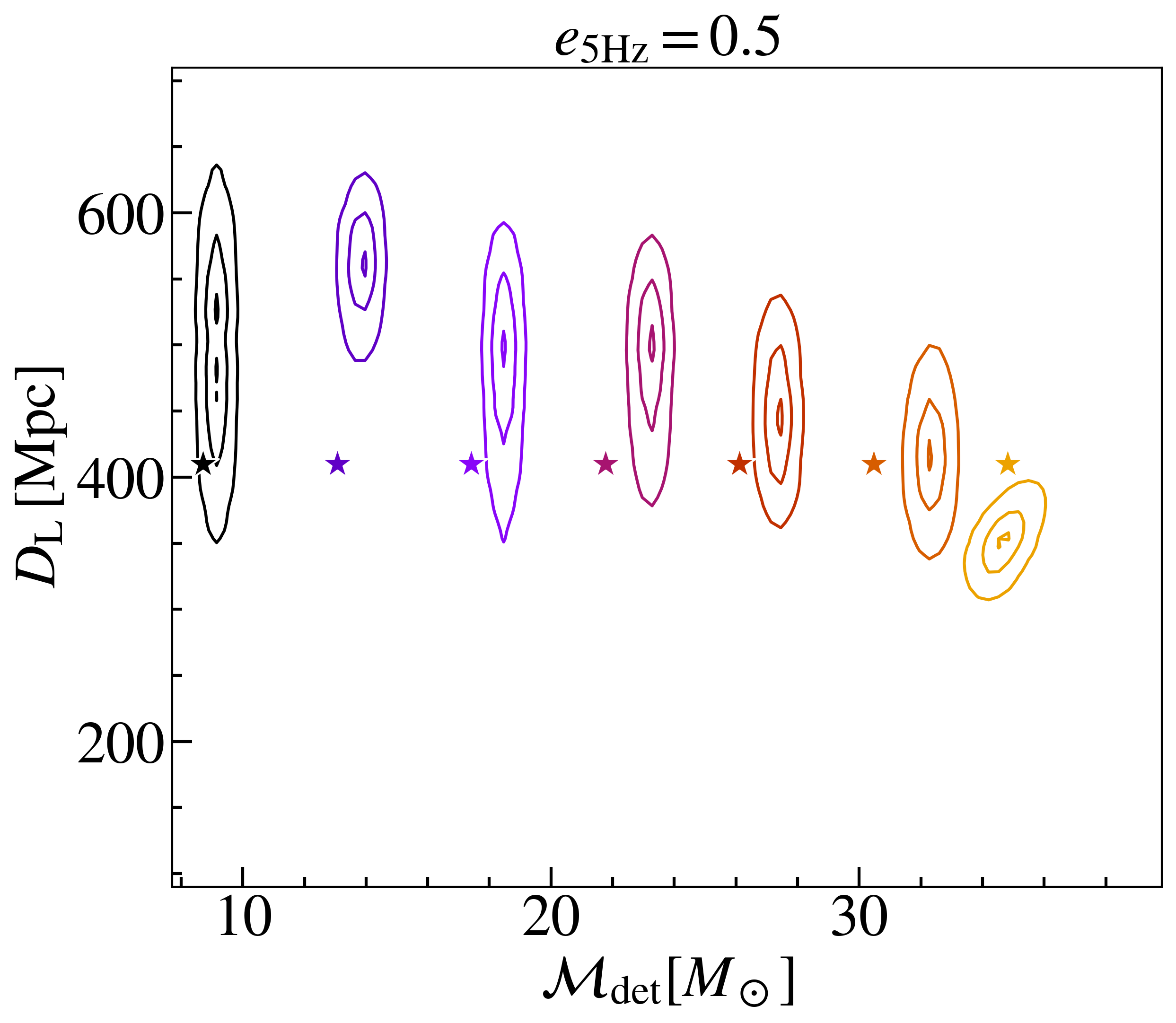}
    \caption{ Posterior distributions of the luminosity distance ($D_{\mathrm{L}}$) versus chirp mass ($\mathcal{M}$) from parameters estimation using a precessing-spin waveform model, for injections with varying initial eccentricities. The star markers indicate the injected chirp mass values ($\mathcal{M}$). The contours represent the 90\%, 50\%, and 10\% credible intervals obtained from parameter estimation. }
    \label{fig:chirp_dl_precess}
\end{figure*}

\begin{figure}[htbp!]
\centering
    \includegraphics[width=0.45\textwidth]{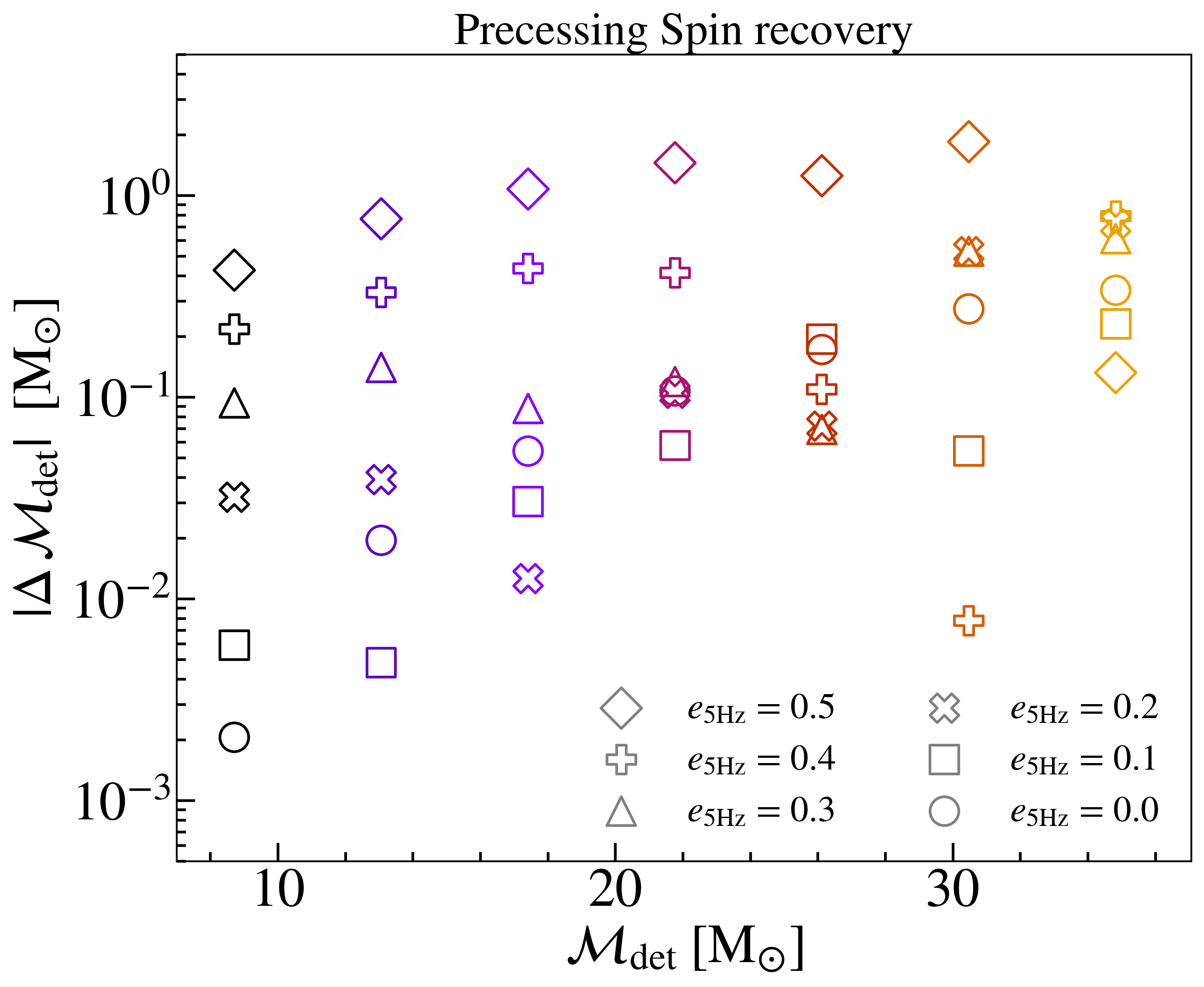}
    \includegraphics[width=0.45\textwidth]{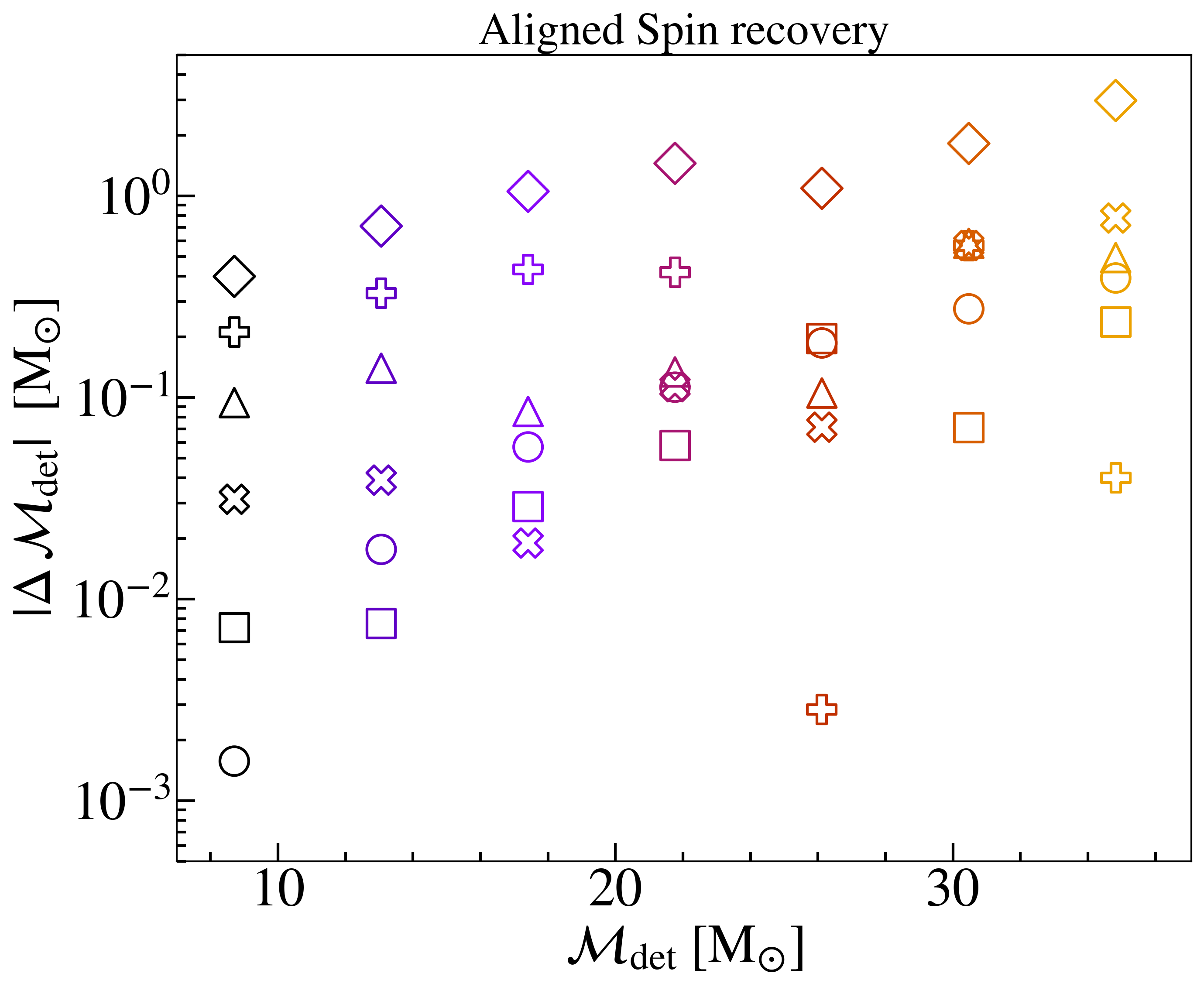}
    \includegraphics[width=0.45\textwidth]{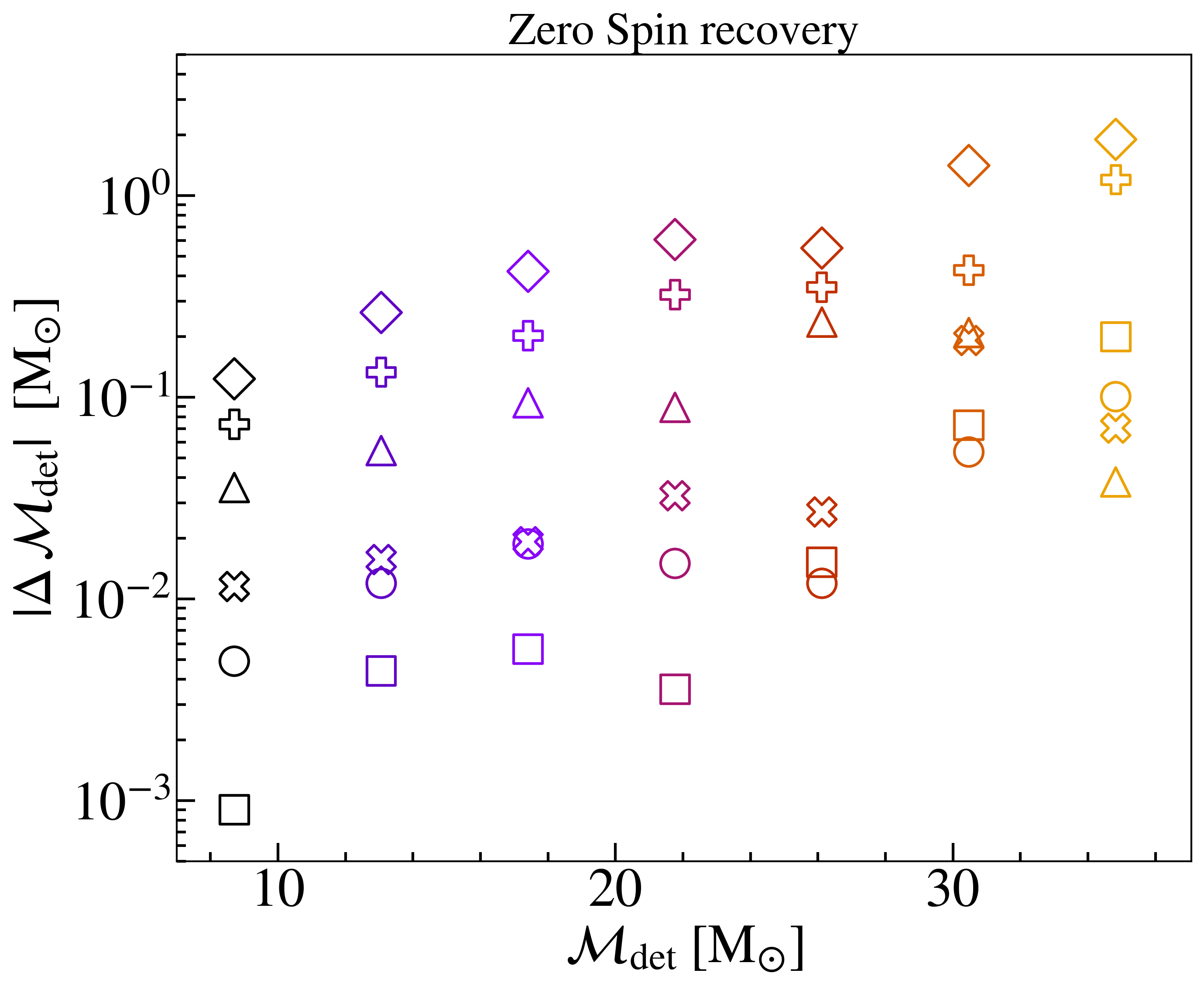}

    \caption{ Absolute differences in chirp mass, $|\Delta \mathcal{M}|$, as a function of the injected chirp mass, $\mathcal{M}$, 
for all injections. Here, $\Delta \mathcal{M}$ is defined as the difference between the injected chirp mass and the median of the recovered posterior distribution. The top, middle, and bottom panels correspond to the precessing-spin, aligned spin, and non-spinning configurations respectively. Different marker styles indicate the injected eccentricity. Larger differences are seen for higher eccentricity, while for a given eccentricity the differences increase with increasing injected chirp mass.}
    \label{fig:abs_delta_chirp_mass}
\end{figure}
%\begin{figure*}[htbp!]
%\centering
%    \includegraphics[width=0.32\textwidth]{plots/chirp_DL_aligned_set_0.png}
%    \includegraphics[width=0.32\textwidth]{plots/chirp_DL_aligned_set_3.png}
%    \includegraphics[width=0.32\textwidth]{plots/chirp_DL_aligned_set_5.png}
    
%    \caption{ Posterior distributions of the luminosity distance ($D_{\mathrm{L}}$) versus chirp mass ($\mathcal{M}$) for the aligned spin recovery setup with varying initial eccentricities of the injected signal. The star markers indicate the injected chirp mass values ($\mathcal{M}$). The contours represent the 90\%, 50\%, and 10\% credible intervals obtained from parameter estimation.}
%    \label{fig:chirp_dl_align}
%\end{figure*}

\begin{figure*}[htbp!]
\centering
    \includegraphics[width=0.32\textwidth]{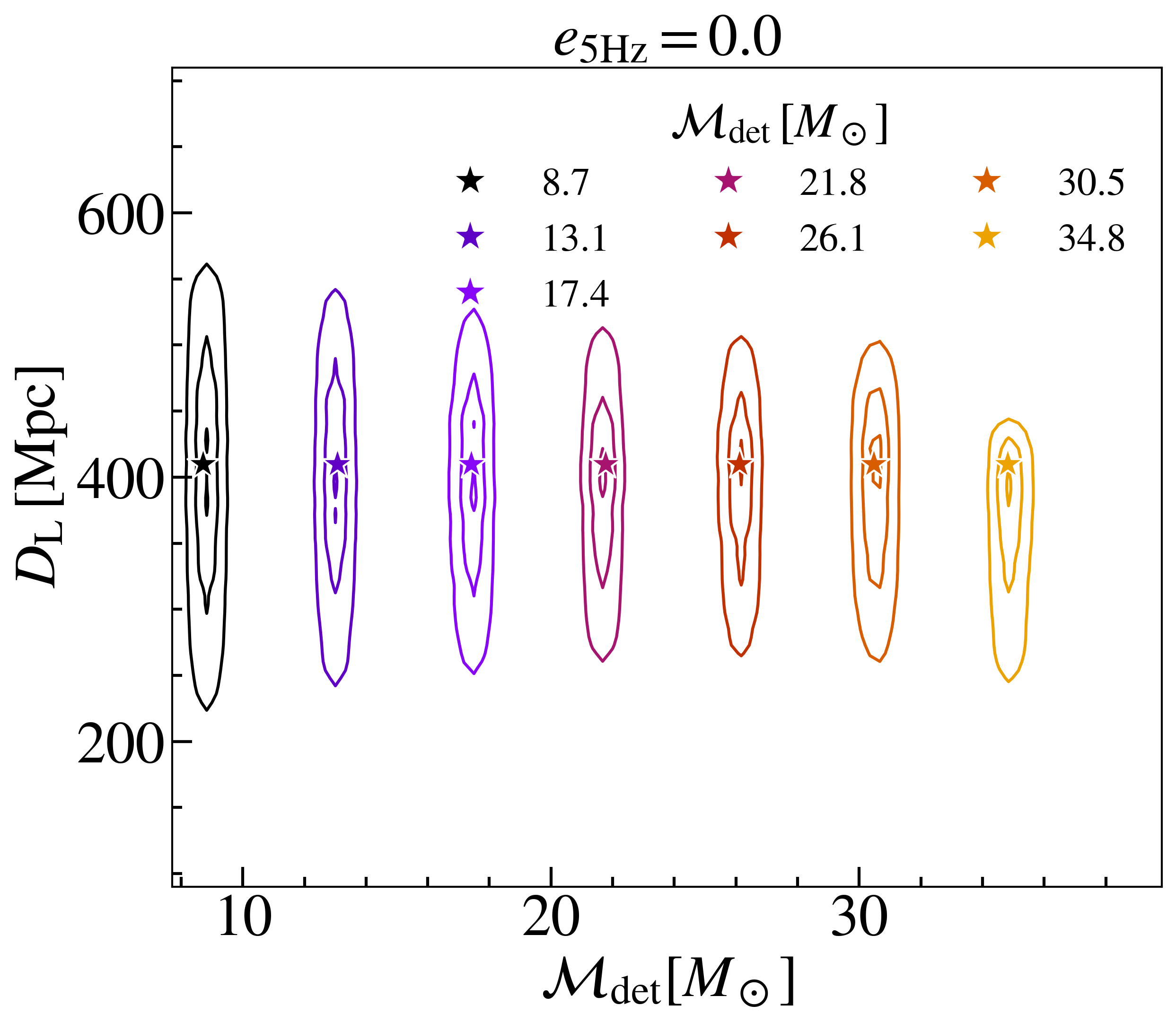}
    \includegraphics[width=0.32\textwidth]{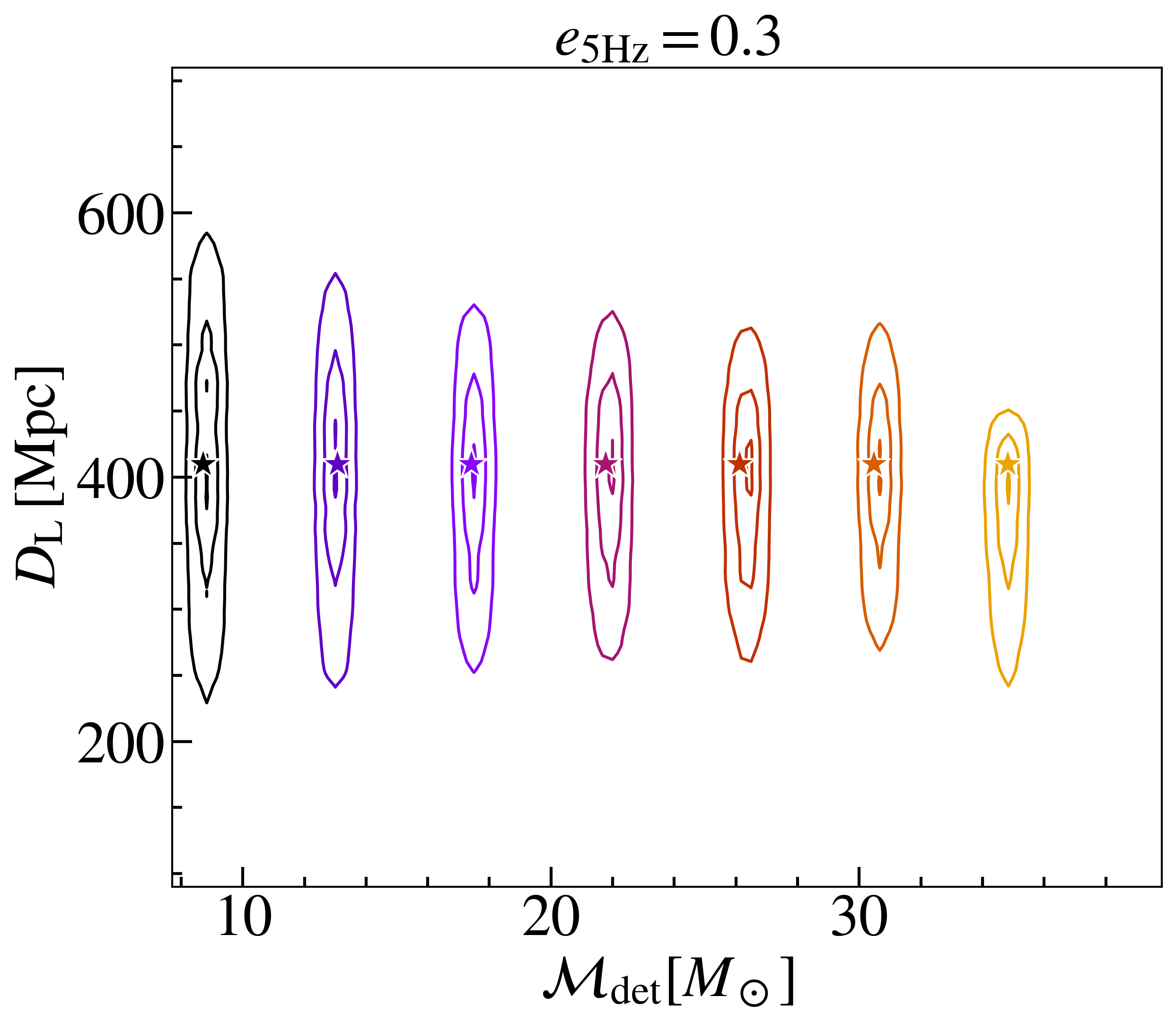}
    \includegraphics[width=0.32\textwidth]{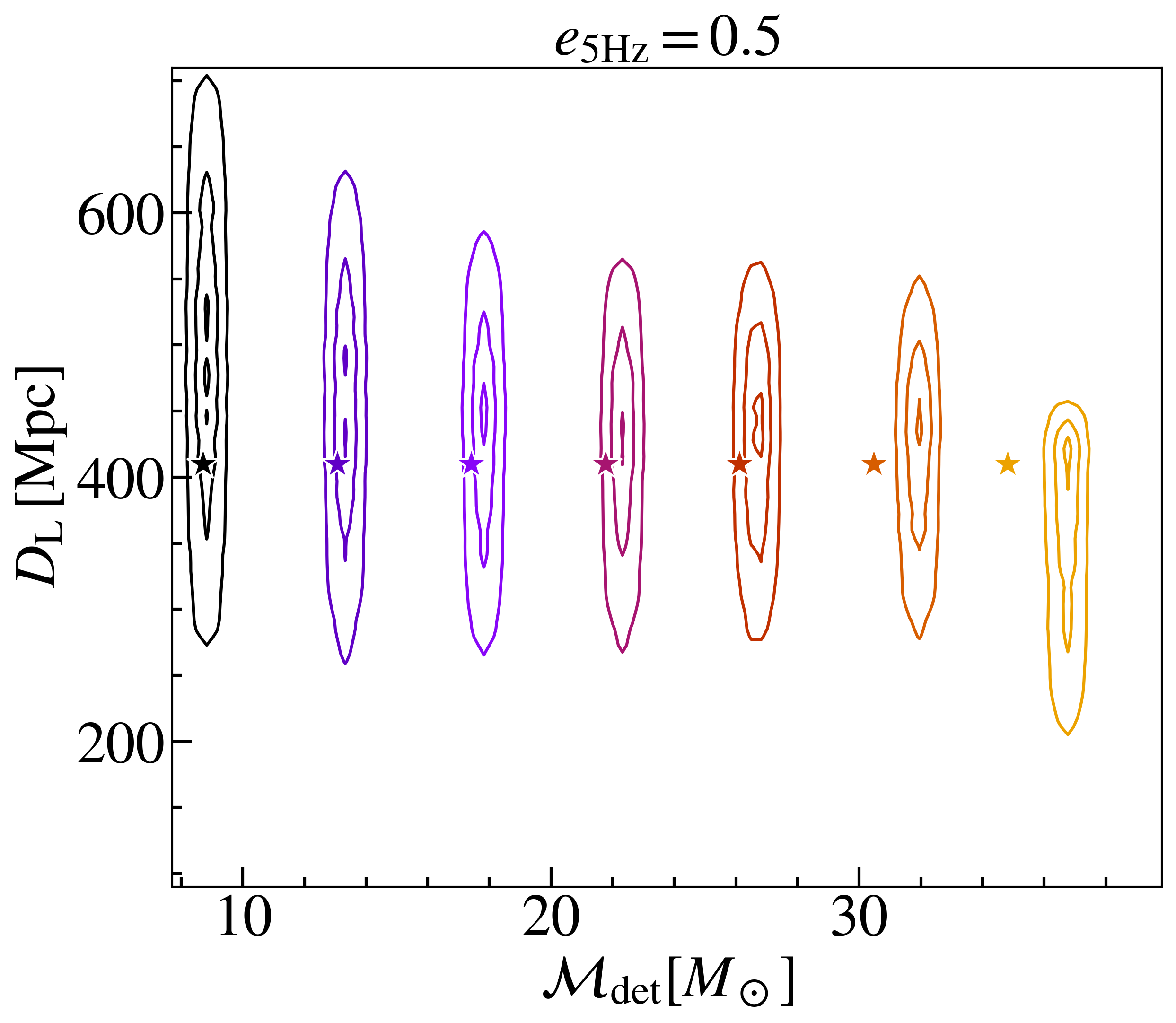}
    \caption{ Posterior distributions of the luminosity distance ($D_{\mathrm{L}}$) versus chirp mass ($\mathcal{M}$) from parameter estimation using a non-spinning waveform model, for injections with varying initial eccentricity. The star markers indicate the injected chirp mass values ($\mathcal{M}$). The contours represent the 90\%, 50\%, and 10\% credible intervals obtained from parameter estimation. }
    \label{fig:chirp_dl_nospin}
\end{figure*}

\begin{figure*}[htbp!]
\centering
    \includegraphics[width=0.32\textwidth]{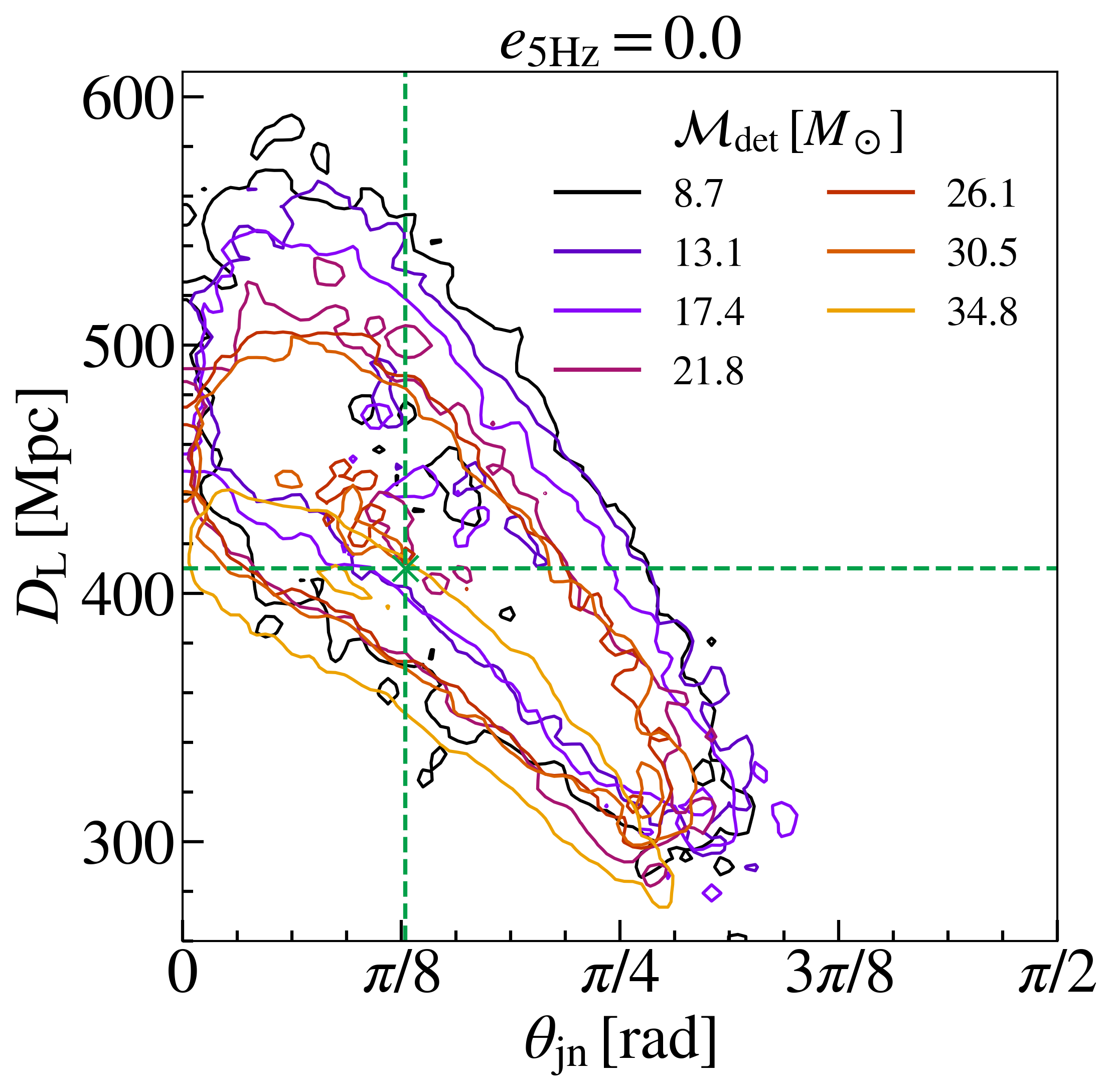} 
    \includegraphics[width=0.32\textwidth]{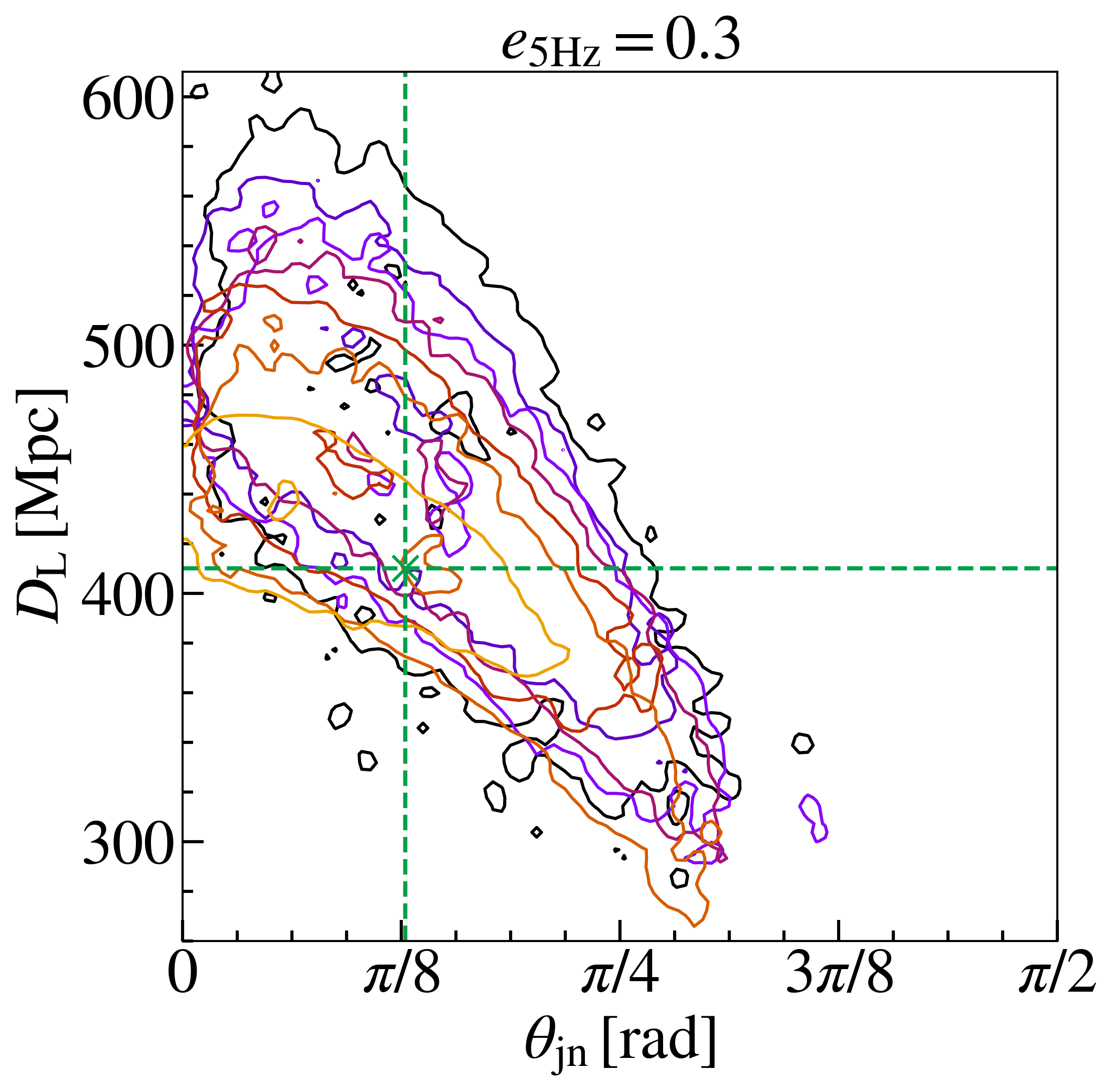}
    \includegraphics[width=0.32\textwidth]{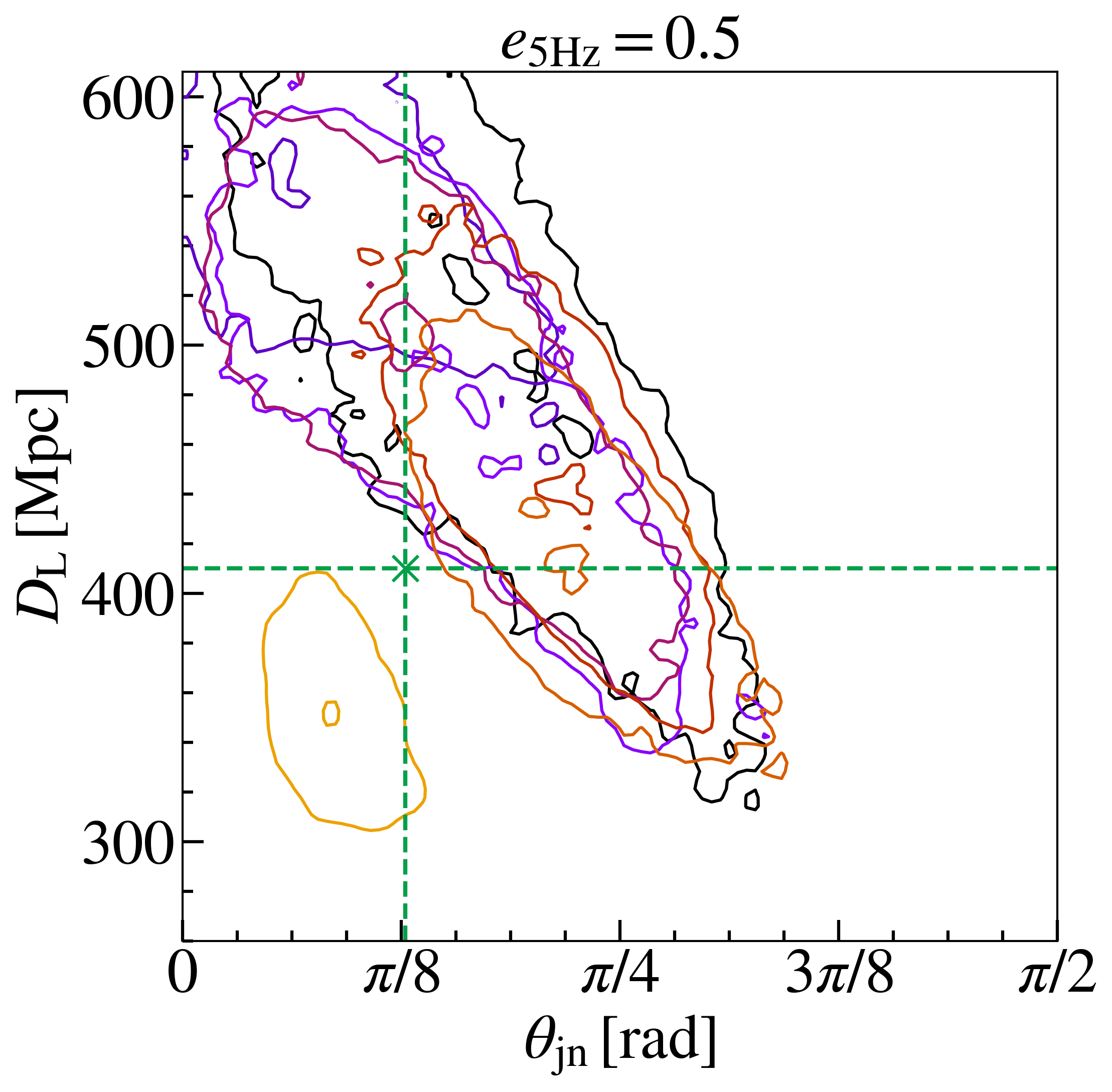}
    
    \caption{Posterior distributions of the luminosity distance ($D_{\mathrm{L}}$) versus inclination angle ($\theta_{\mathrm{jn}}$) from parameter estimation of injected signal with varying eccentricities, using a precessing waveform model. The contours represent $95 \% $  and $5 \%$ credible intervals. The dotted green line indicates the injected values.}
    \label{fig:inc_dl_precess}
\end{figure*}

\section{Results} 
\label{sec:results}

{In order to investigate the systematic effects of orbital eccentricity on parameter estimation in the eBBH parameter space, we perform a set of injections using the \textsc{TEOBResumS--Dal\'{\i}} waveform model~\cite{Chiaramello:2020ehz, Nagar:2021gss, Nagar:2021xnh, Nagar:2023zxh}. Each system is injected with an initial eccentricity in the range $0 \leq e \leq 0.5$ at $5 \, \rm{Hz}$ that varies in steps of $0.1$. The mass ratio is fixed to $1$ for all injections while the total mass varies in the range $20 \, M_\odot \leq \mathrm{M} \leq 80 \, M_\odot$ in steps of $10 M_\odot$. We mainly focus on the mass space, which refers to the component-mass parameter space of the binary system, with particular focus on the $30-40 \, M_\odot$ range where the Rate \& Population distribution exhibits a pronounced peak~\citep{KAGRA:2021duu}. We also estimate eccentricity at $10$ Hz using the standard definition, which is discussed in Section 2.1 in~\cite{O3_eBBH_Collab}. %\RG{Would it also be possible to quote the eccentricity at 20 Hz, i.e. the lower limit of the detector band?} we are planning to add to table.
These signals are injected with zero noise into a two-detector network configuration consisting of Hanford and Livingston with design sensitivities~\cite{LIGOScientific:2014pky}. The sky position for these injections is fixed to an arbitrary location, as listed in Table \ref{tab:paramlist}. }

{We primarily look at recovered posterior distributions for the component masses and the chirp mass $\mathcal{M}= \frac{(m_1 m_2)^{3/5}}{(m_1 + m_2)^{1/5}}$, and the luminosity distance $D_\mathrm{L}$. We also look at the posterior distributions for the two spin parameters $\chi_\mathrm{p}$ and $\chi_\mathrm{eff}$. $\chi_\mathrm{p}$ is the spin precession parameter~\cite{Schmidt:2012rh,Schmidt:2014iyl} which measures the spin effect in the plane of the binary orbit. It is defined as}
\begin{equation}
\chi_\mathrm{p} = \max \left( \chi_{1\perp}, \frac{4q + 3}{4 + 3q} \, q \, \chi_{2\perp} \right)
\end{equation}
{where, $q=m_2/m_1$ is mass ratio, $\chi_{1\perp}=\chi_1 \sin{\theta_{1}}$ and  $\chi_{2\perp}=\chi_2 \sin{\theta_{2}}$. $\chi_{1} =c |\vec{S}_{1}| / G m_{1}^2$ and $\chi_{2} =c |\vec{S}_{2}| / G m_{2}^2$. $\theta_{1}$ and $\theta_{2}$ are the angles between the orbital angular momentum and the spin vectors $\vec{S}_{1}$ and $\vec{S}_{2}$  respectively. The other is the effective spin parameter $\chi_\mathrm{eff}$~\cite{Ajith:2009bn,Santamaria:2010yb} which measures the effect of spin in the direction of the orbital angular momentum and is defined as }

\begin{equation}
    \chi_{\mathrm{eff}} =
\frac{m_1 \chi_{1\parallel} + m_2 \chi_{2\parallel}}
{m_1 + m_2}
\end{equation}
{where, $\chi_{1 \parallel} = \chi_{1} \cos\theta_{1}$ and $\chi_{1 \parallel} = \chi_{1} \cos\theta_{1}$.}

\subsection{Recovery with quasi-circular spinning BBH waveform model (\text{IMRPhenomXPHM})}

 %\TRC{The sub heading says non-aligned but we mention aligned spin in this paragraph. Also, changed it to quasi-circular instead of circular.}
 
 Parameter estimation is carried out using the Bilby framework as discussed in Sec \ref{sec:bayesian_pe} with IMRPhenomXPHM ~\cite{PhysRevD.103.104056,PhysRevD.111.104019}, a precessing frequency-domain waveform model that assumes circular orbits and includes the inspiral, merger, and ringdown phases but no eccentricity. For this study, we limit ourselves to the dominant mode only, excluding the effects of higher modes for both injection and recovery. We performed the analysis using three different recovery parameter setups: one with spin precession, a second with an aligned-spin and a third with a no-spin. 

Each injection is analysed using a prior distribution on the binary parameters, carefully selected to ensure sufficient sampling of the relevant parameter space. Specifically, we adopt priors that are uniform in spin magnitudes and redshifted component masses, isotropic in spin orientations, sky location, and binary orientation, and a power law prior on luminosity distance. %\TRC{We are using a power law prior in luminosity distance.} %The full set of priors used is detailed as follows: 
%The intrinsic source-frame masses are computed by dividing the redshifted (detector-frame) masses by $(1 + z)$, where $z$ is the cosmological redshift. The redshift is found from the luminosity distance assuming a flat $\Lambda$CDM cosmology with a Hubble constant $H_0 = 67.9\, \mathrm{km\,s^{-1}\, Mpc^{-1}}$ and a matter density parameter $\Omega_m = 0.3065$. %\TRC{Since we are not calculating source frame masses, do we need to talk about redshifted masses at all?} 

%We performed parameter estimation for each injection. 
Figures~\ref{fig:comp_mass_precess}  and~\ref{fig:comp_mass_nospin} show the credible interval contours for all injections in the detector frame component-mass plane for the precessing-spin and no-spin recoveries, respectively. Each subplot corresponds to a fixed eccentricity, arranged in increasing order of eccentricity. For each eccentricity, we considered seven different equal mass binaries with total masses {$\mathrm{M}_{\mathrm{det}}=m_{1,\mathrm{det}}+m_{2,\mathrm{det}}$} ranging from $20~M_{\odot}$ to $80~M_\odot$. For each injection, we display the {90\%, 50\%, and 10\%}  probability density credible regions obtained from parameter estimation, with the injected values marked by a star of the same color. The legend indicates the injected {detector-frame {chirp mass $\mathcal{M}_{\mathrm{det}}$.} The first subplot in each plot serves as a reference case, where the injection was performed using the \textsc{TEOBResumS--Dal\'{\i}} waveform with $e_{5\mathrm{Hz}} = 0$, and the recovery was carried out using the IMRPhenomXPHM waveform model. In no-spin parameter-estimation recovery, the inferred masses are generally consistent with the injected masses. Although the recovered values do not exactly match the injected value, the injected parameters lie within the 90\% credible interval in most cases. %The detailed parameter estimation results are listed in Tables \ref{tab:rec_} and \ref{tab:rec_precess}, where we report both the injected and recovered parameters, along with their corresponding 90\% credible intervals.
In both configurations, we can see that the deviation is a function of both eccentricity and total mass. %The high-mass system shows a deviation from low eccentricity compared to the lower-mass system at high eccentricity. 
For high-mass systems, the posterior distributions of the component masses begin to show clear deviations even at relatively low injected eccentricities. In contrast, for lower-mass systems, comparable deviations in the mass posteriors only appear at higher injected eccentricities. We observe larger deviations for precessing recovery compared to the non-spinning configuration. %However, in the case of a no-spin configuration, 
%for eccentricities $e_{5\mathrm{Hz}} < 0.4$, the injected values remain well within the contours, showing no significant deviations. 
{We see a similar trend in the deviations for aligned spin recovery as well.} %These trends are further supported by the numerical results presented in Table \ref{tab:rec_precess}.
The deviation tends to increase for high-mass systems and is smaller for low-mass binaries, with the injected values lying outside the contours for systems with {$\mathrm{M}_\mathrm{det} > 40~M_\odot$}. In the precessing recovery setup (numerical values in Table \ref{tab:rec_aligned}), we observe a significant deviation from the injected values for systems with 
$e_{5\mathrm{Hz}} > 0.3$, corresponding to $e_{10\mathrm{Hz}} \approx 0.15$.  The deviation initially appears in high-mass systems and, with increasing eccentricity, becomes evident in lower-mass systems as well. This trend suggests that the deviation is directly correlated with both the total mass and the eccentricity of the system. The contribution of the mass ratio was not explored in the present study. The observed behavior differs slightly from the no-spin case, primarily due to the influence of the effective precession spin parameter ($\chi_p$). {In Table ~\ref{tab:rec_aligned} of Appendix A, we list the median recovered values along with the $90 \%$ bounds for the various recovered parameters for precessing recovery.}}

For the same parameter estimation runs, we also studied the deviations in the inferred chirp mass in detector frame $\mathcal{M}_\mathrm{det}$ and the inferred luminosity distance  $D_\mathrm{L}$. In Figures ~\ref{fig:chirp_dl_precess} and~\ref{fig:chirp_dl_nospin} we plot the recovered posteriors of ${\mathcal{M}}_{\mathrm{det}}$ and $D_\mathrm{L}$ for precessing and non-spinning configurations, respectively. The labeling and color scheme follow the same convention as in the previous plots. These figures show the credible region contour for all injections, where the stars denote the injected values. We see that for recovery with the precessing waveform model, the injected values of chirp mass and luminosity distance are largely recovered except when $\mathcal{M}_\mathrm{det} \geq$ 13 $M_\odot$ for $e_{5\mathrm{Hz}} = 0.5$. We show only three cases of initial eccentricity $e_{5\mathrm{Hz}}$ to better understand this behavior. For recovery with a {non-spinning} waveform model, the injected values of the chirp mass and luminosity distance are recovered for all cases up to $e_{5\mathrm{Hz}} = 0.4$. For $e_{5\mathrm{Hz}} = 0.5$, the injected values of the chirp mass and luminosity distance for systems with $\mathcal{M}_\mathrm{det} \geq$ 30 $M_\odot$ are not recovered. We observe a significant difference between the precessing and {non-spinning} configurations in this parameter space at high eccentricity.

To understand this clearly, we have plotted the absolute deviation in the mass parameter space, specifically the deviations in the chirp mass ${\cal{M}}_\mathrm{det}$. Figure~\ref{fig:abs_delta_chirp_mass} shows the absolute deviation in the chirp mass ${\cal{M}}_\mathrm{det}$ between the injected value and the median value of the posterior distribution. The top, middle and bottom panels correspond to the {recovery with precessing spin, aligned spin and non-spinning} waveforms, respectively. We observe two key trends associated with changes in mass and eccentricity. Systems with lower eccentricity have an overall smaller deviation compared to systems with larger eccentricity. For a given injected eccentricity, the deviations increase as the total mass of the binary is increased. Larger deviations are observed in the case of aligned/precessing spin configurations compared to the non-spinning case, which is mainly due to the modulation introduced by spin parameters.

%see major difference between t
%However, we observe noticeable discrepancies for $e_{5\mathrm{Hz}} > 0.4$ at a starting frequency of $5$ Hz, while at $10$ Hz, such deviations appear from $e_{10\mathrm{Hz}} > 0.24$. 
%\addnew{ CHECK : T} \addnew{ADD STATEMENT ABOUT ALIGNED SPIN}

{In Figure \ref{fig:inc_dl_precess}, we plot inclination angle versus the luminosity distance posterior samples. We see that the systems with high eccentricity, when recovered using a precessing waveform model, exhibit a shift in the recovered inclination angle $\theta_{\rm{jn}}$ towards an edge-on system.  Note that we use the same labeling and color scheme for the contours as in the other figures; however, the injected values of luminosity distance and inclination angle are the same, and are denoted by green dashed lines.}

{{Figure~\ref{fig:chip_violin} in Appendix B shows the posterior distributions of the precessing component spin parameter $\chi_\mathrm{p}$ for different eccentric injections when recovered using a precessing spin waveform.} There is an overall increase in the support for precessing spins across all injected masses {as the value of the initial eccentricity increases.} As the eccentricity is increased, the $\chi_\mathrm{p}$ posterior distributions shift to larger values and become narrower. For eccentricities up to $0.2$, we see that the distributions of $\chi_\mathrm{p}$ favor a low precessing spin. For $e_{5\mathrm{Hz}} = 0.3$ and above, the posteriors favor high precessing spins. In the case of  $\chi_{\mathrm{eff}}$ recovery as shown in Figure~\ref{fig:chieff_violin} in Appendix B, we see that for low to moderate initial eccentricity, the recovered median values are close to zero. For $e_{5\mathrm{Hz}} \geq 0.4$ we see that the recovered values are non-negligible. 

To quantify these recoveries and determine which one best represents the multi-dimensional parameter space, we calculate $\mathcal{B}$ between two hypotheses, circular precessing recovery ($\mathcal{H}_P$) and circular non-spinning recovery ($\mathcal{H}_{NS}$). Figure~\ref{fig:log_bayes} shows the estimated $\mathcal{B}$ for all simulations with varying masses and eccentricities. {$\mathcal{B} > 1$ indicates precessing recovery has stronger support compared to non-spinning recovery. We also noticed that as eccentricity increases, $\mathcal{B}$ also increases.} {Additionally, we observed that this trend is not strongly dependent on chirp mass.} %\TRC{$\mathcal{B}>1$ means one hypothesis A has more support than hypothesis B. }

%We observe two distinct patterns based on the chirp mass: for $\mathcal{M} \lesssim 21.8 M_{\odot}$, the aligned-spin hypothesis is favoured, whereas for higher chirp masses, the precessing-spin hypothesis is more supported. Additionally, the $\mathcal{B}_{PA}$ depends on eccentricity, with higher eccentricities showing larger differences between the two hypotheses.

\begin{figure}[htbp!]
\centering
    \includegraphics[width=0.49\textwidth]{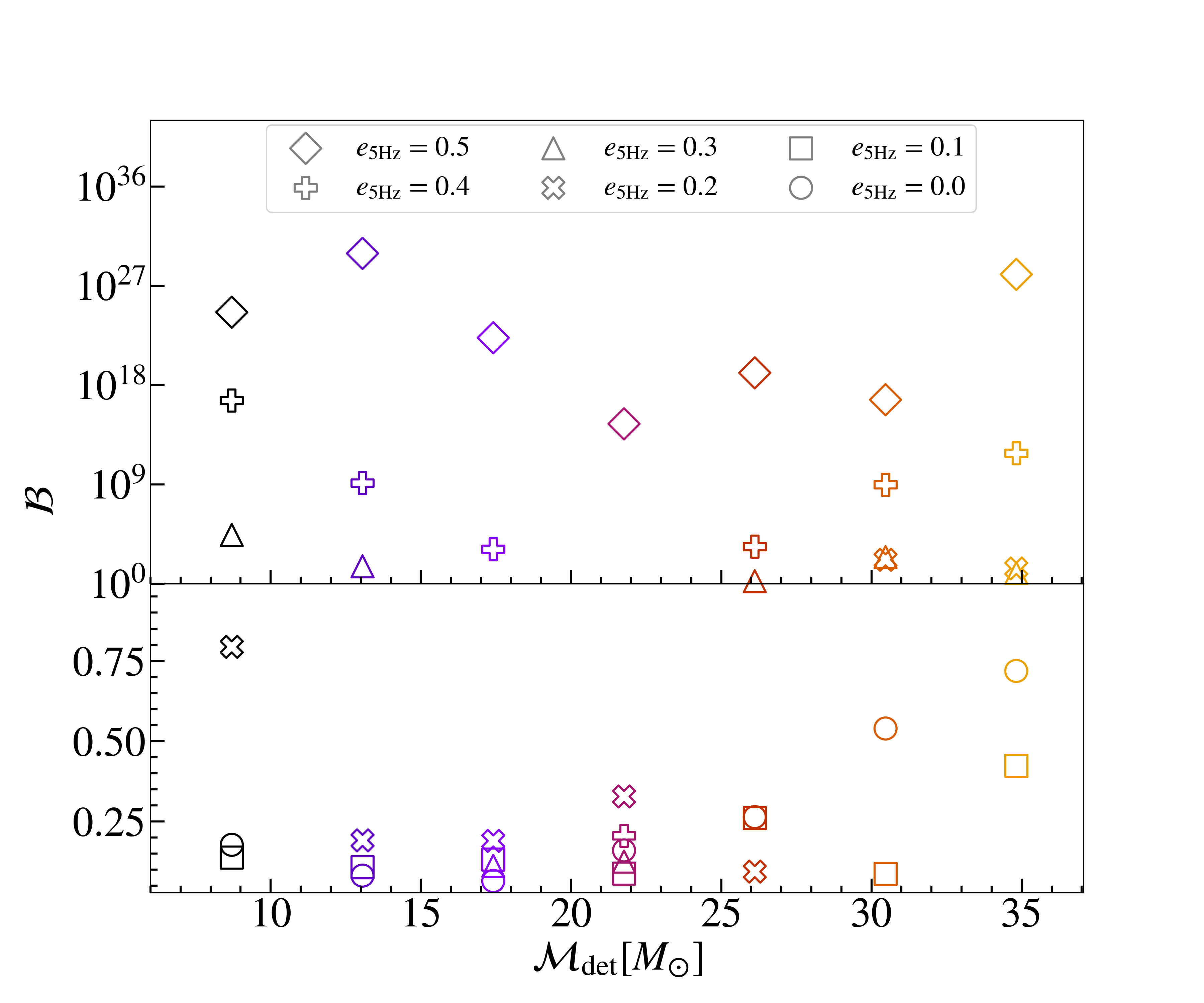}
    %\captionsetup{width=0.4\textwidth}
    \caption{Bayes factor $\mathcal{B}$ calculated between precessing spin and non-spinning recovery of simulated equal-mass non-spinning eccentric systems with various chirp masses (horizontal axis) and eccentricities (symbols). The plot is split into two panels about $\mathcal{B} = 1$. $\mathcal{B} > 1$ indicates that a precessing spin recovery of the injection is preferred over the no spin recovery.}
    \label{fig:log_bayes}
\end{figure}

\begin{figure}[htbp!]
\centering
    \includegraphics[width=0.47\textwidth]{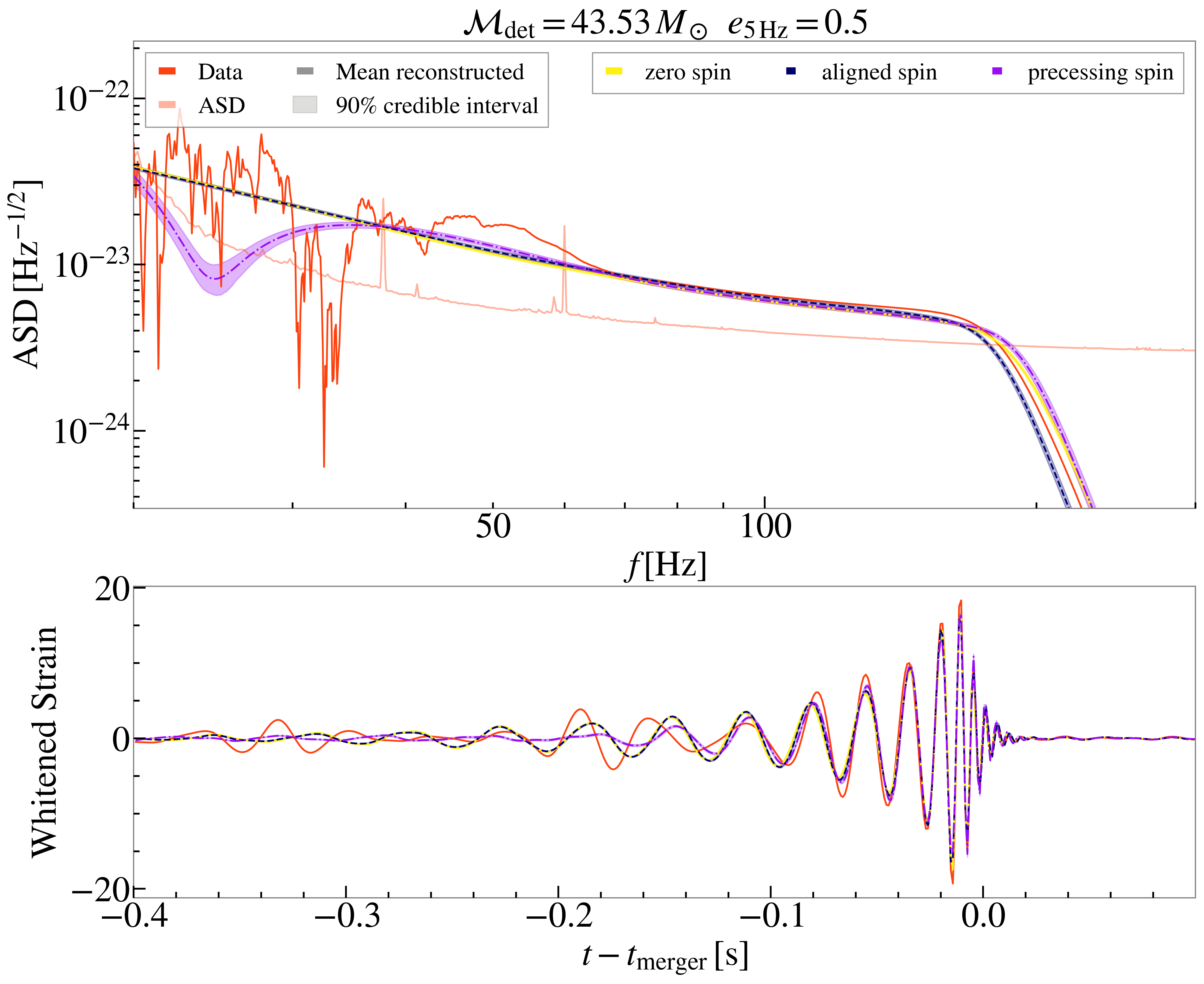}
    
    \caption{ Waveform reconstruction for the three circular recovery with no spin, and spin (aligned and precessing) plotted against the injected signal. All the waveforms recover the late inspiral and merger part of the signal well. }
    \label{fig:wvf_rec_circ}
\end{figure}

\begin{figure*}[htbp!]
\centering
    \includegraphics[width=0.9\textwidth]{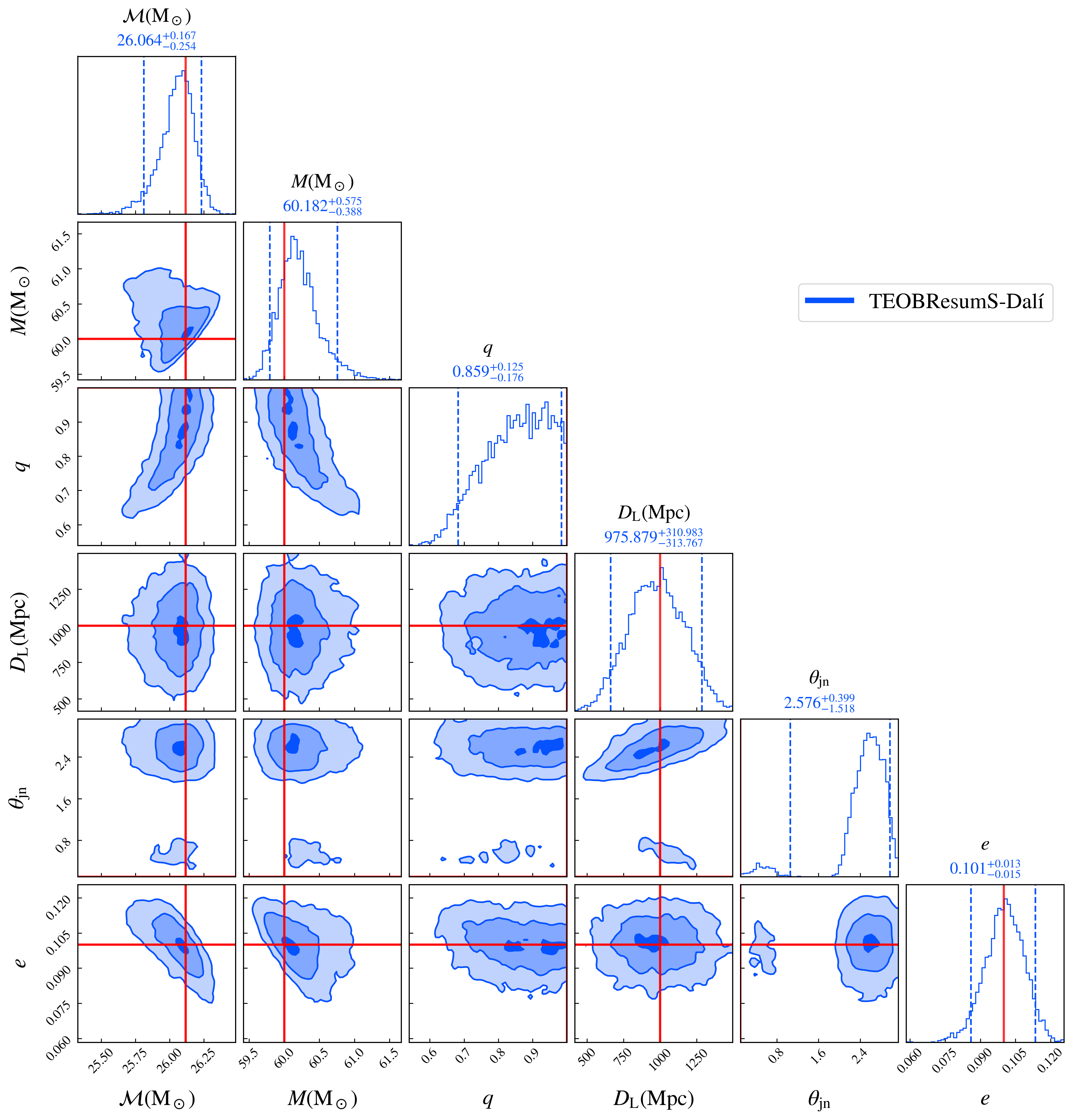}
    
    \caption{Posterior samples from parameter estimation of a non-spinning eccentric signal injected and recovered using \textsc{TEOBResumS--Dal\'{\i}}. Red lines indicate the injected parameter values. Median recovered value with $90\%$ credible intervals quoted above each parameter posterior plot.}
    \label{fig:teob_inj_rec}
\end{figure*}

{In figure \ref{fig:wvf_rec_circ} we show the reconstructed waveform as well as the injected waveform in frequency domain (top panel) and time domain (bottom panel) for an injection with eccentricity $e_{5\mathrm{Hz}}=0.5$. The solid red line represents data with an injected signal. Yellow, dashed black, and violet-colored lines correspond to non-spinning, aligned spin, and precessing spin waveforms, respectively. This plot shows that all recovery configurations recover the late inspiral and merger part of the signal better than the early inspiral part.}      

\subsection{Recovery with eccentric aligned spin binary black hole waveform model:  \textsc{TEOBResumS--Dal\'{\i}}}
%e_{20Hz}=0.1 and e_{5Hz} \apro 0.3
{ We have injected a non-spinning equal mass binary with ${\cal{M}}=26.1 M_{\odot}$ and $e_{20Hz}=0.1$, with all other parameters fixed as the same as previous injections. Here, we have recovered using the eccentric waveform model \textsc{TEOBResumS-Dal\'i}, where we have employed a uniform prior on eccentricity. Note that the recovered eccentricity is estimated at $f_{ref}=20\,\text{Hz}$. For these parameter estimations, we have used a slightly different range of priors {compared to the runs in the previous section}. Figure \ref{fig:teob_inj_rec} shows the posterior distributions for parameters such as chirp mass, mass ratio, eccentricity and distance with \textsc{TEOBResumS--Dal\'{\i}}. %The recovery with IMRPhenomonXPHM is represented by blue color curves and TEOEResumS-Dali as gray color. 
The injected values are indicated by the orange vertical lines. {We also perform parameter estimation on this system with a precessing \textsc{IMRPhenomXPHM} in order to calculate the Bayes factor between the two runs.} We find that the Bayes factor between precession recovery w.r.t eccentricity recovery is  $\mathcal{B}_{PE}=1.18 \times 10^{-7}$. This indicates eccentric recovery is more accurate in multi-dimensional parameters. This example highlights the importance of using eccentric waveform models for reliable parameter estimation of such sources. We plan to analyze selected special events with this model in future work. }%\RG{Oh awesome. Some of the parameters (like polarisation and phase) might be removed from the cornerplot I suppose, given that the two are (i) completely degenerate for (2,$\pm$2) waveforms and (ii) as such, not measured}

\begin{comment}

\begin{table*}[t]
\centering

\begin{tabular}{p{1.5cm}p{1.5cm}|p{1.5cm}p{1.5cm}p{1.5cm}p{1.5cm}p{1.5cm}p{1.5cm}p{1.5cm}p{1.5cm}}
\hline
\hline
\centering

\multicolumn{2}{c|}{\textbf{Injection Parameters}} & \multicolumn{8}{c}{\textbf{$\Delta$ Recovery Parameters}} \\\hline
$\underset{(M_\odot)}{M_\text{inj}}$ & $\underset{(5\,\mathrm{Hz})}{e}$ &
$\underset{(M_\odot)}{\Delta \mathcal{ M}}$ & $\underset{(M_\odot)}{\Delta m_1}$ & $\underset{(M_\odot)}{\Delta m_2}$ &
$\Delta \chi_{\mathrm{ eff}}$ & $\underset{(\mathrm{Gpc})}{\Delta D_{\mathrm{L}}}$ &
$\Delta \chi_{\mathrm{ p}}$ & 
$\underset{(\mathrm{deg}^2)}{\Delta\Omega}$ & $\mathrm{SNR}$ \\\hline
\endhead
\hline
\hline
\endfoot
\endlastfoot

\end{tabular}
\caption{Injection and Recovery Parameters}
\label{tab:injection_recovery}
\end{table*}
\end{comment}

%\begin{table*}[ht]
%\centering
%\setlength{\tabcolsep}{5pt}
%\renewcommand{\arraystretch}{1.2}

%\input{rec_table_aligned}  % This includes the table content

%\caption{Recovery of parameters using an aligned spin IMRPhenomXPHM waveform for various eccentric injections. Median value of recovered parameters quoted along with the $90\%$ credible levels. }
%\label{tab:rec_precess}
%\end{table*}

\section{Conclusion}
\label{sec:conclusion}
In this study, we systematically investigated the impact of orbital eccentricity on parameter estimation and the biases introduced when this effect is ignored. We utilized the \textsc{TEOBResumS--Dal\'{\i}} model to simulate eccentric non-spinning systems as injections and the \textsc{IMRPhenomXPHM} model of circular (non-eccentric) systems for recovery. We use three different configurations, namely, the precessing case (involving non-aligned component spins), the non-precessing case (involving spins aligned with the orbital angular momentum) and the non-spinning case (where all spin componenets are set to be zero). Our analysis covered a broad range of binary black hole masses ($20~M_{\odot}$–$80~M_{\odot}$) and eccentricities ($0 \leq e_{5\mathrm{Hz}} \leq 0.5$) and revealed that the accuracy of recovered parameters strongly depends on both total mass and eccentricity.  Especially for high-mass systems, the posterior distributions of the component masses begin to show clear deviations even at relatively low injected eccentricities. But in the case of lower-mass systems, comparable deviations in the mass posteriors only appear at higher injected eccentricities. For low eccentricities ($e_{5\mathrm{Hz}} < 0.3$), the injected parameters were generally recovered within the 90\% credible intervals. However, we observe deviations in the recovered parameters as the eccentricity is increased. In particular, systems with $e_{5\mathrm{Hz}} > 0.4$ and total masses above $40~M_{\odot}$ showed significant discrepancies in the recovered chirp mass, spin precession, and luminosity distance. 
 
In this paper, we also demonstrated that using the \textsc{TEOBResumS--Dal\'{\i}} waveform model (which is applicable for eccentric binary systems) within the Bilby framework allows us to recover eccentricity well. Using this setup, we can reliably estimate the source parameters of events, including the eccentricity. 

These findings suggest that waveform models that assume circular orbits, such as IMRPhenomXPHM, lose accuracy when applied to gravitational-wave signals from highly eccentric mergers, particularly in the high-mass regime. The observed bias in parameters such as inclination and precessing spin further underscores the interplay between eccentricity and spin precession, which can lead to misinterpretation of astrophysical parameters if eccentricity is neglected.

In future work, we plan to extend this systematic study to higher masses, as well as the high-spin and high-mass-ratio space. Additionally, we plan to conduct a study to evaluate the effect of these biases when including higher-order modes. This will result in a better understanding of where we can ignore eccentricity in parameter estimation for real events. Incorporating next-generation eccentric waveform models into parameter estimation pipelines will be crucial to improve source characterization for upcoming observing runs as well as to understand the origin of these binaries. %Additionally, systematic injection campaigns in realistic noise environments will help evaluate the detectability and parameter bias of eccentric signals in actual detector data, thereby paving the way for the robust identification of eccentric binary black hole mergers in the LIGO–Virgo–KAGRA network.

\section{Acknowledgement}
We gratefully acknowledge the support of LIGO and Virgo for the provision of computational resources. G.V. and T.R.C acknowledge the support of the National Science Foundation under grant PHY-2207728 and PHY-2513124.  I.B. acknowledges the support of the National Science Foundation under grants \#1911796, \#2110060 and \#2207661 and of the Alfred P. Sloan Foundation. We would like to thank Pratyusava Baral for useful comments on the manuscript.
This research has made
use of data, software and/or web tools obtained from
the Gravitational Wave Open Science Center (https:
//www.gw-openscience.org), a service of LIGO Laboratory, the LIGO Scientific Collaboration and the Virgo Collaboration.
LIGO is funded by the U.S. National Science Foundation. Virgo is funded by the French Centre National de Recherche Scientifique (CNRS), the Italian Istituto Nazionale della Fisica Nucleare (INFN) and the
Dutch Nikhef, with contributions by Polish and Hungarian institutes. This material is based upon work supported by NSF's LIGO Laboratory, which is a major facility fully funded by the National Science Foundation. Computations were performed in the CIT cluster provided by LIGO Laboratory and Hawk computing center at Cardiff University supported by STFC grant.

This document has a LIGO preprint number LIGO-P2600058.

%\newpage
\bibliography{bib.bib} 

\newpage
\appendix
\section{Parameter estimation using the precessing IMRPhenomXPHM waveform model }
%%%%%%%%%%%%%%%%%%%%%%%%%%%%%%%%%%%%%%%%%%%%%%%%%%
This appendix presents tables summarizing the recovery of source parameters obtained using the precessing IMRPhenomXPHM waveform model for a set of eccentric signal injections with zero component spins and equal component masses. For each injection, parameter estimation is performed and the median values of the recovered parameters are reported, together with their associated 
90\% credible intervals. These results quantify the accuracy and robustness of the waveform model in recovering intrinsic and extrinsic source properties across different eccentricity configurations.

\begin{table*}[ht!]
\centering
\setlength{\tabcolsep}{5pt}
\renewcommand{\arraystretch}{1.2}

%\begin{tabular}{p{1cm}p{1cm}p{1cm}p{1cm}|p{1.5cm}p{1.5cm}p{1.5cm}p{1.5cm}p{1.5cm}p{1.5cm}}
\begin{tabular}{ccccc|ccccccc}
%{p{1.2cm}p{1.2cm}p{1.2cm}p{1.2cm}p{1.2cm}p{1.2cm}p{1.2cm}p{1.2cm}p{1.2cm}p{1.2cm}p{1.2cm}p{1.2cm}}
%20 & 8.7 & 0.00 & 0.0000 & 27.9 & $8.7$ & $11.2$ & $8.9$ & $438.5$ & $0.1$ & $0.0$ & $27.7$ \\
%\hline
%\hline
%\centering

%\multicolumn{4}{r}{\textbf{ Injection Parameters}} & \multicolumn{6}{r}{\textbf{Recovery Parameters}} \\\hline
\hline
\hline
\multicolumn{5}{c}{\textbf{Injection Parameters}} & \multicolumn{7}{c}{\textbf{Recovery Parameters}} \\
\hline
$\underset{(\mathrm{M}_\odot)}{M}$ & $\underset{(\mathrm{M}_\odot)}{\mathcal{M}}$ & $\underset{(5\,\mathrm{Hz})}{e}$ & $\underset{(10\,\mathrm{Hz})}{e}$ &
$\mathrm{SNR}$ & $\underset{(\mathrm{M}_\odot)}{\mathcal{M}}$ & $\underset{(\mathrm{M}_\odot)}{m_1}$ & $\underset{(\mathrm{M}_\odot)}{m_2}$ & $\underset{(\mathrm{Mpc})}{ D_{\mathrm{L}}}$ &
$\chi_{\mathrm{p}}$ & 
$\chi_{\mathrm{eff}}$ &
$\mathrm{SNR}$ \\\hline

20 & 8.7 & 0.00 & 0.0000 & 27.9 & $8.7_{-0.0}^{+0.0}$ & $11.2_{-1.1}^{+2.4}$ & $8.9_{-1.4}^{+1.0}$ & $438.5_{-103.1}^{+97.6}$ & $0.1_{-0.1}^{+0.2}$ & $0.0_{-0.0}^{+0.1}$ & $27.7_{-1.8}^{+1.8}$\\ 
30 & 13.1 & 0.00 & 0.0000 & 38.2 & $13.0_{-0.0}^{+0.0}$ & $16.6_{-1.4}^{+2.7}$ & $13.5_{-1.8}^{+1.2}$ & $448.6_{-105.2}^{+74.7}$ & $0.1_{-0.1}^{+0.2}$ & $0.0_{-0.0}^{+0.0}$ & $38.0_{-1.8}^{+1.8}$\\ 
40 & 17.4 & 0.00 & 0.0000 & 48.0 & $17.4_{-0.1}^{+0.1}$ & $21.4_{-1.3}^{+2.2}$ & $18.6_{-1.7}^{+1.2}$ & $442.8_{-103.7}^{+69.2}$ & $0.1_{-0.1}^{+0.2}$ & $-0.0_{-0.0}^{+0.0}$ & $47.8_{-1.9}^{+1.9}$\\ 
50 & 21.8 & 0.00 & 0.0000 & 56.8 & $21.7_{-0.1}^{+0.1}$ & $26.6_{-1.5}^{+2.6}$ & $23.3_{-2.0}^{+1.5}$ & $417.8_{-82.9}^{+67.9}$ & $0.1_{-0.1}^{+0.2}$ & $-0.0_{-0.0}^{+0.0}$ & $56.6_{-1.9}^{+1.9}$\\ 
60 & 26.1 & 0.00 & 0.0000 & 65.1 & $25.9_{-0.2}^{+0.2}$ & $31.6_{-1.6}^{+2.8}$ & $28.1_{-2.3}^{+1.5}$ & $420.8_{-78.8}^{+60.1}$ & $0.1_{-0.1}^{+0.2}$ & $-0.0_{-0.0}^{+0.0}$ & $65.0_{-2.0}^{+1.9}$\\ 
70 & 30.5 & 0.00 & 0.0000 & 73.1 & $30.2_{-0.3}^{+0.3}$ & $36.6_{-1.7}^{+2.7}$ & $32.9_{-2.4}^{+1.6}$ & $415.8_{-74.7}^{+59.1}$ & $0.1_{-0.1}^{+0.2}$ & $-0.0_{-0.0}^{+0.0}$ & $73.0_{-2.0}^{+1.9}$\\ 
80 & 34.8 & 0.00 & 0.0000 & 80.8 & $34.5_{-0.4}^{+0.4}$ & $41.8_{-2.0}^{+3.0}$ & $37.6_{-2.8}^{+1.9}$ & $386.6_{-75.0}^{+36.0}$ & $0.1_{-0.1}^{+0.2}$ & $-0.0_{-0.0}^{+0.0}$ & $80.7_{-1.8}^{+1.8}$\\ 
\hline  
20 & 8.7 & 0.10 & 0.0479 & 27.9 & $8.7_{-0.0}^{+0.0}$ & $11.1_{-1.0}^{+2.2}$ & $9.0_{-1.4}^{+0.9}$ & $452.7_{-119.5}^{+89.5}$ & $0.1_{-0.1}^{+0.2}$ & $0.0_{-0.0}^{+0.1}$ & $27.6_{-1.8}^{+1.8}$\\ 
30 & 13.1 & 0.10 & 0.0475 & 38.3 & $13.1_{-0.0}^{+0.0}$ & $16.7_{-1.6}^{+2.6}$ & $13.5_{-1.7}^{+1.3}$ & $450.6_{-95.4}^{+76.5}$ & $0.1_{-0.1}^{+0.2}$ & $0.0_{-0.0}^{+0.0}$ & $38.0_{-1.8}^{+1.8}$\\ 
40 & 17.4 & 0.10 & 0.0472 & 48.0 & $17.4_{-0.1}^{+0.1}$ & $21.5_{-1.4}^{+2.2}$ & $18.6_{-1.7}^{+1.2}$ & $437.4_{-94.9}^{+66.9}$ & $0.1_{-0.1}^{+0.2}$ & $-0.0_{-0.0}^{+0.0}$ & $47.8_{-1.8}^{+1.8}$\\ 
50 & 21.8 & 0.10 & 0.0468 & 56.8 & $21.7_{-0.1}^{+0.1}$ & $26.4_{-1.3}^{+2.4}$ & $23.5_{-1.9}^{+1.2}$ & $430.8_{-77.8}^{+64.3}$ & $0.1_{-0.1}^{+0.2}$ & $-0.0_{-0.0}^{+0.0}$ & $56.5_{-1.8}^{+1.9}$\\ 
60 & 26.1 & 0.10 & 0.0465 & 65.1 & $25.9_{-0.2}^{+0.2}$ & $31.3_{-1.4}^{+2.6}$ & $28.3_{-2.1}^{+1.3}$ & $427.2_{-77.1}^{+59.6}$ & $0.1_{-0.1}^{+0.2}$ & $-0.0_{-0.0}^{+0.0}$ & $64.8_{-1.9}^{+1.9}$\\ 
70 & 30.5 & 0.10 & 0.0461 & 73.3 & $30.4_{-0.3}^{+0.3}$ & $36.8_{-1.7}^{+2.8}$ & $33.2_{-2.5}^{+1.6}$ & $421.1_{-71.1}^{+59.4}$ & $0.1_{-0.1}^{+0.2}$ & $-0.0_{-0.0}^{+0.0}$ & $73.1_{-2.0}^{+2.1}$\\ 
80 & 34.8 & 0.10 & 0.0458 & 80.9 & $34.6_{-0.4}^{+0.4}$ & $41.9_{-1.9}^{+3.1}$ & $37.8_{-2.9}^{+1.8}$ & $383.0_{-70.2}^{+32.7}$ & $0.1_{-0.1}^{+0.2}$ & $-0.0_{-0.0}^{+0.0}$ & $80.7_{-1.7}^{+1.7}$\\ 
\hline  
20 & 8.7 & 0.20 & 0.0999 & 27.9 & $8.7_{-0.0}^{+0.0}$ & $11.4_{-1.2}^{+2.7}$ & $8.9_{-1.5}^{+1.0}$ & $471.9_{-118.2}^{+76.3}$ & $0.2_{-0.1}^{+0.3}$ & $0.0_{-0.0}^{+0.1}$ & $27.6_{-1.8}^{+1.8}$\\ 
30 & 13.1 & 0.20 & 0.0991 & 38.3 & $13.1_{-0.0}^{+0.0}$ & $16.2_{-1.1}^{+2.8}$ & $14.0_{-2.0}^{+1.0}$ & $456.9_{-101.4}^{+73.7}$ & $0.1_{-0.1}^{+0.2}$ & $0.0_{-0.0}^{+0.0}$ & $37.9_{-1.7}^{+1.8}$\\ 
40 & 17.4 & 0.20 & 0.0984 & 48.1 & $17.4_{-0.1}^{+0.1}$ & $21.4_{-1.3}^{+2.2}$ & $18.7_{-1.7}^{+1.2}$ & $429.5_{-88.8}^{+75.8}$ & $0.1_{-0.1}^{+0.2}$ & $-0.0_{-0.0}^{+0.0}$ & $47.7_{-1.9}^{+1.9}$\\ 
50 & 21.8 & 0.20 & 0.0977 & 56.9 & $21.7_{-0.1}^{+0.1}$ & $26.3_{-1.3}^{+2.2}$ & $23.5_{-1.8}^{+1.2}$ & $420.5_{-77.7}^{+68.1}$ & $0.1_{-0.1}^{+0.2}$ & $-0.0_{-0.0}^{+0.0}$ & $56.4_{-1.8}^{+1.9}$\\ 
60 & 26.1 & 0.20 & 0.0969 & 65.3 & $26.2_{-0.2}^{+0.2}$ & $31.4_{-1.2}^{+2.6}$ & $28.8_{-2.2}^{+1.2}$ & $415.1_{-75.1}^{+70.8}$ & $0.1_{-0.1}^{+0.2}$ & $0.0_{-0.0}^{+0.0}$ & $64.8_{-2.1}^{+1.9}$\\ 
70 & 30.5 & 0.20 & 0.0962 & 73.2 & $29.9_{-0.3}^{+0.3}$ & $36.7_{-2.0}^{+2.8}$ & $32.3_{-2.4}^{+1.9}$ & $418.1_{-60.2}^{+50.3}$ & $0.2_{-0.1}^{+0.2}$ & $-0.1_{-0.0}^{+0.0}$ & $72.7_{-1.9}^{+1.8}$\\ 
80 & 34.8 & 0.20 & 0.0955 & 80.9 & $34.1_{-0.4}^{+0.4}$ & $41.1_{-1.8}^{+2.8}$ & $37.3_{-2.6}^{+1.7}$ & $389.1_{-68.8}^{+35.9}$ & $0.1_{-0.1}^{+0.2}$ & $-0.1_{-0.0}^{+0.0}$ & $80.3_{-1.8}^{+1.8}$\\ 
\hline  
20 & 8.7 & 0.30 & 0.1599 & 28.0 & $8.8_{-0.0}^{+0.0}$ & $11.9_{-1.6}^{+2.7}$ & $8.6_{-1.4}^{+1.3}$ & $459.5_{-107.5}^{+92.1}$ & $0.2_{-0.1}^{+0.2}$ & $0.1_{-0.0}^{+0.1}$ & $27.3_{-1.8}^{+1.7}$\\ 
30 & 13.1 & 0.30 & 0.1588 & 38.4 & $13.2_{-0.0}^{+0.0}$ & $16.6_{-1.3}^{+2.7}$ & $13.9_{-1.9}^{+1.1}$ & $466.4_{-85.1}^{+70.2}$ & $0.2_{-0.1}^{+0.2}$ & $0.0_{-0.0}^{+0.0}$ & $37.7_{-1.8}^{+1.6}$\\ 
40 & 17.4 & 0.30 & 0.1577 & 48.3 & $17.5_{-0.1}^{+0.1}$ & $21.3_{-1.1}^{+2.2}$ & $19.0_{-1.7}^{+1.0}$ & $433.0_{-88.1}^{+75.5}$ & $0.1_{-0.1}^{+0.2}$ & $-0.0_{-0.0}^{+0.0}$ & $47.4_{-1.9}^{+1.8}$\\ 
50 & 21.8 & 0.30 & 0.1566 & 57.1 & $21.9_{-0.1}^{+0.1}$ & $26.3_{-1.0}^{+2.0}$ & $24.1_{-1.7}^{+0.9}$ & $441.9_{-97.0}^{+65.7}$ & $0.1_{-0.1}^{+0.2}$ & $0.0_{-0.0}^{+0.0}$ & $56.0_{-1.9}^{+1.8}$\\ 
60 & 26.1 & 0.30 & 0.1555 & 65.7 & $26.0_{-0.2}^{+0.2}$ & $31.7_{-1.5}^{+2.8}$ & $28.3_{-2.3}^{+1.4}$ & $444.0_{-70.0}^{+54.3}$ & $0.2_{-0.1}^{+0.2}$ & $-0.0_{-0.0}^{+0.0}$ & $64.5_{-1.8}^{+1.9}$\\ 
70 & 30.5 & 0.30 & 0.1545 & 73.3 & $29.9_{-0.3}^{+0.3}$ & $36.0_{-1.4}^{+2.6}$ & $32.9_{-2.3}^{+1.4}$ & $412.1_{-84.9}^{+58.9}$ & $0.2_{-0.1}^{+0.2}$ & $-0.1_{-0.0}^{+0.0}$ & $71.9_{-1.9}^{+1.9}$\\ 
80 & 34.8 & 0.30 & 0.1534 & 81.6 & $35.4_{-0.4}^{+0.4}$ & $42.0_{-0.8}^{+0.9}$ & $39.4_{-0.7}^{+0.6}$ & $430.1_{-42.5}^{+26.9}$ & $0.5_{-0.3}^{+0.1}$ & $0.0_{-0.0}^{+0.0}$ & $80.2_{-1.7}^{+1.6}$\\ 
\hline  
20 & 8.7 & 0.40 & 0.2321 & 28.1 & $8.9_{-0.0}^{+0.0}$ & $12.0_{-1.2}^{+2.8}$ & $8.8_{-1.5}^{+0.9}$ & $505.0_{-98.0}^{+66.6}$ & $0.5_{-0.3}^{+0.3}$ & $0.1_{-0.0}^{+0.1}$ & $26.8_{-1.6}^{+1.5}$\\ 
30 & 13.1 & 0.40 & 0.2306 & 38.8 & $13.4_{-0.0}^{+0.1}$ & $16.7_{-1.2}^{+2.9}$ & $14.2_{-2.0}^{+1.0}$ & $485.1_{-92.3}^{+79.6}$ & $0.2_{-0.1}^{+0.4}$ & $0.1_{-0.0}^{+0.0}$ & $37.2_{-1.7}^{+1.8}$\\ 
40 & 17.4 & 0.40 & 0.2291 & 48.6 & $17.8_{-0.1}^{+0.1}$ & $21.8_{-1.1}^{+2.4}$ & $19.3_{-1.8}^{+1.1}$ & $479.4_{-87.0}^{+68.9}$ & $0.2_{-0.1}^{+0.2}$ & $0.1_{-0.0}^{+0.0}$ & $46.8_{-1.8}^{+1.7}$\\ 
50 & 21.8 & 0.40 & 0.2276 & 57.5 & $22.2_{-0.1}^{+0.1}$ & $26.8_{-1.2}^{+2.2}$ & $24.2_{-1.8}^{+1.1}$ & $439.2_{-91.7}^{+75.4}$ & $0.1_{-0.1}^{+0.2}$ & $0.0_{-0.0}^{+0.0}$ & $55.4_{-1.9}^{+1.9}$\\ 
60 & 26.1 & 0.40 & 0.2262 & 65.6 & $26.2_{-0.2}^{+0.2}$ & $31.6_{-0.7}^{+0.7}$ & $28.7_{-0.5}^{+0.6}$ & $470.1_{-41.9}^{+38.6}$ & $0.7_{-0.2}^{+0.1}$ & $-0.0_{-0.0}^{+0.0}$ & $63.0_{-1.6}^{+1.7}$\\ 
70 & 30.5 & 0.40 & 0.2248 & 73.9 & $30.5_{-0.4}^{+0.4}$ & $36.5_{-0.7}^{+0.7}$ & $33.5_{-0.6}^{+0.6}$ & $446.3_{-43.4}^{+34.1}$ & $0.5_{-0.1}^{+0.1}$ & $-0.0_{-0.0}^{+0.0}$ & $71.2_{-1.6}^{+1.7}$\\ 
80 & 34.8 & 0.40 & 0.2234 & 81.9 & $34.0_{-0.6}^{+0.8}$ & $41.5_{-0.6}^{+0.9}$ & $36.8_{-0.9}^{+1.2}$ & $380.0_{-37.9}^{+37.6}$ & $0.9_{-0.2}^{+0.1}$ & $-0.2_{-0.1}^{+0.1}$ & $79.0_{-1.7}^{+1.7}$\\ 
\hline  
20 & 8.7 & 0.50 & 0.3204 & 28.4 & $9.1_{-0.0}^{+0.0}$ & $13.4_{-1.1}^{+0.8}$ & $8.3_{-0.4}^{+0.7}$ & $493.9_{-109.1}^{+106.8}$ & $0.6_{-0.1}^{+0.4}$ & $0.2_{-0.0}^{+0.0}$ & $25.3_{-1.6}^{+1.6}$\\ 
30 & 13.1 & 0.50 & 0.3185 & 39.3 & $13.8_{-0.1}^{+0.1}$ & $18.0_{-0.8}^{+1.9}$ & $14.0_{-1.2}^{+0.6}$ & $560.8_{-58.8}^{+50.1}$ & $0.7_{-0.1}^{+0.1}$ & $0.2_{-0.0}^{+0.0}$ & $36.2_{-1.5}^{+1.6}$\\ 
40 & 17.4 & 0.50 & 0.3167 & 49.5 & $18.5_{-0.1}^{+0.1}$ & $22.6_{-1.2}^{+1.7}$ & $20.0_{-1.4}^{+1.1}$ & $485.9_{-102.1}^{+78.3}$ & $0.3_{-0.2}^{+0.2}$ & $0.1_{-0.0}^{+0.0}$ & $45.5_{-1.7}^{+1.7}$\\ 
50 & 21.8 & 0.50 & 0.3149 & 58.7 & $23.2_{-0.2}^{+0.2}$ & $27.7_{-0.7}^{+1.3}$ & $25.6_{-1.1}^{+0.7}$ & $490.8_{-89.1}^{+69.3}$ & $0.6_{-0.4}^{+0.2}$ & $0.1_{-0.0}^{+0.0}$ & $54.0_{-1.7}^{+1.8}$\\ 
60 & 26.1 & 0.50 & 0.3131 & 66.8 & $27.4_{-0.3}^{+0.3}$ & $34.1_{-0.6}^{+0.6}$ & $29.1_{-0.4}^{+0.5}$ & $447.7_{-64.9}^{+70.0}$ & $0.8_{-0.1}^{+0.1}$ & $0.1_{-0.0}^{+0.0}$ & $61.2_{-2.0}^{+2.0}$\\ 
70 & 30.5 & 0.50 & 0.3113 & 76.4 & $32.3_{-0.4}^{+0.4}$ & $39.6_{-0.6}^{+0.6}$ & $34.9_{-0.4}^{+0.4}$ & $416.5_{-59.0}^{+62.8}$ & $0.8_{-0.1}^{+0.1}$ & $-0.0_{-0.0}^{+0.0}$ & $70.6_{-2.0}^{+2.1}$\\ 
80 & 34.8 & 0.50 & 0.3096 & 84.1 & $34.7_{-0.8}^{+0.9}$ & $55.9_{-2.6}^{+2.4}$ & $29.1_{-1.5}^{+1.3}$ & $352.2_{-32.8}^{+34.7}$ & $1.0_{-0.0}^{+0.0}$ & $-0.2_{-0.1}^{+0.1}$ & $77.5_{-1.7}^{+1.8}$\\ 

\hline  
\hline 
%\endhead
%\hline
%\hline
%\endfoot
%\endlastfoot

\end{tabular}  % This includes the table content

\caption{Recovery of parameters using a precessing IMRPhenomXPHM waveform model for various eccentric injections with zero spin and equal component masses. Median value of recovered parameters quoted along with the $90\%$ credible levels. }
\label{tab:rec_aligned}
\end{table*}
%\newpage
\section{The $\chi_{\mathrm{eff}}$ and $\chi_\mathrm{p}$ recovery from precessing configuration :}
This appendix presents posterior distributions of the effective inspiral spin parameter, $\chi_{\mathrm{eff}}$, and the effective precession spin parameter, $\chi_{\mathrm{p}}$, obtained from parameter estimation analyses using the precessing IMRPhenomXPHM waveform model. The results are shown as violin plots for injections with varying chirp masses $\cal{M}$ and initial eccentricities. The shaded regions represent the full posterior distributions, while the coloured lines within each violin denote the corresponding 
90\% credible intervals. These figures illustrate the dependence of spin parameter recovery on the injected source properties and on the source's eccentricity.
\begin{figure*}[htbp!]
\centering
    \includegraphics[width=0.45\textwidth]{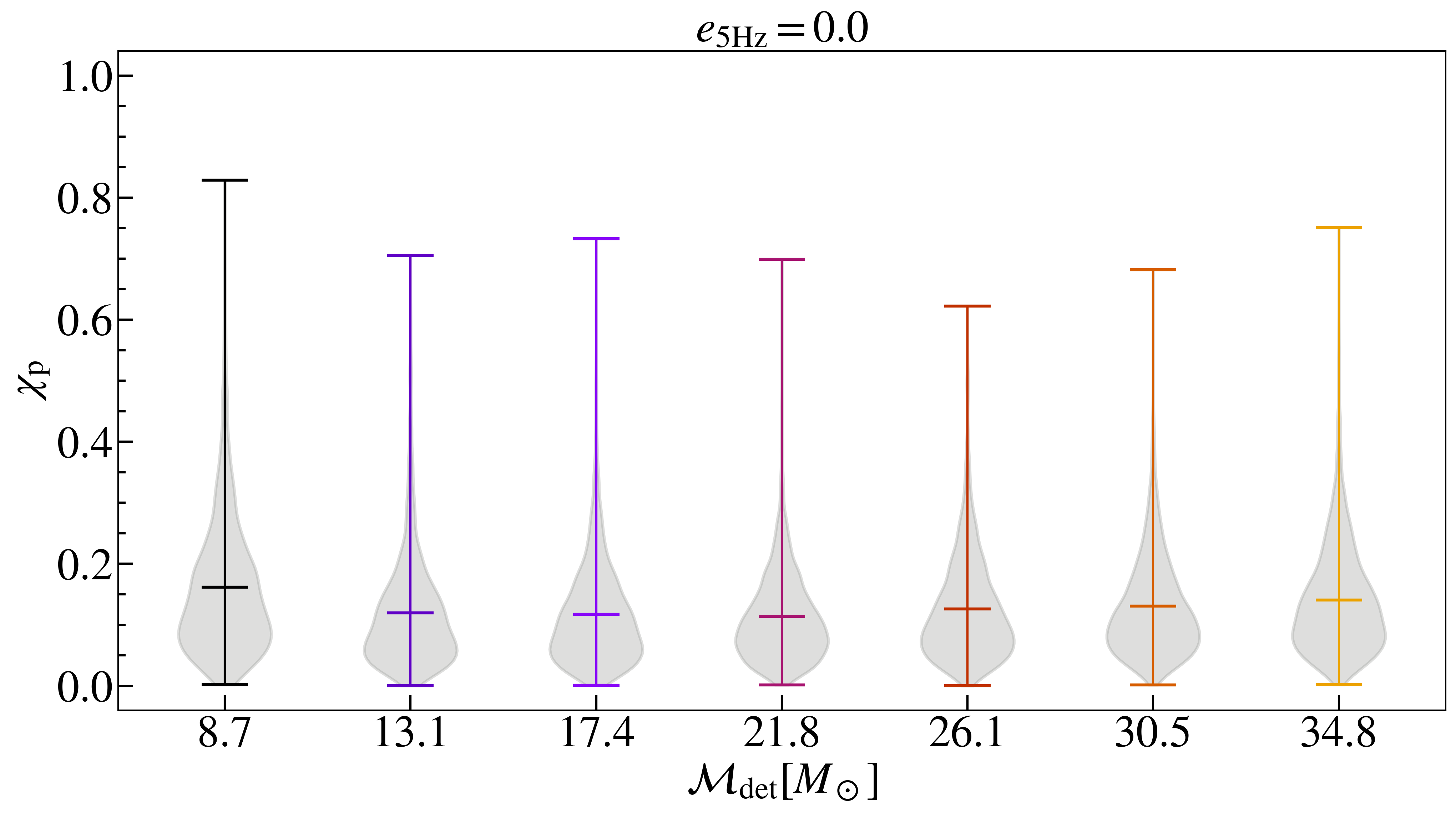}
    \includegraphics[width=0.45\textwidth]{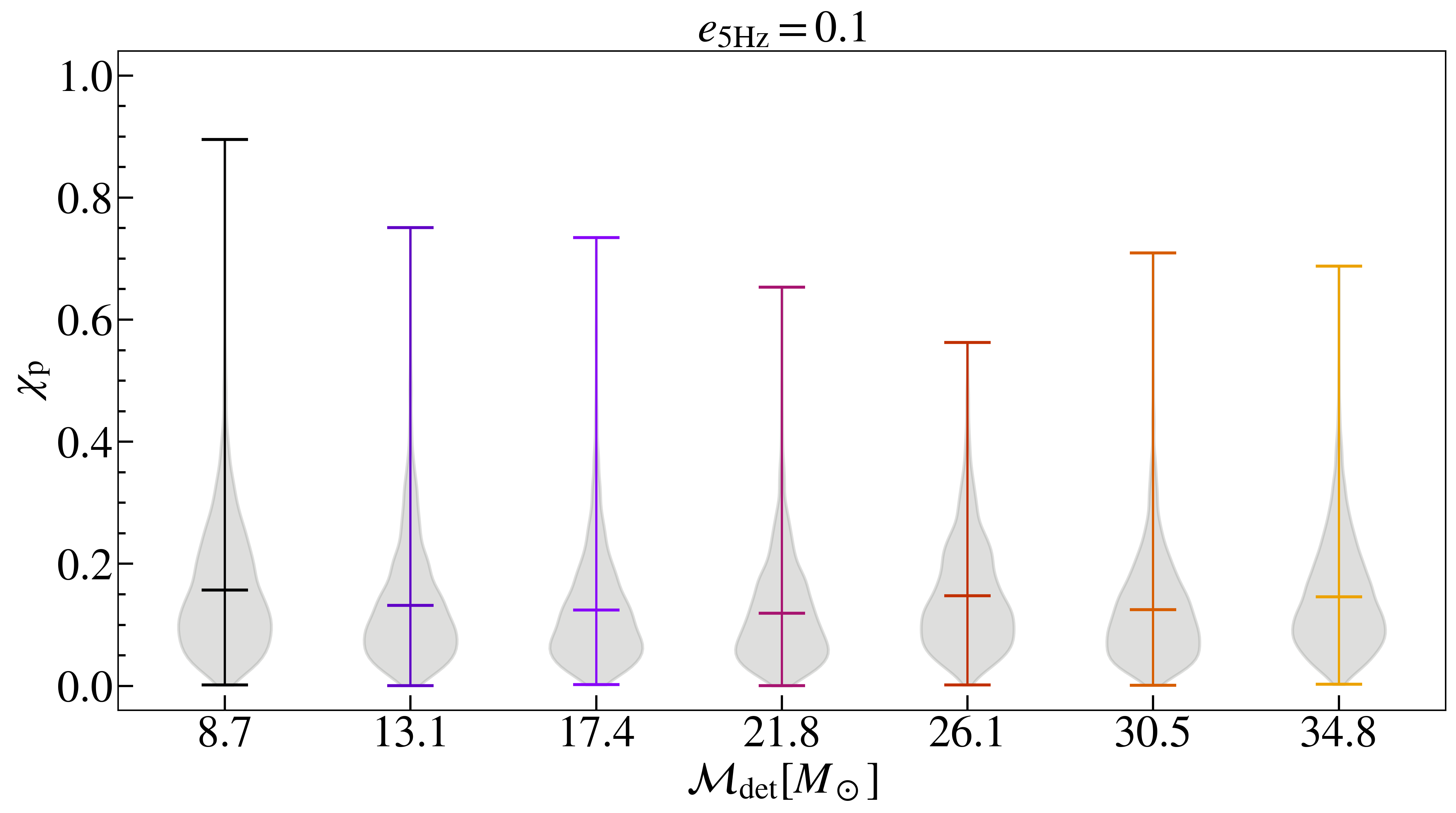}
    \includegraphics[width=0.45\textwidth]{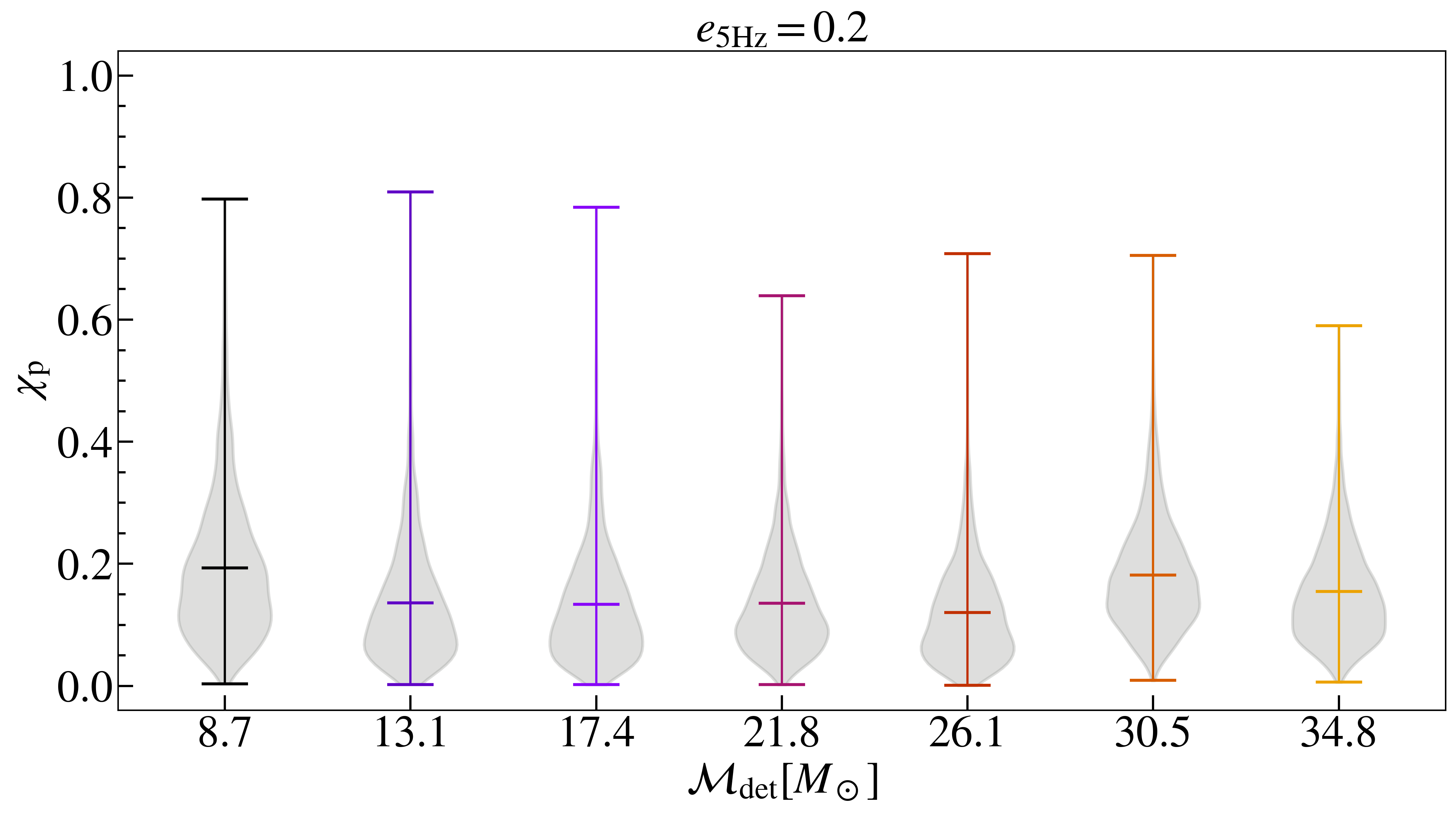}
    \includegraphics[width=0.45\textwidth]{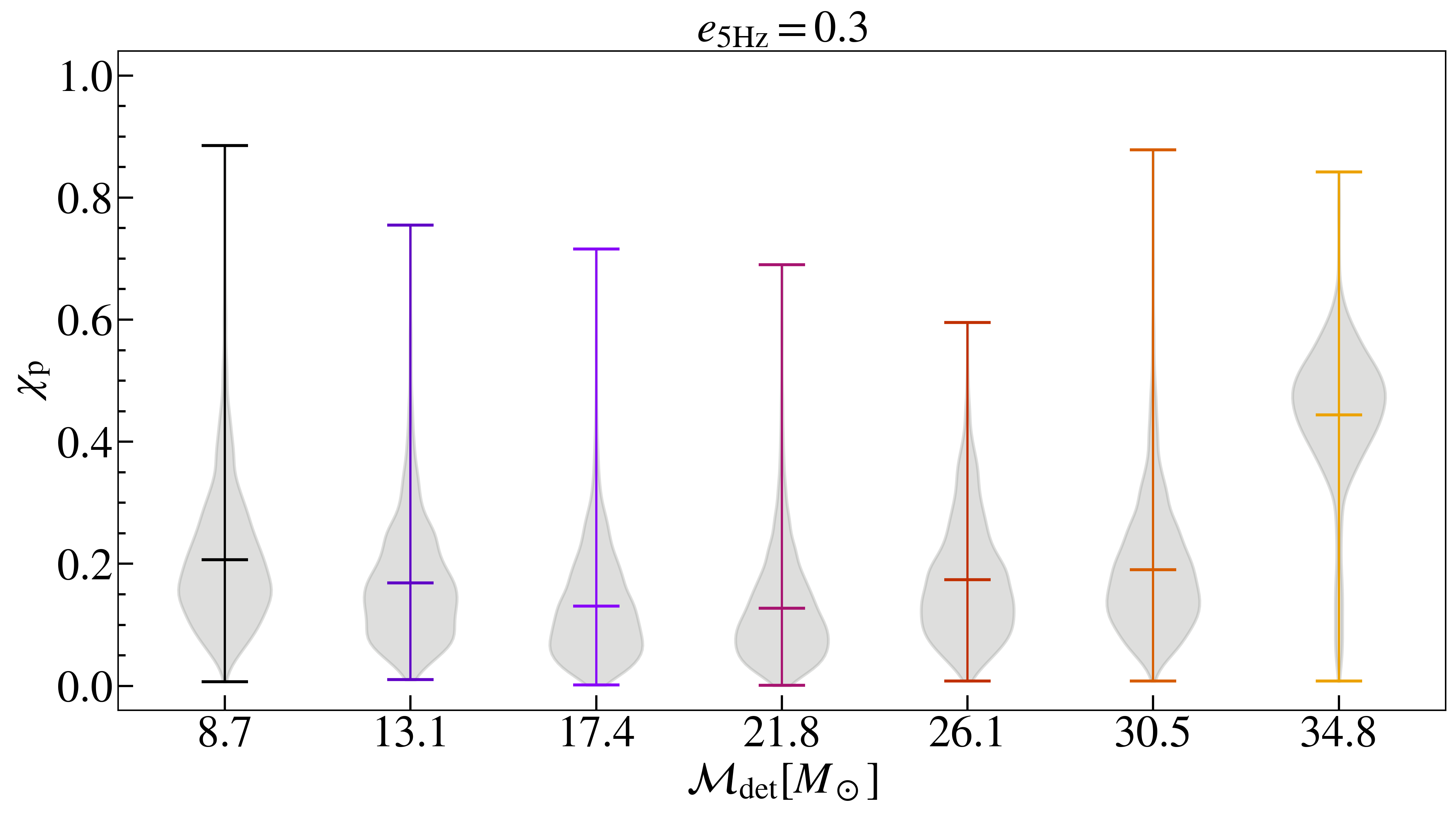}
    \includegraphics[width=0.45\textwidth]{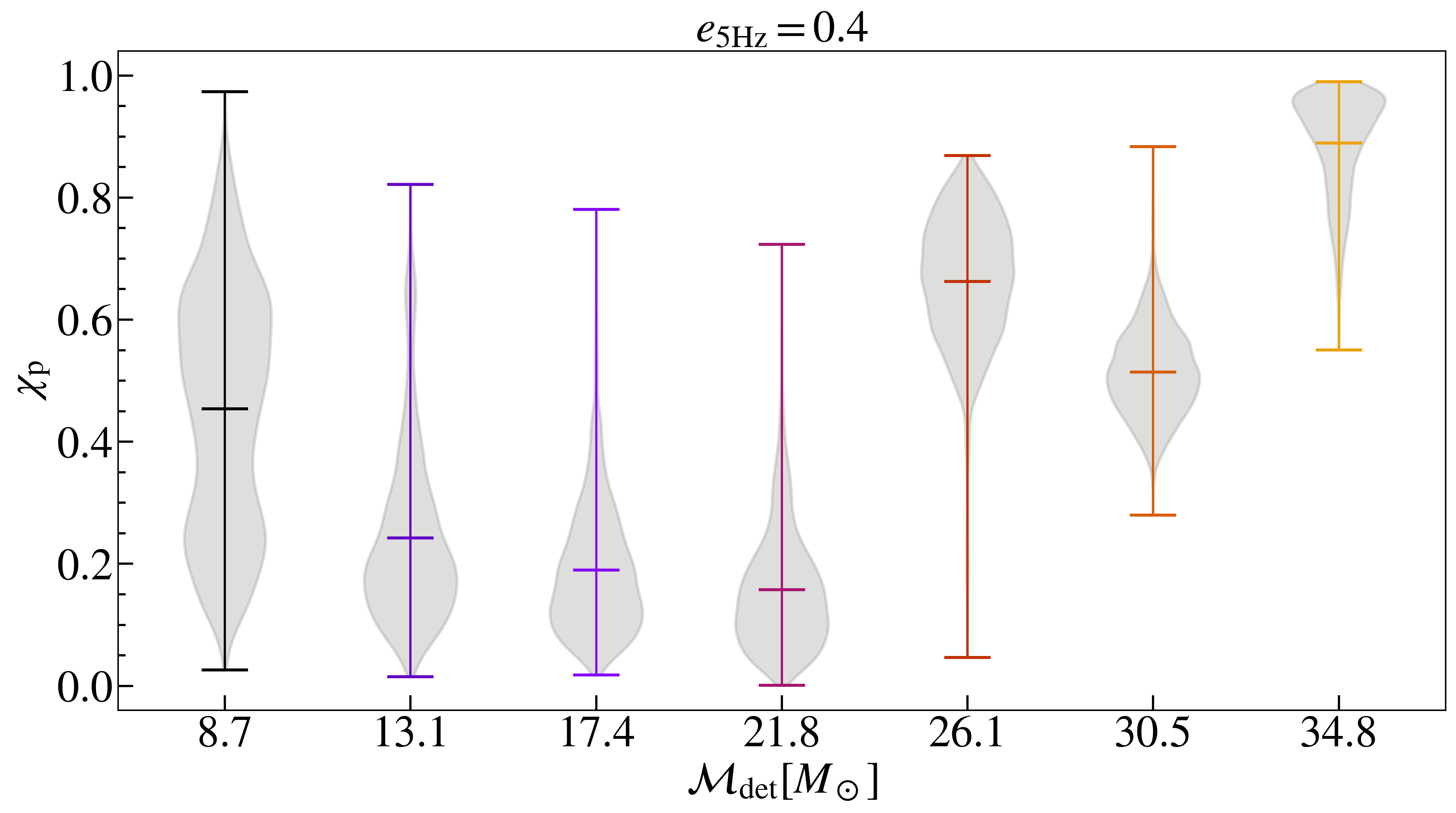}
    \includegraphics[width=0.45\textwidth]{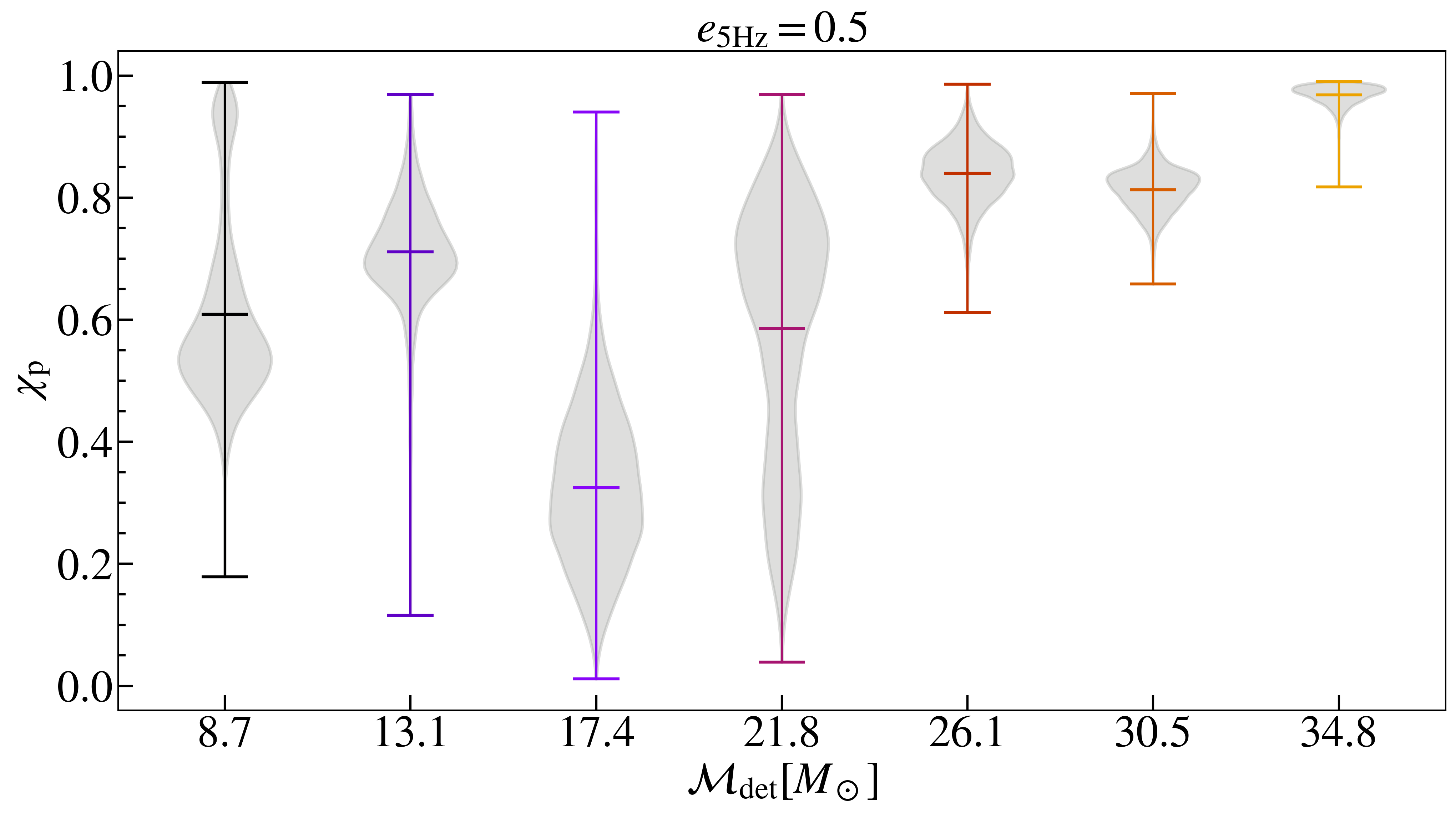}
    
    \caption{The $\chi_{\mathrm{p}}$ posterior violin plots for varying injected $\mathcal{M}$ and initial eccentricities.The colored lines inside the shaded posteriors indicate 90\% credible interval. Posterior distributions largely support zero precessing spins for all cases except for $e \geq 0.4$.}
    \label{fig:chip_violin}
\end{figure*}

\begin{figure*}[htbp!]
\centering
    \includegraphics[width=0.45\textwidth]{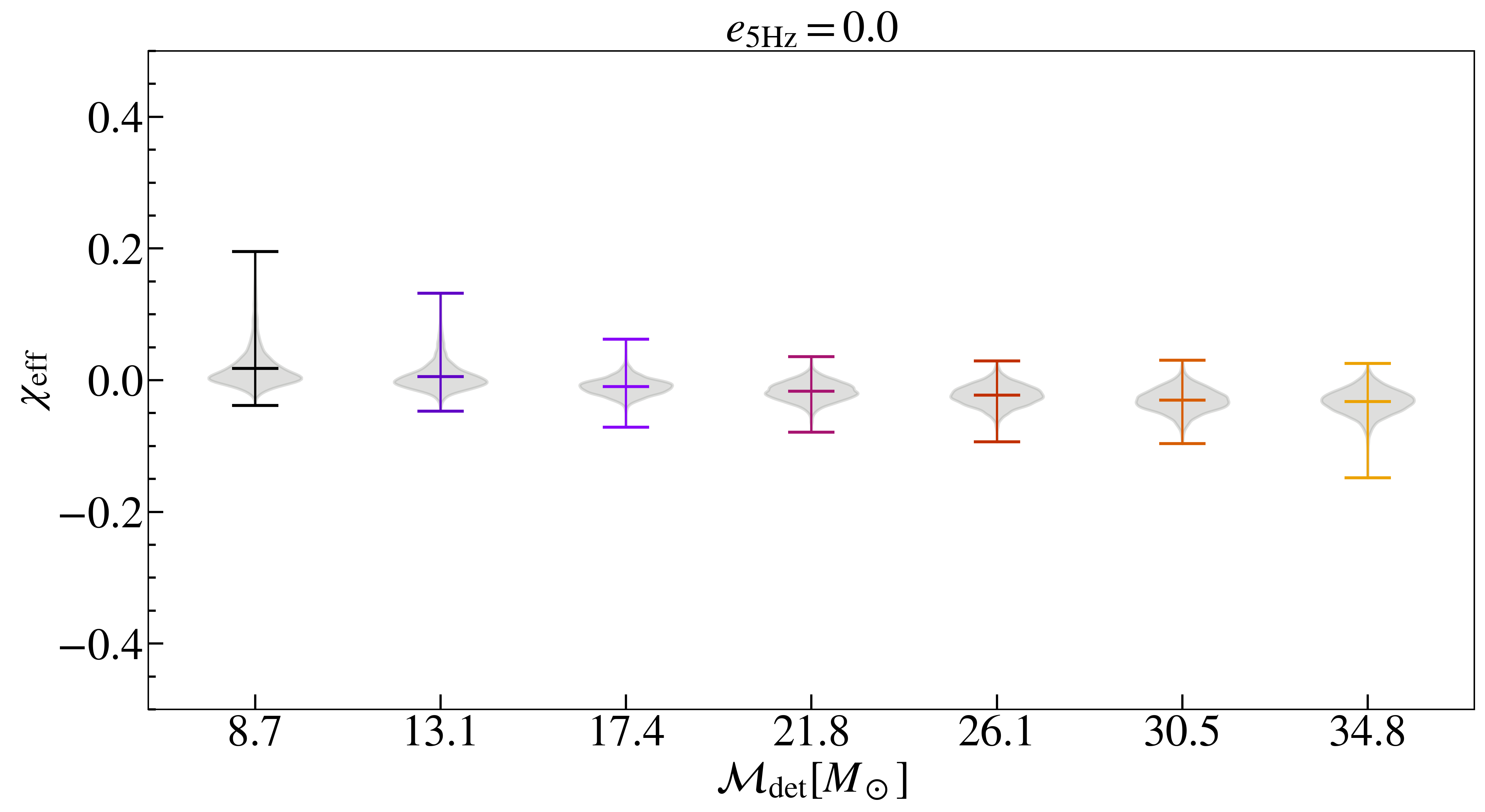}
    \includegraphics[width=0.45\textwidth]{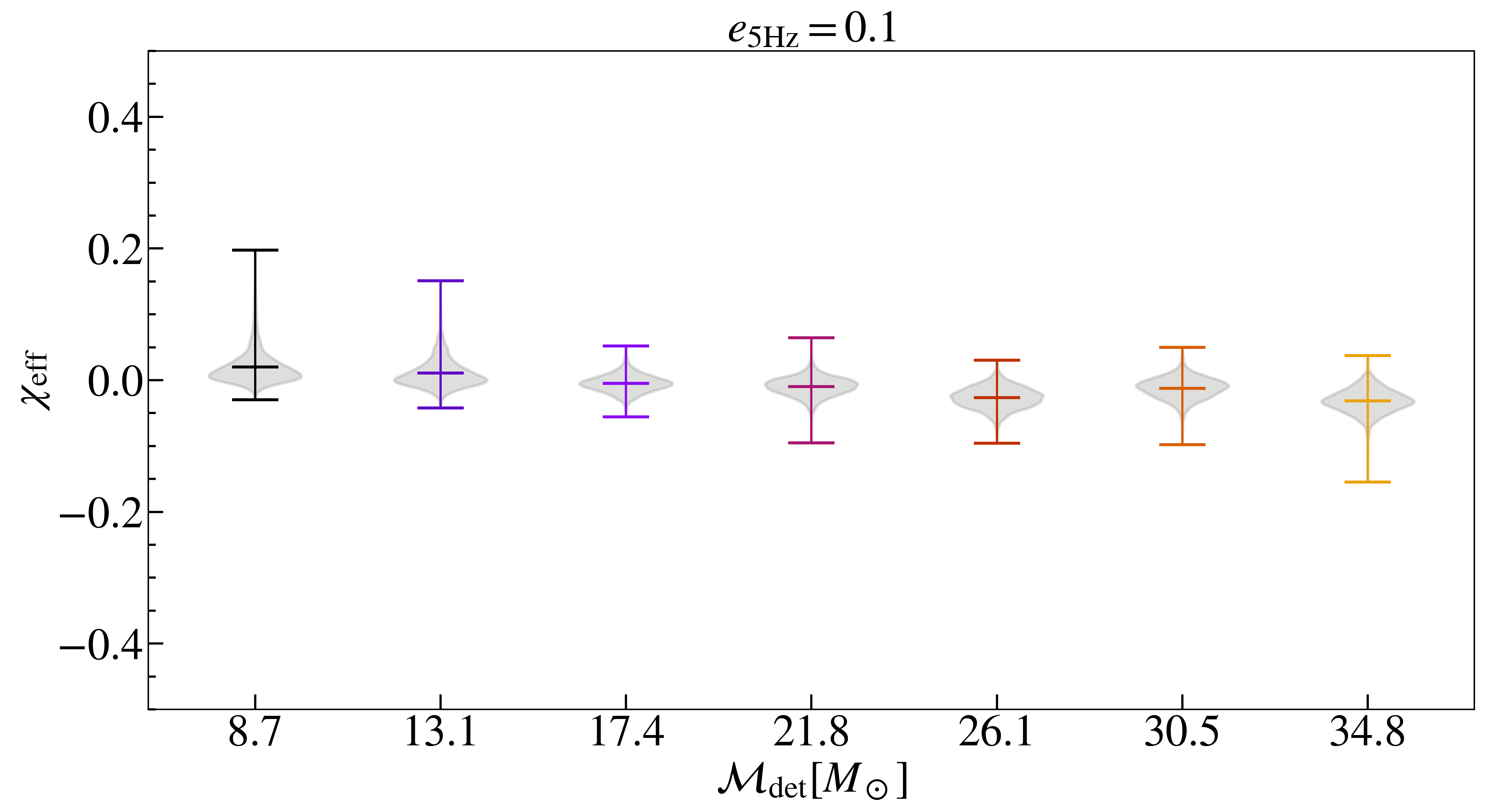}
    \includegraphics[width=0.45\textwidth]{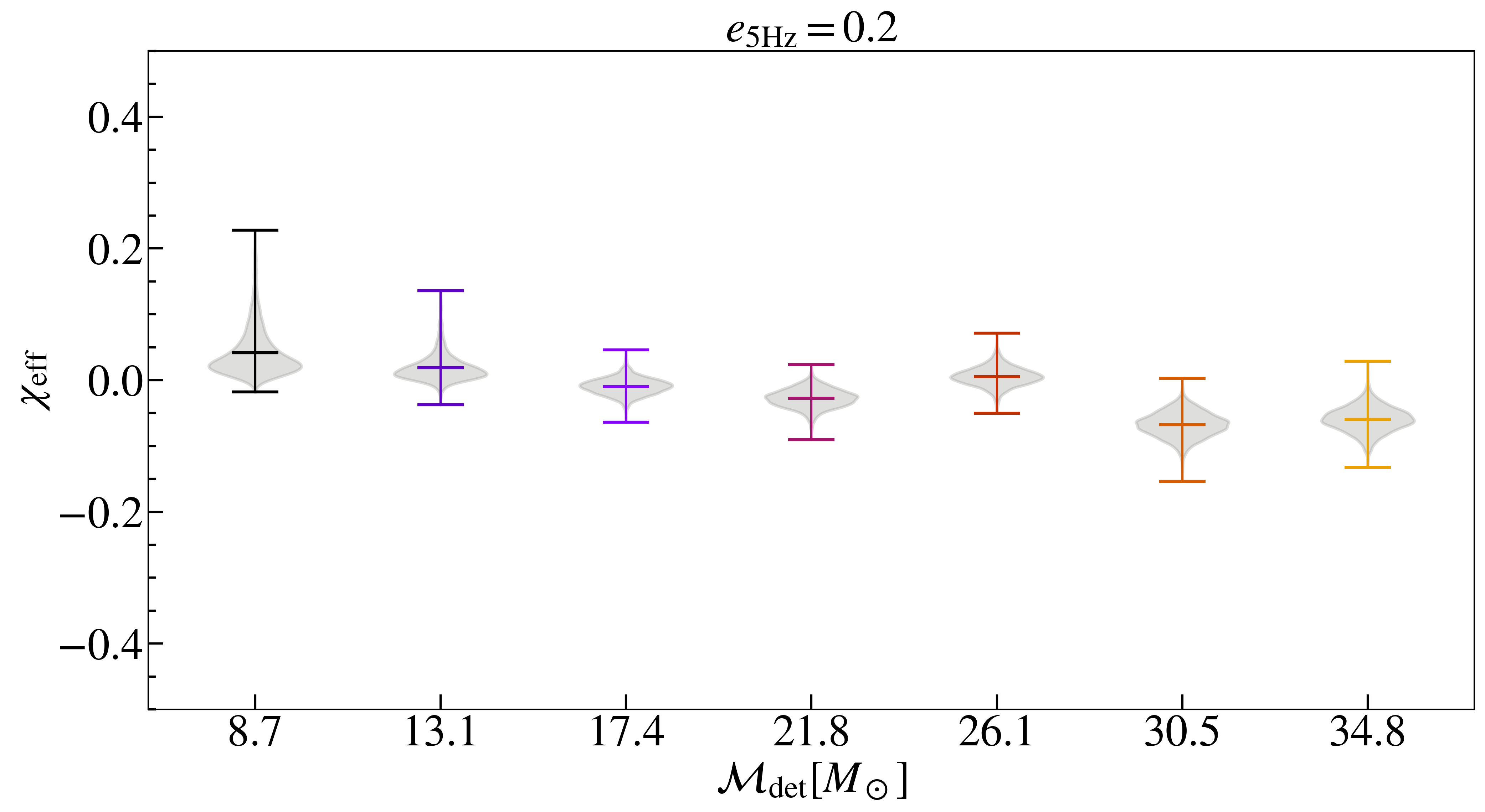}
    \includegraphics[width=0.45\textwidth]{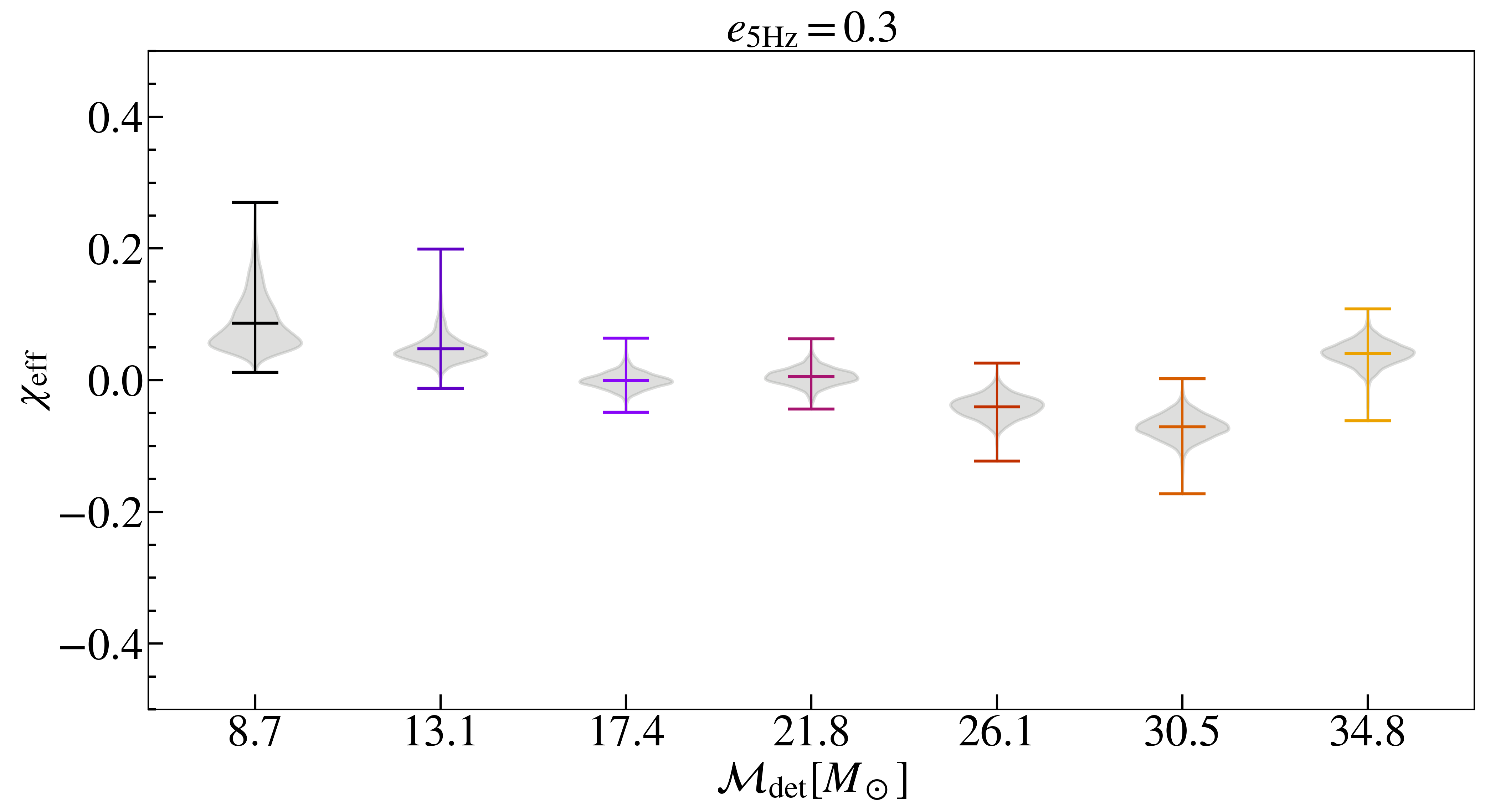}
    \includegraphics[width=0.45\textwidth]{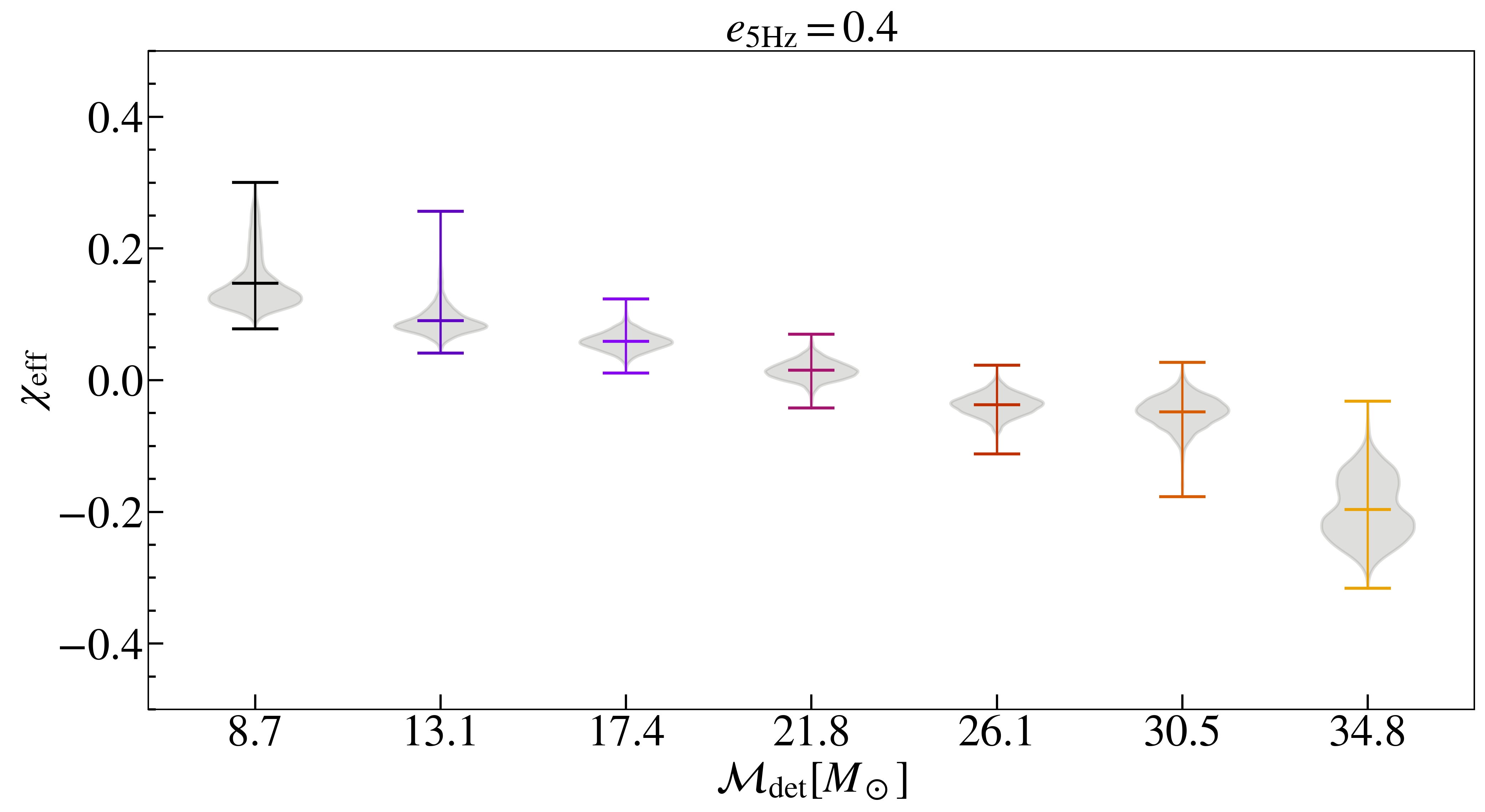}
    \includegraphics[width=0.45\textwidth]{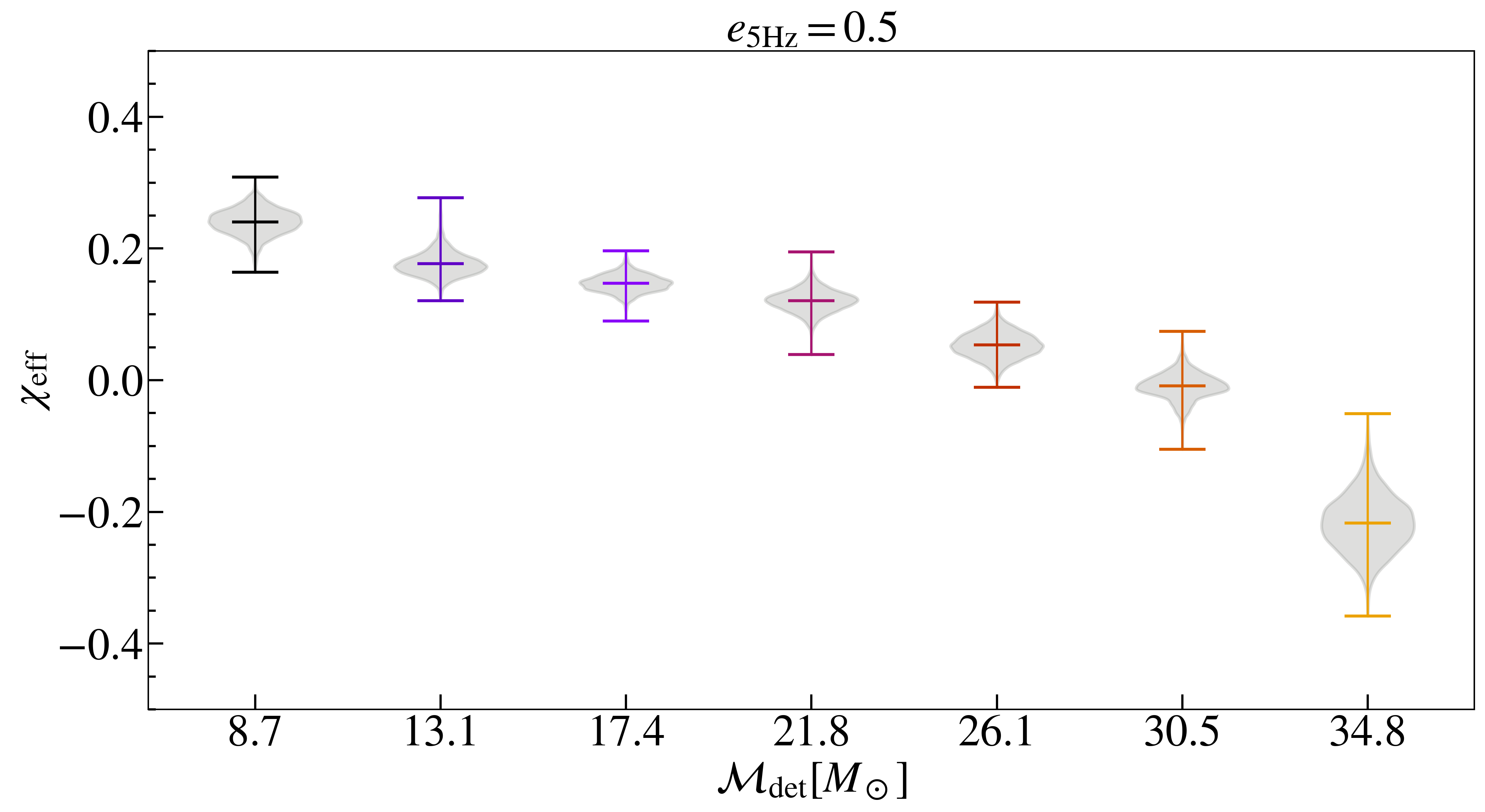}
    
    \caption{The $\chi_{\mathrm{eff}}$ posterior violin plots for varying injected $\mathcal{M}$ and initial eccentricities. The colored lines inside the shaded posteriors indicate 90\% credible interval. Posterior distributions show zero effective spins for all cases except for $e \geq 0.4$ where there are deviations from zero.}
    \label{fig:chieff_violin}
\end{figure*}
\end{document}